\pdfoutput=1
\documentclass[PAPER, atlasdraft=false, UKenglish, cernpreprint, texmf, orcidlogo]{atlasdoc}
\usepackage{atlaspackage}
\usepackage{atlasbiblatex}

\usepackage{atlasphysics}
\usepackage{lipsum}

\graphicspath{{logos/}{figures_paper/}}

\usepackage{ANA-HIGP-2025-09-PAPER-defs}

\AtlasTitle{Measurements of the Higgs boson production, fiducial and differential cross-sections in the four lepton decay channel using 164 fb$^{-1}$ of data collected at $\sqrt{s}$ = 13.6 TeV with the ATLAS detector}

\AtlasAbstract{%
Inclusive, differential and production-mode cross-section measurements of the Higgs boson are performed in the $H \to ZZ^{*} \to 4\ell$ decay channel. The analysis uses proton--proton collision data produced at the Large Hadron Collider at a centre-of-mass energy of $\sqrt{s}=13.6$~TeV and recorded with the ATLAS detector, corresponding to an integrated luminosity of 164~fb$^{-1}$. The inclusive fiducial cross-section for the $H \to ZZ^{*} \to 4\ell$ process is measured to be $\sigma_{\textrm{fid}} = 3.65^{+0.35}_{-0.33}~\textrm{fb}$, in agreement with the Standard Model prediction of $\sigma_{\textrm{fid}}^{\textrm{SM}} =  3.68 \pm 0.17~\textrm{fb}$. Differential cross-sections are measured as a function of key kinematic observables of the Higgs boson and the four-lepton final state. Cross-sections are measured for the main production-modes in several exclusive regions of the Higgs boson production phase space and combined to measure an overall Higgs boson signal strength, defined as the measured cross-section normalised to the SM prediction, of $\mu = 0.99 \pm 0.13$.
The results are interpreted in terms of modifications of Higgs boson couplings using the $\kappa$ framework, within the Standard Model Effective Field Theory, and in scenarios probing the Higgs boson self-coupling. All the results are consistent with Standard Model expectations.
}

\AtlasRefCode{HIGP-2025-09}
\PreprintIdNumber{CERN-EP-2026-112}

\AtlasJournalRef{JHEP 08 (2026) 052}
\AtlasDOI{10.1007/JHEP08(2026)052}

\AtlasCoverEgroupAnalysisTeam{atlas-higp-2025-09-analysis-team@cern.ch}

\hypersetup{pdftitle={ATLAS document},pdfauthor={The ATLAS Collaboration}}

\begin{document}

\maketitle

\section{Introduction}
\label{sec:intro}

%
%

The observation of the Higgs boson by the ATLAS and CMS Collaborations~\cite{HIGG-2012-27,CMS-HIG-12-028} using proton--proton ($pp$) collisions at the Large Hadron Collider (LHC) marked a major milestone in the understanding of the electroweak (EW) symmetry-breaking mechanism~\cite{Englert:1964et,Higgs:1964pj,Guralnik:1964eu}. Subsequent studies of the Higgs boson properties have established the observed particle as a scalar boson consistent with the Standard Model (SM) predictions for a Higgs boson with a mass of $125.09 \pm 0.24\GeV$~\cite{HIGG-2014-14}.

The $H \to ZZ^* \to 4\ell$ decay channel, where $\ell = e,\mu$, provides a particularly clear experimental signature, featuring an excellent signal-to-background ratio and full reconstruction of the final state enabled by the high-precision lepton momentum measurements of the ATLAS detector. Background contributions are small and can be precisely estimated by using data-driven techniques. As a result, this channel has played a central role in precision measurements of Higgs boson properties by both the ATLAS and CMS experiments using the full LHC Run~2 integrated luminosity of $140~\mathrm{fb}^{-1}$. At a centre-of-mass energy of $\sqrt{s}=13~\TeV$, measurements of the Higgs boson production and differential cross-sections ~\cite{HIGG-2013-22,HIGG-2018-28,HIGG-2018-29,HIGG-2022-04,CMS-HIG-14-028,CMS-HIG-21-009,CMS-HIG-19-001,CMS-HIG-21-009}, and derived constraints on couplings, as well as its mass~\cite{HIGG-2020-07,HIGG-2022-20,CMS-HIG-21-019}, tensor structure~\cite{HIGG-2018-30, CMS-HIG-19-009}, and width~\cite{HIGP-2024-14,CMS-HIG-21-019}, have been performed. More recently, first measurements at $\sqrt{s}=13.6~\TeV$ have been reported by ATLAS~\cite{HIGG-2022-12} and CMS~\cite{CMS-HIG-24-013}.

This paper presents measurements of Higgs boson production using $H \to ZZ^* \to 4\ell$ decay candidates recorded during Run~3 with the ATLAS experiment in $pp$ collisions at $\sqrt{s}=13.6~\TeV$ between the 2022 and 2024 data-taking periods, corresponding to an integrated luminosity of $\intLum$. Inclusive and differential cross-sections for eight observables sensitive to the production and decay kinematics of the Higgs boson, are measured within a well-defined fiducial phase space. In addition, cross-sections per Higgs boson production-mode are extracted using the Simplified Template cross-section (STXS) framework~\cite{YR4,Berger:2019wnu}.

The cross-sections are determined using binned likelihood fits to the reconstructed four-lepton invariant mass spectrum for the inclusive and differential cross-sections measurements, and are corrected for detector efficiencies and resolution effects through an unfolding procedure. Neural-network-based (NN) discriminant distributions are used for the production-mode cross-sections measurements.

The measured cross-sections are compared with SM predictions and are further interpreted to constrain possible anomalous Higgs boson couplings. In particular, the differential production cross-sections are interpreted in the context of the Standard Model effective field theory (SMEFT), while the cross-sections per production-mode are interpreted in terms of modified Higgs boson couplings to vector bosons. Constraints on the Higgs boson trilinear self-coupling are derived by exploiting both the Higgs boson transverse momentum spectra and the production-mode cross-sections. Furthermore, a statistical combination of Run~2 and Run~3 data is performed in the context of the coupling-modifier interpretation, significantly improving the precision of the constraints.

Compared with previous results in this analysis channel~\cite{HIGG-2018-28,HIGG-2018-29,HIGG-2022-12}, the results presented in this paper benefit from additional data collected in 2023 and 2024. A new transformer-based~\cite{Vaswani:2017lxt} machine learning (ML) algorithm is used to identify jets from the hadronisation of b-quarks, providing substantial improvements in background rejection~\cite{FTAG-2023-05} for measuring the top-quark-associated Higgs production-mode. A data-driven fake-factor method~\cite{EGAM-2019-01} is used to estimate backgrounds with non-prompt leptons, such as those arising from $Z$+jets, $t\bar{t}$, and $WZ$ processes. The double-differential distribution of the transverse momentum of the four-lepton system as function of the invariant mass of the sub-leading dilepton pair is measured for the first time, owing to its high sensitivity to possible beyond-the-Standard-Model (BSM) effects in Higgs boson couplings. A new neural-network architecture is also employed to construct discriminants that enhance the sensitivity to distinguish between the various Higgs boson production-modes. In addition, the azimuthal angle between the decay planes of the two $Z$ bosons, $\phi$, was found to improve sensitivity to SM-BSM interference effects and is reinterpreted in the context of SMEFT for the first time in this channel.

The paper is organised as follows. Section~\ref{sec:detector} provides a brief description of the ATLAS detector. The data used in the analysis, together with the simulated signal and background samples, is described in Section~\ref{sec:simulation}. The event selection for Higgs boson candidates is presented in Section~\ref{sec:selection}, while the estimation of background contributions is discussed in Section~\ref{sec:background}. An overview of the analysis methodology used to measure inclusive and differential cross-sections in a fiducial phase space, and production-mode cross-sections in the STXS framework, is given in Section~\ref{sec:strategy}. The experimental and theoretical systematic uncertainties affecting the measurements are summarised in Section~\ref{sec:systematics}. The cross-section results are presented in Section~\ref{sec:results}, and their interpretation is discussed in Section~\ref{sec:interpretations}. Finally, the conclusions are given in Section~\ref{sec:conclusion}.

%

%

%
%

%

%


%

%

%
\section{ATLAS detector}
\label{sec:detector}

%

%
%

The ATLAS experiment~\cite{PERF-2007-01,GENR-2019-02} at the LHC is a multipurpose particle detector
with a forward--backward symmetric cylindrical geometry and a near \(4\pi\) coverage in
solid angle.\footnote{\AtlasCoordFootnote}
It consists of an inner tracking detector (ID) surrounded by a thin superconducting solenoid
providing a \qty{2}{\tesla} axial magnetic field, electromagnetic and hadronic calorimeters, and a muon spectrometer.
The inner tracking detector covers the pseudorapidity range \(|\eta| < 2.5\).
It consists of silicon pixel, silicon microstrip, and transition radiation tracking (TRT) detectors.
Lead/liquid-argon (LAr) sampling calorimeters provide electromagnetic (EM) energy measurements
with high granularity within the region \(|\eta|< 3.2\).
A steel/scintillator-tile hadronic calorimeter covers the central pseudorapidity range (\(|\eta| < 1.7\)).
The endcap and forward regions are instrumented with LAr calorimeters
for EM and hadronic energy measurements up to \(|\eta| = 4.9\).
The muon spectrometer (MS) comprises separate trigger and
high-precision tracking chambers measuring the deflection of muons in a magnetic field generated by the superconducting air-core toroidal magnets.
The field integral of the toroids ranges between \num{2.0} and \qty{6.0}{\tesla\metre}
across most of the detector.
Three layers of precision chambers, each consisting of layers of monitored drift tubes, cover the region \(|\eta| < 2.7\),
except in the innermost layer of the end-cap region, where layers of small-strip thin-gap chambers and Micromegas chambers both provide precision tracking in the region \(1.3 < |\eta| < 2.7\).
The muon trigger system covers the range \(|\eta| < 2.4\) with resistive-plate chambers in the barrel, thin-gap chambers in the endcap regions, and the aforementioned small-strip thin-gap chambers and Micromegas chambers in the innermost layer of the endcap.
The luminosity is measured mainly by the LUCID--2 detector which is located close to the beampipe.
A two-level trigger system is used to select events~\cite{TRIG-2022-01}.
The first-level trigger is implemented in hardware and uses a subset of the detector information
to accept events at a rate close to \qty{100}{\kHz}.
This is followed by a software-based trigger that
reduces the accepted rate of complete events to \qty{3}{\kHz} on average,
depending on the data-taking conditions.
A software suite~\cite{SOFT-2022-02} is used in data simulation, in the reconstruction
and analysis of real and simulated data, in detector operations, and in the trigger and data acquisition
systems of the experiment.


%
%
\section{Signal and background simulation}
\label{sec:simulation}


\subsection{Data}
\label{subsec:data}

The measurements presented in this paper are performed using \pp collision data recorded with the ATLAS detector between 2022 and 2024, corresponding to an integrated luminosity of $\intLum$. Data are selected only if all detector components relevant for each analysis channel are in good operating condition. The uncertainty in the integrated luminosity is 1.9\%~\cite{ATL-DAPR-PUB-2025-001}, following the methodology described in Ref.~\cite{DAPR-2021-01}.
The average number of interactions per bunch crossing (pile-up), averaged over all colliding bunch pairs, is about 42 in 2022, increases to about 51 in 2023, and reaches approximately 57 in 2024.

Events used in the measurement are recorded using a combination of single-lepton, dilepton, and trilepton triggers~\cite{TRIG-2018-01,TRIG-2018-05}. The electron triggers used for the 2023 and 2024 data-taking periods exploit the new electromagnetic feature extractors deployed for LHC Run~3~\cite{TRIG-2022-01}. The transverse momentum thresholds of the single-lepton triggers are 24~\GeV\ for muons and 26~\GeV\ for electrons. The dielectron triggers require a minimum transverse momentum of 17~\GeV\ per electron, while the dimuon triggers require at least 14~\GeV\ per muon. Mixed lepton flavour $e\mu$ triggers combine these thresholds, and trilepton triggers allow for lower thresholds in exchange for a higher lepton multiplicity.
For simulated Higgs boson candidates produced via the gluon--gluon fusion process, the resulting total trigger efficiency is 98.1\% for signal events satisfying the final selection described in Section~\ref{sec:selection}.

\subsection{Simulated event samples}
\label{sec:simulation:samples}

This analysis uses simulated event samples to model the Higgs boson signal and all relevant background processes, and to derive corrections for detector effects.

\subsubsection{Signal modelling}
\label{sec:simulation:signal}

Higgs boson production and decay are simulated for the primary production-modes: gluon--gluon fusion (\ggF), vector-boson fusion (\VBF), associated production with a vector boson (\VH, where $V = W,Z$), top- or bottom-associated production\footnote{\ttH and \bbH are also denoted by $ttH$ and $bbH$ in the figures.} (\ttH and \bbH), and associated production with a single top quark ($t$-channel for $\tH$ and $\tHW$)\footnote{The $t$-channel processes $\tH$ and $\tHW$ are denoted by $\tH$ in the figures.}. Hard-scattering matrix-elements (MEs) for the nominal signal samples are generated with the \POWHEGBOX[v2]~\cite{Alioli:2010xd,Nason:2004rx,Frixione:2007vw} Monte Carlo (MC) generator. Next-to-leading-order (NLO) accuracy in quantum chromodynamics (QCD) is achieved for the \ggF and \VH processes with up to one additional parton by employing the \MINLO~\cite{Hamilton:2012np,Campbell:2012am,Hamilton:2012rf} approach. In a subsequent step, the \ggF simulation is improved to next-to-next-to-leading-order (NNLO) \cite{,Hamilton:2013fea,Hamilton:2015nsa} accuracy for inclusive observables by reweighting the Higgs boson rapidity spectrum in \textsc{Hj}-\MINLO~\cite{Hamilton:2012np,Campbell:2012am,Hamilton:2012rf} to that obtained with HNNLO~\cite{Catani:2007vq}. The \VBF and \ttH processes are simulated at NLO accuracy in QCD, while the \bbH sample is generated in the four-flavour scheme. The loop-induced $\ggZH$ process is generated at leading-order in QCD. $\tH$ and $\tHW$ processes are generated at NLO using the \MGNLO generator~\cite{Alwall:2014hca}. All signal samples are generated assuming a Higgs boson mass of 125~\GeV and a Higgs boson width of 4.07~\MeV. A shift of 90~\MeV\ is applied to the signal parameterisation used in the fit to the data to match the measured Higgs boson mass of $125.09 \pm 0.24\GeV$~\cite{HIGG-2014-14}.

The signal events are then interfaced to a parton-shower generator for showering, hadronisation, modelling of the underlying event and decay of the Higgs boson. All signal samples use the \PDFforLHC[21] Parton Density Function (PDF) set~\cite{Ball:2022oua} for the ME calculations. The nominal signal samples are interfaced to \PYTHIA[8.309] to simulate the $H\to ZZ^*\to4\ell$ decays and use the A14 tune for the parton shower~\cite{ATL-PHYS-PUB-2014-021} with the \NNPDF[2.3lo] PDF set~\cite{Ball:2012cx}.

Alternative signal samples are produced to assess uncertainties related to parton-shower and hadronisation modelling. These are obtained by interfacing the same \POWHEG{} events to \HERWIG[7] ~\cite{Bellm:2017jjp} with the H7.1 event tune, using the same PDF setup for the ME calculation. In addition, a simulation of the \ggF process is generated using the \MGNLO generator and showered with \PYTHIA[8.312]~\cite{Bierlich:2022pfr} using the A14 tune with the \NNPDF[2.3lo] PDF set. This sample is accurate to NLO in QCD for up to two additional partons merged using the \FXFX~\cite{Frederix:2012ps,Alwall:2014hca} merging scheme.

The decays of bottom and charm hadrons are simulated using \EVTGEN~\cite{Lange:2001uf}.

\subsubsection{Signal normalisation and branching ratios}
\label{sec:simulation:signal:norm}

The signal samples are normalised to state-of-the-art theoretical cross-section predictions~\cite{deFlorian:2016spz} interpolated to a centre-of-mass energy of 13.6~\TeV~\cite{karlberg2024adinterimrecommendationshiggs}. The \ggF sample is normalised to a N$^{3}$LO QCD calculation including NLO EW corrections~\cite{Aglietti:2004nj,Actis:2008ug,Actis:2008ts,Anastasiou:2008tj,Pak:2009dg,Harlander:2009bw,Harlander:2009mq,Harlander:2009my,Anastasiou:2015ema,Anastasiou:2016cez,Dulat:2018rbf,Bonetti:2018ukf}. The \VBF and \VH samples are normalised to an approximate NNLO QCD cross-section with NLO EW corrections~\cite{Ciccolini:2007jr,Ciccolini:2007ec,Bolzoni:2010xr, Ciccolini:2003jy,Brein:2003wg,Brein:2011vx,Altenkamp:2012sx,Denner:2014cla,Brein:2012ne,Harlander:2014wda,Harlander:2018yio}.
The \bbH contribution is normalised using a calculation that combines NLO contributions with the resummation of collinear logarithms up to NNLO~\cite{Bonvini:2015pxa, Bonvini:2016fgf}, and is cross-checked using an alternative matching scheme~\cite{Forte:2015hba,Forte:2016sja}.
The \ttH cross-section is taken from a calculation accurate to NLO in QCD with NLO EW corrections~\cite{Beenakker:2002nc,Dawson:2003zu,Yu:2014cka,Frixione:2015zaa}. $\tH$ and $\tHW$ are normalized to cross-sections that are accurate to NLO in QCD~\cite{Demartin:2015uha, Demartin:2016axk}.

The Standard Model prediction for the $H \to ZZ^* \to 4\ell~(\ell = e,\mu)$ branching ratio is obtained from \progname{Prophecy4f}~\cite{Bredenstein:2006rh,Bredenstein:2006nk}. This prediction includes the full NLO EW corrections and interference effects that lead to a branching ratio approximately 10\% higher for same-flavour final states ($4\mu$ and $4e$) than for different-flavour final states ($2e2\mu$ and $2\mu2e$). The predicted SM branching ratio is $(1.24 \pm 0.03)\times10^{-4}$ for a Higgs boson mass of 125~\GeV.

\subsubsection{Background modelling}
\label{sec:simulation:backgrounds}

The dominant background contribution arises from continuum $(Z^{(*)}/\gamma^{*})(Z^{(*)}/\gamma^{*})$ production, and is denoted by $\zzstar$. The quark-initiated process ($qqZZ^*$) is the dominant component, while the corresponding gluon-initiated process ($ggZZ^*$) provides a smaller contribution of the order of 10\%. Additional background contributions arise from events with non-prompt or misidentified leptons, including those from $Z$ boson production in association with jets ($Z$+jets), top-quark pair-production ($t\bar{t}$), and $WZ$ production. Additional minor backgrounds include $t\bar{t}+\ell\ell$ and triboson ($WWZ$, $WZZ$, $ZZZ$) production.

\paragraph*{Continuum $\zzstar \to4\ell$ production.}
The nominal $qqZZ^*$ sample is simulated using the \SHERPA[2.2.14]~\cite{Bothmann:2019yzt} generator, including off-shell effects. Matrix elements for fully leptonic final states are calculated at NLO accuracy in QCD for up to one additional parton and at LO accuracy for up to three additional partons. These matrix elements are matched and merged with the \SHERPA parton shower, based on Catani--Seymour dipole factorisation~\cite{Gleisberg:2008fv,Schumann:2007mg}, using the MEPS@NLO prescription~\cite{Hoeche:2011fd,Hoeche:2012yf,Catani:2001cc,Hoeche:2009rj}. Virtual QCD corrections are provided by the \OPENLOOPS library~\cite{Buccioni:2019sur,Cascioli:2011va,Denner:2016kdg}. The \NNPDF[3.0nnlo] PDF set~\cite{Ball:2014uwa} is used, together with the dedicated tune.%

An alternative $qqZZ^*$ sample is simulated using the \MGNLO generator with the \NNPDF[3.0nnlo] PDF set, accurate to NLO in QCD for zero and one additional parton merged using the \FXFX scheme. In this case, parton showering, hadronisation, and underlying-event modelling are performed with \PYTHIA[8.308] using the A14 tune with the \NNPDF[2.3lo] PDF set.

The loop-induced $ggZZ^*$ contribution is simulated at NLO accuracy in QCD matched to a parton shower using the \POWHEG{} approach~\cite{Alioli_2021} with the \NNPDF[3.0nlo] PDF set. The NLO calculation includes two-loop virtual corrections and loop-squared real contributions, with heavy-quark mass effects included in the virtual corrections. The real matrix elements retain the full top-quark mass dependence~\cite{Alioli_2021}.

Production of \zzstar\ via electroweak vector-boson-scattering-like topologies is simulated at LO using \SHERPA[2.2.14] with \NNPDF[3.0nnlo] PDF set, with Higgs-mediated contributions removed at generation level.

Since the overall normalisation of the \zzstar\ background is derived from collision data, the MC predicted cross-sections are used only to determine the relative fraction of the three different processes.

\paragraph*{top-quark-associated and triboson backgrounds.}
Minor background contributions include top-quark-associated multilepton production and triboson processes. The production of $tWZ$, $tZq$, $t\bar{t}Z$, and $t\bar{t}VV$ is modelled using the \MGNLO~\cite{Alwall:2014hca} generator at NLO accuracy with the \NNPDF[3.0nlo] PDF set. The events are interfaced to \PYTHIA[8.230] using the A14 tune and the \NNPDF[2.3lo] PDF set. The $t\bar{t}W$ process is simulated using \SHERPA[2.2.14] with \NNPDF[3.0nnlo] PDF set. The combination of these processes is referred to as \tXX.
Triboson processes ($WWW$, $WWZ$, $WZZ$, $ZZZ$) are simulated using \SHERPA[2.2.16], with matrix elements evaluated at NLO accuracy in QCD for up to one additional jet and at LO accuracy for up to three additional jets, merged with the \SHERPA parton shower.

\paragraph*{Samples for non-prompt lepton background estimates.}
Simulated samples of $Z$+jets, $t\bar{t}$, and $WZ$ production are used exclusively to support the data-driven estimate of backgrounds arising from non-prompt or misidentified leptons. The $Z$+jets samples are generated with \SHERPA[2.2.14] following the setup described in Ref~\cite{PMGR-2021-01}. These samples use the \NNPDF[3.0nnlo] PDF set and are normalised to a NNLO QCD prediction~\cite{Anastasiou:2003ds}.

The $t\bar{t}$ process is simulated using the \POWHEGBOX[v2] generator at NLO accuracy with the \NNPDF[3.0nlo] PDF set and the \hdamp\footnote{The \hdamp parameter is a resummation damping factor and one of the parameters that controls the matching of \POWHEG matrix elements to the parton shower and thus effectively regulates the high-\pt radiation against which the \ttbar system recoils.} parameter set to $1.5\,m_{\text{top}}$~\cite{ATL-PHYS-PUB-2016-020}. Events are interfaced to \PYTHIA[8.230] using the A14 tune and the \NNPDF[2.3lo] PDF set. The decays of bottom and charm hadrons are simulated using \EVTGEN[2.1.1]~\cite{Lange:2001uf}. The $WZ$ process is modelled using the \SHERPA[2.2.16] generator with the same setup as used for the $qqZZ^*$ sample.

\subsubsection{Detector simulation and pile-up modelling}
\label{sec:simulation:detector}

All generated signal and background events are passed through a full simulation of the ATLAS detector based on \GEANT~\cite{Agostinelli:2002hh,SOFT-2010-01}. Scale factors are applied to the simulation to match the performance of algorithms used to reconstruct physics objects to the one observed in data.

The effects of multiple interactions in the same and neighbouring bunch crossings (pile-up) are modelled by overlaying simulated hard-scattering events with inelastic $pp$ collisions generated using a mixture of \EPOS[2.0.1.4]~\cite{Werner:2005jf} and \PYTHIA[8.308]~\cite{Bierlich:2022pfr}. The \EPOS events are generated with the EPOS LHC tune~\cite{Pierog:2013ria}, while the \PYTHIA events use the A3 tune~\cite{ATL-PHYS-PUB-2016-017} and the \NNPDF[2.3lo] PDF set. The \PYTHIA pile-up events include either a high-\pt jet, a prompt photon, or a lepton from a $b$-hadron decay, whereas \EPOS is used to model the remaining pile-up activity. The individual samples are first reweighted to ensure a smooth transition across jet \pt, and the combined pile-up distribution is subsequently reweighted to match the distribution of the average number of interactions per bunch crossing observed in data.


%

%
\section{Event selection}
\label{sec:selection}

%
%
%

Higgs boson candidates are formed by selecting quadruplets of reconstructed and identified charged leptons (electrons and muons). The selection procedure follows that of Ref.~\cite{HIGG-2022-12} and is briefly summarised below.

\subsection{Event reconstruction}
Event reconstruction begins with the formation of tracks in the ID~\cite{IDTR-2022-04} from measurements in the silicon detectors and the TRT, and in the MS from measurements in the muon chambers. ID tracks are reconstructed for $\pt > 500~\MeV$ and are used to form vertex candidates along the beam line~\cite{ATL-PHYS-PUB-2019-015,PERF-2015-01}. Candidate events are required to contain at least one vertex reconstructed from at least two ID tracks.
Among all reconstructed vertices in an event, the primary interaction vertex is defined as the one with the largest summed $\pt^2$ of its associated tracks.

Electron candidates are reconstructed by matching electromagnetic calorimeter clusters to ID tracks~\cite{EGAM-2018-01}. A refit of the ID track, allowing for bremsstrahlung energy losses, is performed to improve the impact-parameter\footnote{The transverse impact parameter $d_0$ of a charged-particle track is defined in the transverse plane as the distance of closest approach to the beam spot. The longitudinal impact parameter $z_0$ is defined as the distance along the beam axis between this point and the primary vertex.} resolution. Electron identification is based on a likelihood discriminant~\cite{EGAM-2021-01}, using \textit{loose} criteria that combine track properties, electromagnetic shower-shape variables, and the quality of the track--cluster matching. Electrons are required to satisfy $\pt > 7~\GeV$ and $|\eta|<2.47$. The electron energy is calibrated following the procedure described in Ref.~\cite{EGAM-2018-01}. Photons are reconstructed from EM calorimeter signals exceeding noise
thresholds using dynamic topological clustering, with additional clusters included for photon conversion
products when applicable~\cite{EGAM-2018-01}. The photon candidates are calibrated~\cite{EGAM-2021-02} and identified by an algorithm that relies on the shape of the EM showers reconstructed in the calorimeter~\cite{EGAM-2021-01}.

Muon candidates are reconstructed in the pseudorapidity range $|\eta|<2.5$ by matching tracks reconstructed in the MS to ID tracks. In the region $|\eta|<0.1$, where the MS is only partially instrumented, muons are additionally reconstructed from ID tracks matched either to partial MS track segments or to calorimeter energy deposits consistent with a minimally ionising particle (calorimeter-tagged muons)~\cite{MUON-2018-03}.
Muons are required to have $\pt > 5~\GeV$, except for calorimeter-tagged muons, which must satisfy $\pt > 15~\GeV$. A set of \textit{Loose} identification criteria is applied to suppress muons from hadron decays and misidentified tracks~\cite{MUON-2018-03}. The muon momentum is calibrated using the procedure described in Ref.~\cite{MUON-2022-01}.

Electron and muon candidates are associated with the primary vertex using their impact parameters.
Muon candidates with $|d_0|>1~\text{mm}$ are rejected to suppress cosmic-ray background. Both electron and muon candidates are required to satisfy $|z_0 \sin\theta| < 0.5~\text{mm}$.

Jet candidates are reconstructed using the particle-flow (PFlow) algorithm~\cite{PERF-2015-09}, which combines calorimeter energy deposits and ID tracks into particle-flow objects that are clustered into jets using the using the \antikt~\cite{Cacciari:2008gp} jet algorithm with radius parameter $R=0.4$. Jets are calibrated using simulation with corrections obtained by using \textit{in situ} techniques in data~\cite{JETM-2018-05}.
After calibration, jets are required to satisfy $\pt > 30~\GeV$ and $|\eta|<4.5$. To suppress jets originating from pile-up interactions, a neural-network-based algorithm~\cite{PERF-2014-03,ATL-PHYS-PUB-2019-026} is used to associate jets with $\pt<60~\GeV$ to the primary vertex.

Jets containing $b$-hadrons are identified ($b$-tagged) using the GN2 transformer-based neural network algorithm~\cite{FTAG-2023-05}. Compared with previous algorithms, GN2 does not rely on lower-level flavour-tagging algorithms and uses a single training to optimise different downstream tasks, including jet identification. It demonstrates improved background rejection at a given efficiency compared with the previous tagger~\cite{FTAG-2023-05}. Two $b$-tagging working points are used, one with a b-tagging efficiency of approximately $85\%$ and a tighter working point with an efficiency of about $65\%$. The corresponding light-jet rejections are approximately $6\times 10^{1}$ and $3\times 10^{3}$, respectively.

Ambiguities are resolved if electron, muon, or jet candidates overlap in geometry or share the same detector information.
If the two calorimeter energy clusters from the two electron candidates overlap, the electron with the higher \pt is retained. If a reconstructed electron and muon share the same ID track, the muon is rejected if it is calorimeter-tagged; otherwise the electron is rejected. Reconstructed jets geometrically overlapping in a cone of radial size $\Delta R$ = 0.1 (0.2) with a muon (an electron) are also removed.

The missing transverse momentum vector, $\vec{p}_{\text{T}}^{\text{miss}}$, is defined as the negative vector sum of the transverse momenta of all the identified and calibrated leptons and jets and the remaining unclustered energy, where the latter is estimated from low-$\pt$ tracks associated with the primary vertex but not assigned to any lepton, photon, hadronically decaying $\tau$-lepton or jet candidate~\cite{JETM-2020-03}. The missing transverse momentum ($\MET$) is defined as the magnitude of $\vec{p}_{\text{T}}^{\text{miss}}$.

\subsection{Higgs boson candidate formation}
\label{subset:hcand}
Higgs boson candidates are formed from pairs of same-flavour, opposite-charge (SFOC) electron or muon pairs. Each such combination, referred to as a quadruplet, is required to include the leptons that triggered the event, as described in Section~\ref{subsec:data}. The three highest-$\pt$ leptons in the quadruplet are required to satisfy $\pt > 20$, 15, and 10~\GeV, respectively.

Further requirements are imposed on the invariant masses of the SFOC lepton pairs. For each quadruplet, the lepton pair with invariant mass $m_{12}$ closest to the $Z$-boson mass~\cite{Zyla:2020zbs} is defined as the leading pair, while the remaining SFOC pair, with invariant mass $m_{34}$ second-closest to the $Z$-boson mass, is defined as the subleading pair. On the selected SFOC lepton pairs, additional requirements are applied on the angular separation between leptons, a $J/\psi$ veto, the lepton transverse impact-parameter significance ($|d_{0}|/\sigma(d_{0})$), and the quality of the four-lepton vertex. The four leptons are required to be fitted to a common vertex and to be compatible with originating from the same interaction point. In addition, track- and calorimeter-based isolation requirements are applied on the lepton candidates in the pairs to suppress the non-prompt contributions. Those requirements are detailed in Ref.~\cite{HIGG-2022-12}. The full set of requirements is summarised in Table~\ref{tab:evtSel}.

If more than one quadruplet with different lepton-flavour combination in an event satisfies all above selection criteria, the candidates containing the leading lepton pair closest to the $Z$-boson mass are selected. If more than one quadruplet with the same leading lepton pair but different subleading pairs satisfies this criterion, the quadruplet with the lepton-flavour combination corresponding to the highest total expected selection efficiency, as estimated from simulation, is chosen. The expected efficiencies are approximately $28\%$, $23\%$, $14\%$, and $12\%$ for the $4\mu$, $2e2\mu$, $2\mu2e$, and $4e$ final states, respectively. The $2e2\mu$ and $2\mu2e$ final states are distinguished by the flavour of the leading lepton pair.

If after this selection, an additional lepton, not used in the quadruplet, with $\pt>12 \GeV$ is present in the event, it may originate from leptonic decays of vector bosons and top quarks in \VH and \ttH production-modes. To improve the mass resolution in this scenario, a decay-side ME discriminant is calculated at leading-order using \progname{MadGraph5} for each quadruplet in the event, with the reconstructed lepton four-momenta used as inputs~\cite{HIGG-2018-28}, and the quadruplet with the highest ME value is selected. The improvement in correctly identifying the leptons in the quadruplet is about 50\% for these cases.

The selected Higgs boson candidates are further refined by accounting for final-state radiation (FSR) photons, using the procedure described in Refs.~\cite{HIGG-2018-28,HIGG-2018-29}. For electrons, the reconstruction already includes the recovery of nearby FSR photons within $\Delta R<0.15$~\cite{ATL-PHYS-PUB-2017-022}; therefore, only the recovery of far FSR photons is applied to electron candidates. For muons, both nearby and far FSR photons are explicitly recovered.

\begin{table}[!ht]
\centering
\caption{Summary of the event selection requirements for leptons and Higgs boson candidates. SFOC lepton pairs are same-flavour opposite-charge lepton pairs. For the subleading lepton-pair mass requirement, $m_{\text{threshold}}$ is 12~\GeV\ for $m_{4\ell}<140$~\GeV, and increases linearly to 50~\GeV\ at $m_{4\ell}=190$~\GeV, remaining constant after ~\cite{HIGG-2018-28}.}
\label{tab:evtSel}
\vspace{0.2cm}
\begin{tabular}{ll}
\toprule
\multicolumn{2}{c}{\textbf{Leptons and Jets}} \\
\midrule
Muons & $\pt>5$~\GeV, $|\eta|<2.5$  \\
Electrons & $\pt>7$~\GeV, $|\eta|<2.47$  \\
Jets & $\pt > 30$~\gev, $\abseta< 4.5$ \\
\midrule
\multicolumn{2}{c}{\textbf{Lepton selection and pairing}} \\
\midrule
Lepton kinematics & $\pt > 20, 15, 10$~\GeV  \\
Leading pair ($m_{12}$) & SFOC lepton pair with smallest $|m_Z-m_{\ell\ell}|$  \\
Subleading pair ($m_{34}$) & remaining SFOC lepton pair with smallest $|m_Z-m_{\ell\ell}|$  \\
\midrule
\multicolumn{2}{c}{\textbf{Event selection (at most one Higgs boson candidate per event)}} \\
\midrule
Mass requirements & $50~\GeV < m_{12} < 106~\GeV$ and $m_{\text{threshold}} < m_{34} < 115~\GeV$  \\
Lepton separation & $\Delta R(\ell_i,\ell_j) > 0.1$  \\
$J/\psi$ veto & $m(\ell_i,\ell_j) > 5$~\GeV\ for all SFOC lepton pairs  \\
Impact parameter & $|d_{0}|/\sigma(d_{0}) < 5~(3)$ for electrons (muons) \\
Isolation & Track- and calorimeter-based isolation for all four leptons \\
Vertex selection & $\chi^{2}/N_{\mathrm{dof}} < 6~(9)$ for $4\mu$ (other channels) \\
If extra lepton with $\pt > 12$~\GeV & quadruplet with largest matrix-element value \\
\bottomrule
\end{tabular}
\end{table}


%
\section{Background estimation}
\label{sec:background}

%
\subsection{Background process with prompt leptons}

The dominant background contribution arises from continuum $\zzstar$ production initiated by quarks or gluons. Unlike the Higgs boson signal, the $\mfourl$ spectrum of the $\zzstar$ background does not exhibit a resonant structure near the Higgs boson mass, $m_H$. This feature is exploited to constrain the normalisation of the $\zzstar$ background directly from data.

The shapes of the $m_{4\ell}$ templates and the differential distributions for the $\zzstar$ background are taken from the simulation described in Section~\ref{sec:simulation}. The relative contributions of the individual $\zzstar$ sub-processes are fixed to their simulated expectations, while the overall normalisation is left free in the statistical model through one or more scale factors defined relative to the MC-predicted yields. The binning of these scale factors differs between the two measurements and is described below.

The $\zzstar$ normalisation is constrained using signal-depleted sideband regions defined by $105 < \mfourl < 115~\GeV$ and $130 < \mfourl < 160~\GeV$. These regions are representative of the continuum background process under the Higgs boson peak and avoid the resonant $Z\to4\ell$ process before 105 \GeV\ and the fully resonant $ZZ$ production after 160 \GeV.

For the fiducial cross-section measurements, including differential observables, the full fit is performed over the full mass range $105 < \mfourl < 160~\GeV$. For differential measurements, the $\zzstar$ normalisation is determined independently in each bin of the measured observable. To reduce the impact of statistical fluctuations in categories with low expected event yields, the normalisation parameters of neighbouring bins are combined together.
For the production-mode cross-section measurements in the STXS framework, the low and high mass sideband regions are merged together in $\mfourl$ but are further categorised according to the reconstructed jet multiplicity ($N_{\mathrm{jets}} = 0$, $1$, and $\geq 2$) to reduce theoretical assumptions on extrapolating results between different regions. Signal-region categories with the same jet multiplicity are fitted simultaneously with their corresponding sideband regions, using a common $\zzstar$ normalisation parameter for each jet-multiplicity category.

This strategy substantially reduces theoretical uncertainties in the $\zzstar$ background prediction for both measurements, as the statistical model primarily relies on data to constrain the normalisations, leaving only the relative normalisation between the signal and sideband regions and the modelling of the differential shapes dependent on theoretical assumptions.

Top-quark-associated background processes are treated differently in the two measurement strategies. In the fiducial and differential cross-section measurements, their contribution is small and is therefore taken directly from simulation. In the production-mode cross-section measurements analysis, where sensitivity to \ttH production is enhanced, a dedicated control region is included in the fit to constrain the normalisation of these processes. A single normalisation parameter is determined in a simultaneous fit to the \ttH signal regions and the control region, defined following the Run~2 analysis strategy described in Ref.~\cite{HIGG-2018-28}. In this control region, the upper bound of the $\mfourl$ selection is extended to 350~\GeV\ to improve the statistical precision of the constraint, while avoiding the fully resonant top-associated production.

Contributions from triboson production are taken directly from simulation in all regions considered.

\subsection{Background process with non-prompt leptons}
\label{subsec:reducible}

A small contribution to the selected $4\ell$ sample arises from events with two or three prompt leptons accompanied by additional non-prompt leptons from heavy-flavour hadron decays, light-jet misidentification, or photon conversions (including final-state radiation), primarily from $Z$+jets, $\ttbar$, and $WZ$ production. Non-prompt muons typically originate predominantly from heavy-flavour hadron decays, while non-prompt electrons originate from light-jet misidentification and photon conversions.
Such non-prompt leptons are typically produced around or nearby hadronic activity and therefore have a higher probability to fail the isolation and/or impact-parameter requirements used in the signal selection. This background is referred to as \emph{reducible} and is estimated with a data-driven fake-factor method, following approaches used in other four-lepton final-state analyses~\cite{STDM-2021-05,STDM-2020-02}.

The method is based on two nested lepton definitions: \emph{baseline} leptons satisfy the full signal selection described in Section~\ref{sec:selection}, except for the isolation and impact-parameter requirements, while \emph{tight} leptons additionally satisfy these requirements. The fake efficiency $f$ is defined as the probability for a non-prompt (fake) baseline lepton to also satisfy the tight requirements, and is measured in data after subtracting the expected prompt-lepton contamination.
The fake factor (F) is then defined as
\begin{equation}
F = \frac{f}{1-f},
\end{equation}
relating the yield of events with tight fake leptons to the yield of events in which the corresponding leptons fail to satisfy the tight requirements.

The fake factors are measured in two control regions enriched in reducible background: a $Z$+jets-dominated region (CR$_Z$) and a $\ttbar$-dominated region (CR$_\text{top}$). In both regions, a tight lepton pair is required together with an additional baseline lepton. In CR$_Z$, the tight pair is required to be same-flavour opposite-sign with an invariant mass close to $m_Z$, while in CR$_\text{top}$ the tight pair is required to be an $e\mu$ pair and the event must contain at least one $b$-tagged jet using the working point with 85\% efficiency.
Prompt-lepton contamination in the measurement regions is dominated by $WZ$ production. Its contribution is estimated in each CR applying a scale factor (SF) to correct the $WZ$ MC estimate with data. This SF is extracted from a dedicated region (CR$_{WZ}$) that requires a SFOC tight lepton pair with an invariant mass close to $m_Z$ and one extra tight lepton.

The fake factors are parameterised separately for electrons and muons in a three-dimensional map of the lepton $\pt$, $|\eta|$, and jet multiplicity. The final fake efficiency in the signal region (SR) is obtained by combining the measurements from CR$_Z$ and CR$_\text{top}$ with equal weights,
\begin{equation}
f^\text{SR} = 0.5\,f^{\text{CR}_Z} + 0.5\,f^{\text{CR}_\text{top}},
\end{equation}
while for the 0-jet category $f^\text{SR}_{0j}=f^{\text{CR}_Z}_{0j}$, since the CR$_\text{top}$ definition requires the presence of $b$-tagged jets. While the compositions for the origin of the non-prompt lepton in the signal region are more similar to those observed in CR$_\text{top}$, an equal-weighted average is used to ensure a robust estimate as weighting according to the composition increases sensitivity to statistical variation from  CR$_\text{top}$, which has significantly lower event yields than CR$_Z$. However, as described in Section~\ref{sec:systematics}, the full difference between two regions is included as a conservative systematic uncertainty to account for the composition difference. This systematic uncertainty has a negligible impact on the final results.

The reducible-background yield in the signal region is estimated by applying fake-factor event weights to data events that satisfy the nominal quadruplet selection but contain at least one baseline lepton failing to meet the tight requirements, after subtracting residual prompt contributions from simulation.

The procedure is validated in two additional regions that are kinematically close to the signal region but enriched in the reducible background: a same-charge region requiring the leptons of the subleading pair to have the same charge, and an $e\mu$ region requiring one lepton pair to be an $e\mu$ pair, with all other selections identical to the signal region.
To reduce statistical fluctuations in the signal region, a smoothing procedure is applied and the corresponding statistical uncertainty is evaluated using pseudo-datasets.


%

%
\section{Analysis overview}
\label{sec:strategy}

%
%

The analysis performs two complementary measurements of the Higgs boson cross-section in the $H\to ZZ^*\to 4\ell$ decay channel. In the first approach, inclusive and differential cross-sections are measured in a fiducial phase space designed to closely mirror the detector-level selection, thereby minimising extrapolation and reducing the dependence on theoretical assumptions. In this measurement, the relative contributions of the different Higgs boson production-modes are fixed to the Standard Model expectation when deriving acceptance and migration corrections. Signal yields are extracted from maximum-likelihood fits to the reconstructed \mFourL spectrum, and detector effects are corrected using a likelihood-based matrix unfolding procedure~\cite{HIGG-2018-29} to obtain particle-level results that are as model independent as possible and sensitive to deviations from the Standard Model.

The second, complementary approach aims to measure the underlying Higgs boson production mechanisms. Production-mode cross-sections are measured using event categories and multivariate discriminants optimised for separation of \ggF, \VBF, \VH, and \ttH contributions, and are interpreted within the STXS framework. The STXS approach provides a common set of phase-space regions that enables increasingly granular measurements,
enhancing sensitivity to differences between production kinematics and potential BSM effects.

The following sections describe these two strategies in detail, first presenting the fiducial inclusive and differential cross-section measurement and then the production-mode measurement in the STXS framework. Although both approaches provide an inclusive Higgs boson production cross-section, they do so using complementary combinations and are impacted by different model dependences. In the fiducial analysis, the inclusive cross-section is obtained by combining measurements performed separately in the four decay final states, and depends on the modelling of the relative effects between the $4\mu$, $2e2\mu$, $2\mu2e$, and $4e$ channels. In the STXS measurement, the inclusive cross-section is obtained by summing the fitted production-mode contributions across the reconstructed categories, and is sensitive to the modelling of the relative signal composition across jet-multiplicity regions and categories.

\subsection{Inclusive and differential cross-sections}
\label{subsec:diffxs}

\subsubsection{Fiducial phase space and observables}
\label{subsec:fiducial}

The fiducial cross-sections are defined at particle level in a phase space chosen to closely match the detector-level event selection, thereby reducing the size of model-dependent acceptance extrapolations and minimising theoretical and experimental uncertainties associated with the unfolding procedure~\cite{HIGG-2018-29}. The fiducial phase space definition is outlined in Table~\ref{tab:FiducialRegion}. Leptons are defined as final-state electrons and muons that do not originate from hadron or $\tau$-lepton decays, after FSR~\cite{ATL-PHYS-PUB-2015-013}. Leptons are ``dressed'' by adding the four-momenta of photons within a cone of $\Delta R=0.1$ around the lepton direction (excluding photons from hadron decays) to the lepton four-momentum~\cite{HIGG-2018-29} to recover energy lost to FSR. Dressed leptons are required to satisfy $|\eta|<2.7$ and $\pt>5~\GeV$ irrespective of flavour to simplify comparisons with theoretical predictions. Particle-level jets are reconstructed from final state neutral and charged particles using the anti-$k_t$ algorithm with radius parameter $R=0.4$ and are required to satisfy $\pt>30~\GeV$ and $|y|<4.4$. Overlapping jets are removed if $\Delta R(\ell,\mathrm{jet})<0.1$ with any selected lepton~\cite{HIGG-2018-29}. Higgs boson candidates are formed from SFOC lepton pairs using the quadruplet selection described in Section~\ref{subset:hcand}, with the exception of the isolation, impact-parameter, and vertex requirements. Events enter the fiducial phase space if a candidate satisfying the mass and kinematic requirements is reconstructed, including $105 < m_{4\ell} < 160~\GeV$~\cite{HIGG-2018-29}.

\begin{table}[th]
\centering
\caption{Event selection requirements at particle level defining the fiducial phase space for the cross-section measurement. SFOC lepton pairs are same-flavour opposite-charge lepton pairs.
\label{tab:FiducialRegion}}
\vspace{0.2cm}
\begin{tabular}{ll}
\hline\hline
\multicolumn{2}{c}{\textbf{Leptons and jets}} \\
\hline
Leptons & $\pt>5$ \GeV, $|\eta|<2.7$  \\
Jets &  \pt~$>$ 30 \gev, $|y|<$ \,4.4 \\
\hline
\multicolumn{2}{c}{\textbf{Lepton selection and pairing}} \\
\hline
Lepton kinematics & $\pt > 20, 15, 10 $ \GeV  \\
Leading pair ($m_{12}$) & SFOC lepton pair with smallest $|m_Z-m_{\ell\ell}|$  \\
Subleading pair ($m_{34}$) & remaining SFOC lepton pair with  smallest $|m_Z-m_{\ell\ell}|$  \\
\hline
\multicolumn{2}{c}{\textbf{Event selection (at most one quadruplet per event)}} \\
\hline
Mass requirements & 50 \GeV$ < m_{12} < 106$ \GeV \, and  12 \GeV $< m_{34} < 115$ \GeV  \\
Lepton separation & $\Delta R(\ell_{i},\ell_{j}) > 0.1$ \\
Lepton/Jet separation & $\Delta R(\ell_{i}, \textrm{jet}) > 0.1$ \\
$J/\psi$~veto & $m(\ell_{i},\ell_{j}) > 5$ \GeV \,  for all SFOC lepton pairs  \\
If extra lepton with $\pt>12\GeV$&  Quadruplet with  largest matrix element value \\
\hline
Mass window &  105~\GeV $< m_{4\ell} < 160 $ \GeV  \\
\hline\hline
\end{tabular}
\end{table}

Fiducial production cross-sections are measured inclusively and separately for each $H\to ZZ^*\to4\ell$ final state ($4\mu$, $2e2\mu$, $2\mu2e$, $4e$). Differential fiducial cross-sections are measured for eight observables sensitive to the production and decay kinematics of the Higgs boson. The transverse momentum of the four-lepton system, $\ptFourL$, provides a stringent test of perturbative QCD calculations and is sensitive to modifications of Higgs boson production dynamics. The rapidity of the four-lepton system, $|y_{4\ell}|$, probes the parton distribution functions of the colliding protons and is also affected by QCD radiative corrections. The invariant masses of the leading and subleading SFOC lepton pairs, $m_{12}$ and $m_{34}$, are sensitive to higher-order electroweak effects and to the decay kinematics of $H\to ZZ^*\to4\ell$. The decay angle between the two Z bosons, $\phi$, is sensitive to the Higgs boson spin and parity and to BSM contributions in the $HZZ$ vertex. Jet activity is characterised by the jet multiplicity, $\Njets$, which is sensitive to QCD radiation and to the relative contributions of different production mechanisms. For events with at least two jets, the azimuthal separation of the two leading jets, $|\Delta\phi_{jj}|$, provides additional sensitivity to production dynamics through the structure of associated QCD radiation. In addition, a double-differential measurement of $\ptFourL$ versus $m_{34}$ is performed, which enhances sensitivity to BSM couplings effects affecting both the production and decay of the Higgs boson.

\subsubsection{Signal extraction and unfolding}

The fiducial cross-section $\sigma_{j}^{\textrm{fid}}$ in each bin $j$ of a differential distribution (or, for the inclusive measurement, in each decay final state) is obtained from a binned maximum-likelihood fit of signal and background templates to the $m_{4\ell}$ spectrum. %
The fit is performed simultaneously over all the reconstruction bins $i$ of the corresponding observable. The expected number of events in reconstruction bin $i$, expressed as a function of $m_{4\ell}$, is given by
\begin{equation}
\begin{split}
N_i\left( m_{4\ell} \right) = \sum_{j}{  \sigma_{j}^{\textrm{fid}}  \cdot \mathcal{P}_i\left( m_{4\ell} \right) \cdot r_{ij} \cdot \left( 1+f_{i}^{\textrm{nonfid}} \right)  \cdot  \mathcal{L} + N_i^{\textrm{bkg}}\left( m_{4\ell} \right) }
\end{split}
\label{eq:diffxs_fit}
\end{equation}
where $\sigma_{j}^{\textrm{fid}}=\sigma_j\cdot a_j \cdot \mathcal{B}$. Here, $\sigma_j$ is the total production cross-section in particle-level bin $j$, $a_j$ is the particle-level acceptance of the fiducial selection in that bin, and $\mathcal{B}$ is the branching ratio. The integrated luminosity is denoted by $\mathcal{L}$, and $N_i^{\textrm{bkg}}(m_{4\ell})$ is the background contribution in reconstruction bin $i$.
The term $\mathcal{P}_i(m_{4\ell})$ is the signal $m_{4\ell}$ template for reconstruction bin $i$ taken from simulation. The response matrix element $r_{ij}$ corresponds to the probability that an event generated in fiducial bin $j$ is reconstructed in bin $i$. The fiducial leakage factor, $f_{i}^{\textrm{nonfid}}$, accounts for the fraction of events produced outside the fiducial phase space that are still reconstructed within the detector-level signal region. The factors $a_j$, $r_{ij}$, and $f_{i}^{\textrm{nonfid}}$ are derived from simulated signal samples, assuming Standard Model production-mode fractions within their associated theoretical uncertainties. %

As outlined in Section~\ref{sec:background}, the background contribution $N_{i}^{\textrm{bkg}}(m_{4\ell})$ includes the continuum $\zzstar$ component and smaller prompt and non-prompt backgrounds. To minimise theory dependence, the overall $\zzstar$ yield is constrained directly from data by including $m_{4\ell}$ sideband regions in the likelihood fit. The remaining prompt and non-prompt contributions are taken from simulation and the data-driven method respectively.

The binning choice mostly follows the previous measurement~\cite{HIGG-2018-29}, which was designed to ensure an expected statistical significance per bin of more than two standard deviations for the SM signal process and to minimise migrations between bins. These same criteria are used to define the binning of the new observable $\ptFourL$ versus $m_{34}$. In most cases, the bin width is more than twice the experimental resolution. The response matrices for all the variables considered are well-conditioned and therefore no regularisation of the unfolding is applied.

\begin{figure}[!htbp]
\centering
\subfloat[$p_T^{4\ell}$]{\includegraphics[width=0.45\linewidth]{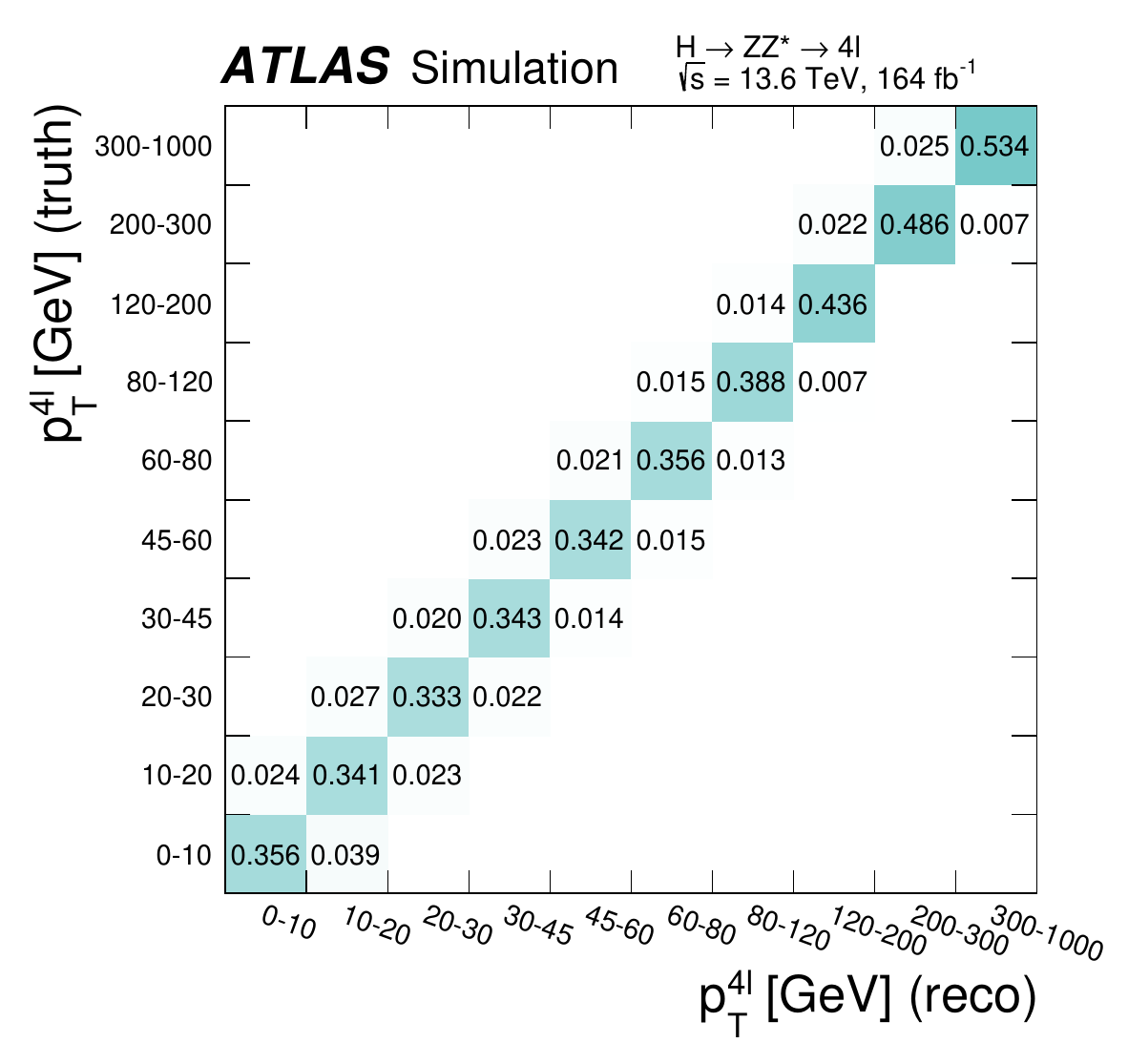}\label{fig:rspmtx_pt}}
\subfloat[$\Njets$]{\includegraphics[width=0.45\linewidth]{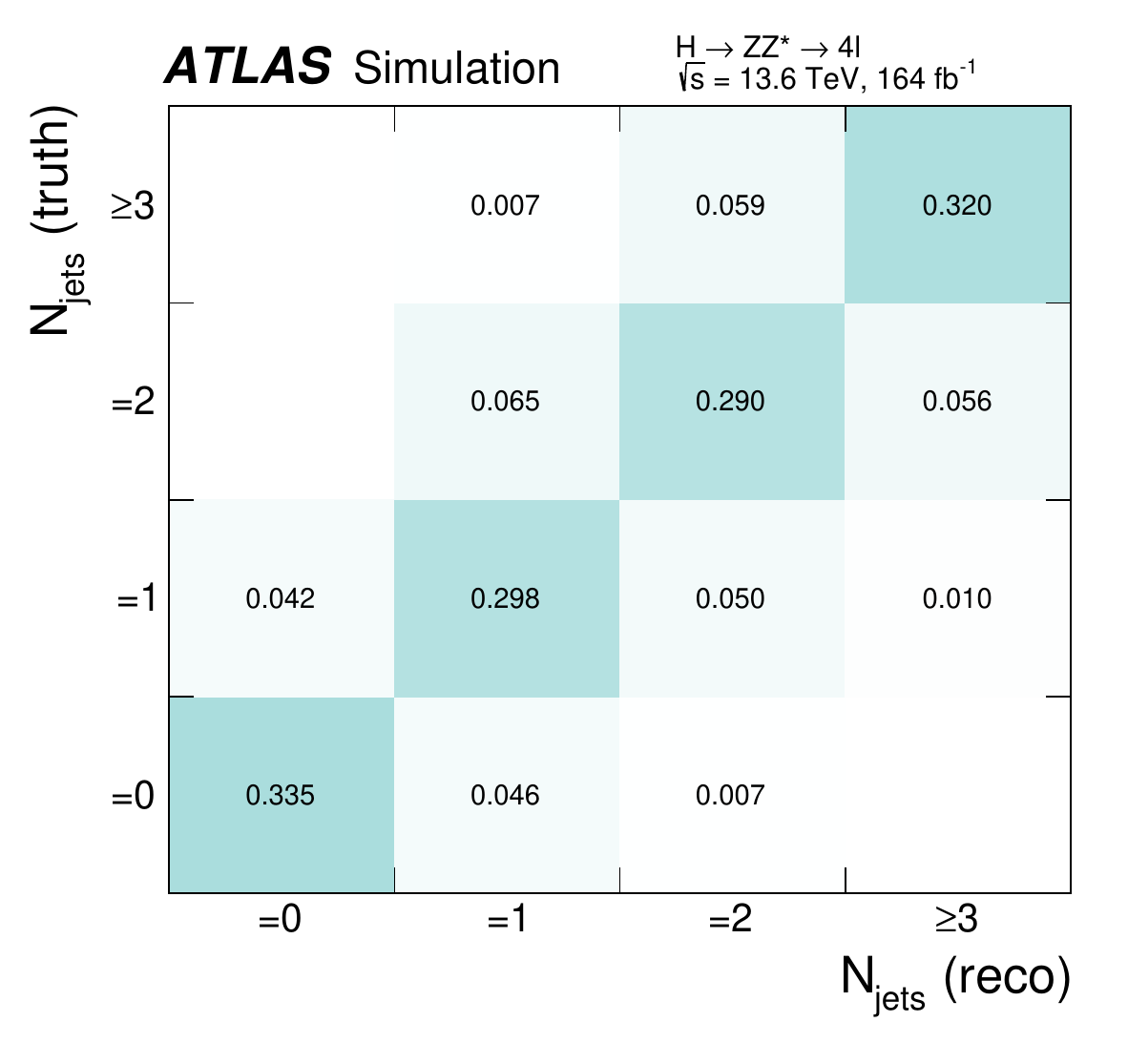}}
\caption{Response matrices, derived from simulation, for (a) the transverse momentum of the four-lepton system $\ptFourL$ and (b) the jet multiplicity $\Njets$. The response matrix is used in the unfolding procedure (Eq.~\ref{eq:diffxs_fit}). The elements $r_{ij}$ correspond to the probability that an event generated in fiducial bin $j$ is reconstructed in bin $i$. Only reconstructed events that are matched to particle-level (``truth'') events are included. Bins with values below 0.005 are omitted for clarity.\label{fig:resp_matrix_main}}

\end{figure}

The response matrices for \ptH~ and \njets~ are shown in Figure \ref{fig:resp_matrix_main}. The bin-to-bin migrations are quantified by the \emph{purity}, defined as the fraction of reconstructed events in a given bin that originate from the corresponding particle-level bin. The purity is mostly larger than 80\% for the observables related to the Higgs decay kinematics, the $\ptFourL$, $y_{4\ell}$, $m_{12}$, $m_{34}$, $\phi$ and $\ptFourL$ versus $m_{34}$. Instead, for the jet-related observables, $\Njets$ and $|\Delta\phi_{jj}|$, migrations are larger due to the jet energy resolution and residual pile-up effects, reducing for example the purity in the highest jet-multiplicity bin to 66\%. The fiducial leakage
mostly ranges from 1\% to 2\%, depending on the bin of the unfolded observable or the decay final state.  Only the first bin of the $m_{12}$ and the last bin of the $m_{34}$ observables reach values of 4\% and 6\% respectively.

\subsection{Production-mode cross-sections}
\label{subsec:stxs}

The cross-sections are also probed separately for four different Higgs boson production-modes:  gluon--gluon fusion, vector-boson fusion and associated production with vector bosons or top quark pairs.

\subsubsection{Simplified template cross-sections}
\label{subsec:stxsdescription}

In the STXS framework~\cite{bendavid2018les,YR4,Berger:2019wnu}, Higgs boson production cross-sections are measured in exclusive particle-level regions (production bins) designed to minimise the impact of theoretical uncertainties that would otherwise enter through acceptance extrapolations, while retaining sensitivity to production dynamics and potential BSM effects. Two alternative STXS binnings are considered: a production-mode (also referred to as Stage-0) measurement, in which cross-sections are extracted for the four dominant production-modes (\ggF, \VBF, \VH, and \ttH), and a Reduced Stage-1.2 measurement that provides additional kinematic information by subdividing production according to jet activity and Higgs boson kinematics. Bins are merged relative to the full binning where required by the statistical precision. All production bins are defined for Higgs bosons with rapidity $|y_H|<2.5$, and no requirements are placed on particle-level leptons. Particle-level jets are used for this categorisation and are built using the anti-$k_t$ algorithm with $R=0.4$ from stable particles (lifetime $c\tau>10$~mm) and are required to satisfy $\pt>30~\GeV$ with no rapidity restriction; Higgs decay products and leptons/neutrinos from leptonic $V$ decays are excluded from the jet clustering, while hadronic $V$ decay products are included~\cite{Berger:2019wnu}.

In the production-mode measurement, contributions from \bbH are included in the \ggF bin and contributions from $\tH$ are included in the \ttH bin, with their relative fractions fixed to SM predictions. \VH events with hadronic $V$ decays are retained in a dedicated \VH bin~\cite{YR4}. For the Reduced Stage-1.2 measurement, the \ggF and $gg\to ZH$ (hadronic $Z$) contributions are split by $\pt^H$ and jet multiplicity, including a dedicated high-$\pt^H$ bin with $\pt^H\ge200~\GeV$ to enhance sensitivity to BSM effects. The electroweak $qqH$ contribution (including \VBF and hadronic \VH) is further partitioned using $\pt^H$ and the dijet invariant mass $m_{jj}$, defining a BSM-sensitive region with $\pt^H\ge200~\GeV$ and $m_{jj}>350~\GeV$, a $\VH$-like region with $60<m_{jj}<120~\GeV$, and a remaining region targeting \VBF, while \VH with leptonic $V$ decays is treated separately as \STXSVHLep. The \ttH production bin definition is unchanged with respect to the production-mode interpretation.

\begin{figure}[htb!]
\begin{center}
\includegraphics[width=0.97\linewidth]{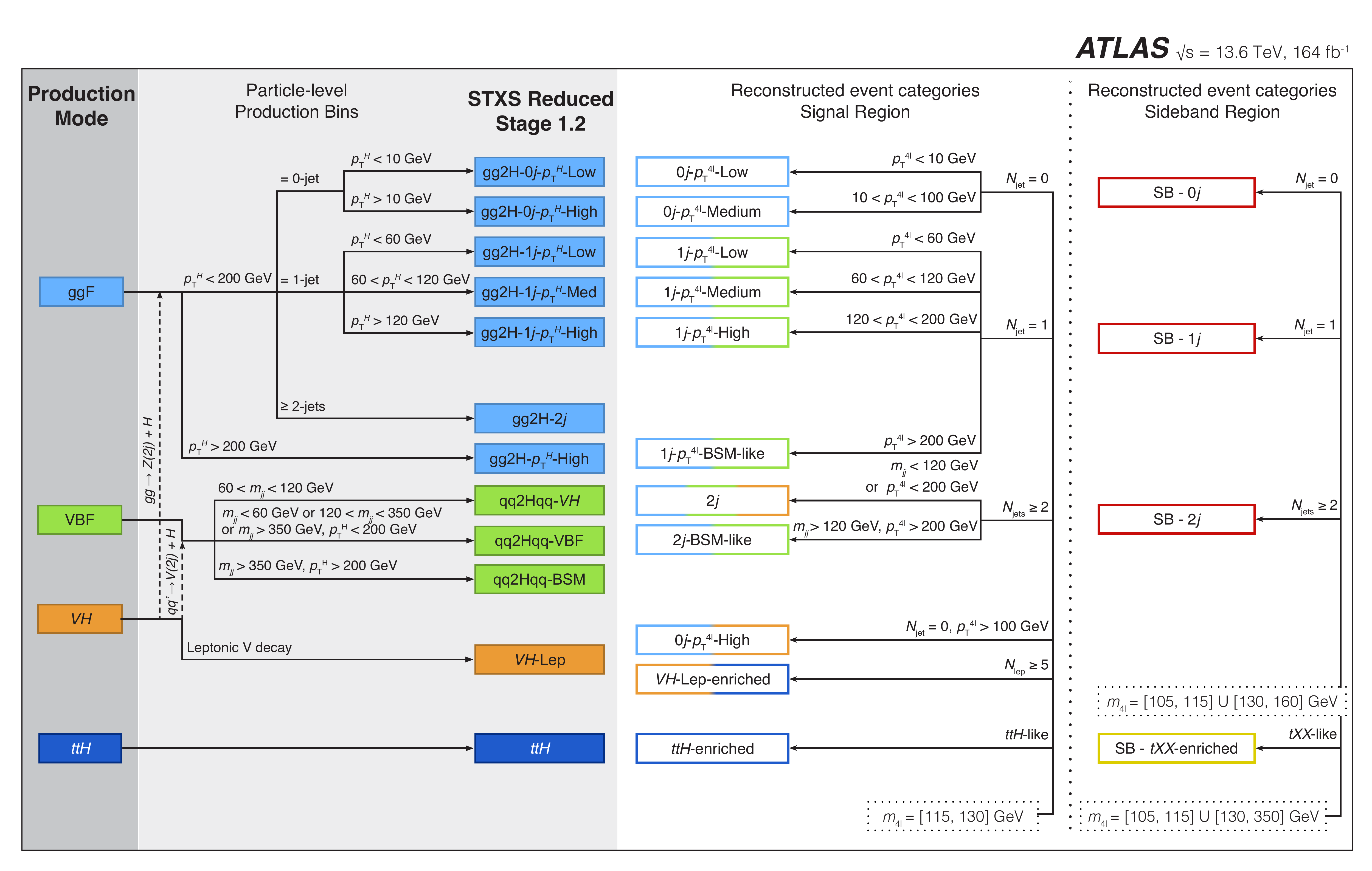}
\caption{Two sets of exclusive phase-space regions (production bins) are defined at particle level for the measurement of the Higgs boson production cross-sections: the production-mode Stage 0 and the Reduced Stage~1.2 scheme (left and middle-left shaded panels). The corresponding reconstructed event categories are shown for signal (middle-right panel) and sidebands (right panel). The description of the production bins is given in Section~\ref{subsec:stxsdescription}, while the reconstructed signal-region and sideband event categories are described in Section~\ref{sec:categorization}. The \bbH ($\tH$) contribution is included in the \ggF (\STXSttH) production bins. The colours of each reconstructed event-category box indicate the relevant production processes that contribute to the reconstructed category. The reconstructed signal categories are defined sequentially from bottom to top, such that each event is assigned to the first category whose selection criteria are satisfied.}

\label{fig:stxs_diagram}
\end{center}
\end{figure}

The exact definitions of the STXS production bins and the corresponding labels used in this analysis are shown on the left grayed panel of Figure~\ref{fig:stxs_diagram}.

\subsubsection{Event categorisation}
\label{sec:categorization}

Signal-region events with a reconstructed Higgs boson candidate in $115<m_{4\ell}<130~\GeV$ are assigned to mutually exclusive categories sensitive to different production bins in the STXS framework, in a sequential procedure, following the ordering shown in Figure~\ref{fig:stxs_diagram}. Events are first tested against the \CatttH selection, which targets $\ttH$ production using requirements on additional leptons and/or high jet and $b$-tag multiplicities. In particular, events are selected if they contain at least one additional isolated lepton with $\pt>12~\GeV$ and satisfy at least one of the following: at least two $b$-tagged jets at the 85\% working point, at least five jets with at least one $b$-tagged jet at the 85\% working point, or at least two jets with at least one $b$-tagged jet at the 65\% working point. Events without an additional isolated lepton can also enter the \CatttH selection if they satisfy tighter jet and $b$-tag requirements, namely at least five jets with at least two $b$-tagged jets at the 85\% working point, or at least four jets with at least one $b$-tagged jet at the 65\% working point.

Events with additional isolated leptons that do not satisfy the \CatttH selection define the \VH category targeting leptonic vector-boson decays (\CatVHLep). The remaining events are then categorised by jet multiplicity ($N_{\mathrm{jets}}=0$, 1, or $\ge2$) and further subdivided using $\ptH$ and, for events with at least two jets, the dijet invariant mass $m_{jj}$ to target specific STXS production bins. Events with at least two jets are separated into a BSM-like category at high $\ptH$ and large $m_{jj}$ and a complementary category that contains the bulk of the \VBF/\VH contributions in addition to \ggF. Both the 0-jet and 1-jet regions are further split into low-, medium-, and high-$\ptH$ intervals.

Sideband categories in $m_{4\ell}$ are defined to constrain the dominant non-resonant \zzstar\ background in different jet multiplicity regions (\CatSBZeroJ, \CatSBOneJ, and \CatSBTwoJ) and the top-quark-associated \tXX background (\CatSBtXX) as described in Section~\ref{sec:background}.

\begin{figure}[htb!]
\begin{center}
\includegraphics[width=0.85\linewidth]{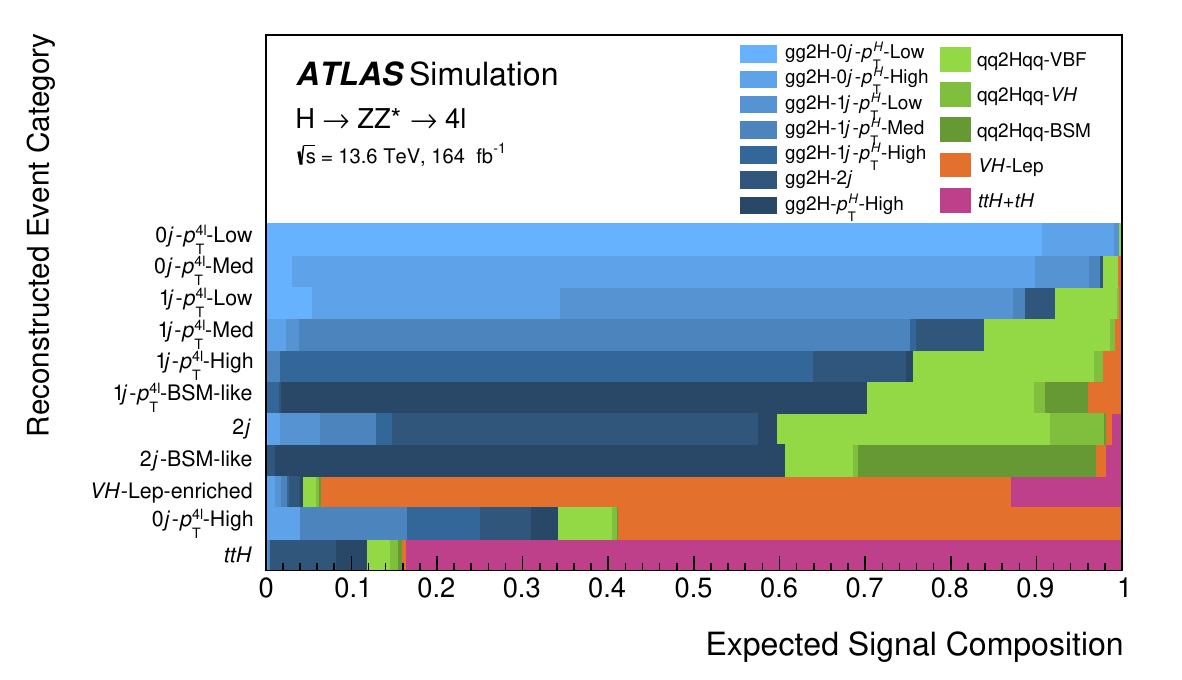}
\caption{Standard Model signal composition in terms of the Reduced Stage-1.2 production bins in each reconstructed event category. The \bbH and \tH contributions are included in the \ggF and \ttH production bins, respectively.}
\label{fig:stxs_composition}
\end{center}
\end{figure}

The definitions of the reconstruction categories and the corresponding names used in this analysis are shown on the right panel of Figure~\ref{fig:stxs_diagram}. The composition of production-modes per reconstructed event category of these categories is shown in Figure~\ref{fig:stxs_composition}.

\subsubsection{Multivariate discriminants and Signal Extraction}

In the STXS measurement, several Higgs boson production mechanisms contribute to each reconstructed event category, together with multiple background processes. To enhance the separation power beyond the cut-based categorisation, multivariate discriminants are constructed using neural networks (NNs) based on a Set Transformer architecture~\cite{setTransformer}.
The networks are trained independently in each reconstructed category on simulated samples of the relevant Higgs-boson production modes and dominant non-Higgs-boson backgrounds, and are optimised to classify events according to their most likely signal-production or background process.
The NN inputs are organised into event-level global observables and sets of reconstructed physics objects. Lepton and jet information is provided as variable-length sets to Set Transformer blocks, which provide permutation-invariant processing through attention mechanisms~\cite{Vaswani:2017lxt}, while global variables are combined with the learned set representations in a classifier head. Each of the NN output nodes is labelled as NN$_{P}$, where $P$ refers to one or multiple targeted processes, and is related to the probability of an event originating from that process.
The specific feature choices, the corresponding training targets and the output nodes follow the strategy outlined in Table~\ref{tab:stxs_nn_training_summary}.
Good agreement between data and simulation is observed for the input variables used in the NN training, as illustrated in Figure~\ref{fig:Cat_observables}. In addition, validation studies are performed in $\zzstar$-enriched control regions to verify that the NN outputs are well modelled and that no overtraining or simulation-specific features are learned.
In categories with limited statistical power, no NN discriminant is applied and only the observed yield enters the fit.

The STXS cross-sections are obtained from a simultaneous binned likelihood fit to the discriminant distributions across all reconstructed signal categories defined in Section~\ref{sec:categorization}, with the dominant background processes constrained as described in Section~\ref{sec:background}. The parameters of interest are collected in a vector $\vec{\sigma}$ and are related to the reconstructed event yields $N_{ik}$ in a histogram bin $k$ of the discriminating observable in reconstructed category $i$ via
\begin{equation}
\begin{split}
N_{ik} =  \vec{A}_{ik} \cdot \vec{\sigma} \cdot \mathcal{B} \cdot \mathcal{L} + N^{\textrm{bkg}}_{ik}\,,
\end{split}
\label{eq:stxs_fit}
\end{equation}
where $\mathcal{B}$ is the $H\to ZZ^{*}\to 4\ell$ branching ratio, $\mathcal{L}$ is the integrated luminosity, and $N^{\textrm{bkg}}_{ik}$ is the expected background contribution. The vector $\vec{A}_{ik}$ contains the signal selection factors that map particle-level STXS bins to reconstructed categories and discriminant bins, and is determined from MC simulation. In the Stage-0 interpretation, the components of $\vec{\sigma}$ (and thus of $\vec{A}_{ik}$) correspond to the chosen set of production-mode cross-sections, while in the Reduced Stage-1.2 interpretation they correspond to bins defined in Section~\ref{subsec:stxsdescription}. In both cases, each factor $A_{ik}^{j}$ is defined as the fraction of generated signal events in STXS bin $j$ that satisfy the reconstruction and event selection and populate histogram bin $k$ of the discriminant observable in reconstructed category $i$.

\begin{table}[htbp]
\centering
\caption{Input variables used for the neural-network trainings in the reconstructed STXS categories that use binned NN discriminants. The inputs are grouped into event-level \emph{global} variables and object-level variables provided as variable-length \emph{lepton} and \emph{jet} sets to Set Transformer blocks. Here, $\ptFourL$ and $\eta_{4\ell}$ are the transverse momentum and pseudo-rapidity of the four-lepton system; $m_{12}$ and $m_{34}$ are the invariant masses of the leading and subleading same-flavour opposite-charge lepton pairs; $D_{ZZ^{*}}$ denotes the kinematic discriminant against non-resonant $ZZ^{*}$ production; $\cos\theta^{*}$, $\cos\theta_{1}$, and $\phi_{ZZ}$ are decay-angle observables defined as in Ref.~\cite{JHU1}; $\ptJLead$ and $\eta_{j}$ are the transverse momentum and pseudorapidity of the leading jet; $m_{jj}$ is the dijet invariant mass; $p_{\mathrm{T}}^{4\ell jj}$ is the transverse momentum of the four-lepton plus dijet system; and $\eta_{ZZ}^{\mathrm{zepp}} = |\eta_{4\ell}-\frac{\eta_{j1}+\eta_{j2}}{2}| $ is the Zeppenfeld variable of the four-lepton system relative to the two leading jets~\cite{Rainwater:1996ud}. $\Delta R_{4\ell,j}$ is the angular separation between the four-lepton system with respect to the leading jet, while $E_{\mathrm{T}}^{\mathrm{miss}}$ is the measured missing transverse momentum.}
\label{tab:stxs_nn_training_summary}
\small
\setlength{\tabcolsep}{6pt}
\renewcommand{\arraystretch}{1.15}
\resizebox{\textwidth}{!}{\begin{tabular}{lccccc}
\toprule
Category & Process & Global inputs & Lepton set inputs & Jet set inputs & Discriminant \\
\midrule

\CatZeroJL & \multirow{2}{*}{\ggF, \zzstar} &
$\eta_{4\ell}$, $\ptFourL$, $D_{ZZ^{*}}$, $m_{12}$ &
\multirow{2}{*}{$p_{\mathrm{T}}^{\ell}$, $\eta_{\ell}$} &
\multirow{2}{*}{--} & \multirow{2}{*}{NN$_{\textrm{ggF}}$} \\
\CatZeroJM & & $m_{34}$, $\cos\theta^{*}$, $\cos\theta_{1}$, $\phi_{ZZ}$ & & &\\
\midrule

\multirow{2}{*}{\CatOneJL} & \multirow{2}{*}{\ggF, \VBF, \zzstar} &
$\ptFourL$, $\ptJLead$, $\eta_{j}$, $D_{ZZ^{*}}$ &
\multirow{2}{*}{$p_{\mathrm{T}}^{\ell}$, $\eta_{\ell}$} &
\multirow{2}{*}{--}  & NN$_{\textrm{ggF+VBF}}$ for NN$_{\textrm{ZZ}}$<0.2  \\
& &$\Delta R_{4\ell,j}$,$\eta_{4\ell}$, $E_{\mathrm{T}}^{\mathrm{miss}}$ & & & NN$_{\textrm{ZZ}}$ for NN$_{\textrm{ZZ}}$>0.2\\

\midrule
\multirow{2}{*}{\CatOneJM} & \multirow{2}{*}{\ggF, \VBF, \zzstar} &
$\ptFourL$, $\ptJLead$, $\eta_{j}$, $D_{ZZ^{*}}$ &
\multirow{2}{*}{$p_{\mathrm{T}}^{\ell}$, $\eta_{\ell}$} &
\multirow{2}{*}{--}  & NN$_{\textrm{ggF+VBF}}$ for NN$_{\textrm{ZZ}}$<0.2  \\
& &$\Delta R_{4\ell,j}$,$\eta_{4\ell}$, $E_{\mathrm{T}}^{\mathrm{miss}}$ & & & NN$_{\textrm{ZZ}}$ for NN$_{\textrm{ZZ}}$>0.2\\

\midrule
\multirow{2}{*}{\CatOneJH} & \multirow{2}{*}{\ggF, \VBF} &
$\ptFourL$, $\ptJLead$, $\eta_{j}$, $D_{ZZ^{*}}$ &
\multirow{2}{*}{$p_{\mathrm{T}}^{\ell}$, $\eta_{\ell}$} &
\multirow{2}{*}{--}  & \multirow{2}{*}{NN$_{\textrm{ggF}}$} \\
& &$\Delta R_{4\ell,j}$,$\eta_{4\ell}$, $E_{\mathrm{T}}^{\mathrm{miss}}$ & & &\\

\midrule

\multirow{6}{*}{\CatTwoJ} &  &
\multirow{6}{*}{$m_{jj}$, $p_{\mathrm{T}}^{4\ell jj}$, $\eta_{ZZ}^{\mathrm{zepp}}$} &
\multirow{6}{*}{$p_{\mathrm{T}}^{\ell}$, $\eta_{\ell}$} &
\multirow{6}{*}{$\ptJLead$, $\eta_{j}$, $m_{j}$} & NN$_{\textrm{ggF+ZZ+tXX}}$ for  \\
& & & & & NN$_{\textrm{VH}}$<0.3, NN$_{\textrm{VBF}}$<0.1\\
& \ggF, \VBF, \VH & & & & NN$_{\textrm{VBF}}$ for\\
& \zzstar, tXX & & & & NN$_{\textrm{VH}}$<0.3, NN$_{\textrm{VBF}}$>0.1\\
& & & & & NN$_{\textrm{VH}}$ for\\
& & & & & NN$_{\textrm{VH}}$>0.3, NN$_{\textrm{VBF}}$<0.1\\
\midrule
\CatTwoJBSM & \ggF, \VBF & $m_{jj}$, $p_{\mathrm{T}}^{4\ell jj}$, $\eta_{ZZ}^{\mathrm{zepp}}$ & $p_{\mathrm{T}}^{\ell}$, $\eta_{\ell}$ & $\ptJLead$, $\eta_{j}$, $m_{j}$ & NN$_{\textrm{VBF}}$  \\
\bottomrule
\end{tabular}}
\end{table}

\subsection{Statistical method}
\label{subsec:statmethod}

All measurements are performed with a binned profile-likelihood-ratio fit~\cite{Cowan:2010js}. The likelihood is built as product of Poisson terms for the fitted discriminant distributions in each differential bin or STXS category, and includes nuisance parameters that account for experimental and theoretical systematic uncertainties affecting the normalisation and shapes of the signal and background templates (Section~\ref{sec:systematics}). The parameters of interest are the cross-sections and the background normalisation factor introduced in Sections~\ref{subsec:diffxs} and~\ref{subsec:stxs} and are extracted by minimising $-2\ln\lambda$, where $\lambda$ is the profile likelihood ratio. The asymptotic approximation~\cite{Cowan:2010js} is validated with pseudo-experiments; where fewer than three events are observed in a bin, the quoted uncertainties are corrected to those obtained from pseudo-experiments.

The compatibility of a measured cross-section and its theoretical prediction is tested by computing a $p$-value based on the difference between the value of $-2 \ln \lambda$ at the best-fit value and the value obtained by fixing the cross-section in each bin to that predicted by SM theory. These $p$-values do not include the uncertainties in the theoretical predictions.

%


%

%
\section{Systematic uncertainties}
\label{sec:systematics}

The dominant uncertainty in the measurement arises from statistical uncertainty.
In addition, many systematic uncertainties are considered.
Experimental uncertainties arise from object reconstruction, identification, isolation, energy and momentum scale, resolution, and trigger efficiencies, the determination of the integrated luminosity and the data-driven background estimate.
Additional theoretical uncertainties are related to the modelling of the signal and background processes.

\subsection{Experimental uncertainties}

The uncertainty in the integrated luminosity is $1.9$\%~\cite{ATL-DAPR-PUB-2025-001}. %
This affects the signal and background yields predicted by simulation if no normalisation to data is used. The uncertainty due to pile-up modelling is below 2\% in the entire $m_{4\ell}$ range considered.

The reconstruction and identification efficiencies for muons and electrons are determined using the methods described in Refs.~\cite{MUON-2018-03,EGAM-2018-01}. For muons, these have associated uncertainties between $0.1\%$ and $1.3\%$. For electrons with $\pT<15\GeV$ conservative uncertainties up to $17\%$ are included to cover the change in detector conditions between Runs 2 and 3 and updates to the simulation.
The isolation efficiency uncertainty is approximately $1\%$ for electrons and $0.5-2\%$ for muons.

Lepton energy and momentum scale and resolution are corrected to describe the data using the procedures documented in Refs.~\cite{EGAM-2021-02,MUON-2022-01}, and the associated uncertainties have an impact at the most of $1-1.5\%$ on the measured cross-sections.

Jet energy scale and resolution are calibrated using the procedure described in Ref.~\cite{JETM-2018-05}. An additional uncertainty is assigned to jets in the kinematic region of $|\eta|>3.2$ to account for an observed mismodelling on the jet energy calibration in the forward calorimeters.

Uncertainties in the $b$-jet identification efficiency are taken from 2022 and 2023 data and applied also to 2024 data and conservatively increased by ten times to account for differences between different data-taking periods. This is done only in the \ttH category and $tXX$ sideband, where $b$-jets are required and the impact is at the level of 1\%.

Uncertainties in the lepton trigger efficiency have a negligible impact on the analysis, as does the uncertainty on the measured Higgs boson mass of $m_{H}=125.09 \pm 0.24~\GeV$~\cite{HIGG-2020-07}.

For the data-driven measurement of the reducible background, several sources of uncertainty are considered. The dominant contribution arises from the statistical uncertainty in the region containing a quadruplet with at least one baseline lepton failing to meet the tight requirements, where the fake factors are applied. An additional significant contribution comes from the systematic uncertainty associated with the choice of control regions used to derive the fake factors. This uncertainty reflects the variation of the result when the fake factors are extracted using only CR$_Z$ or only CR$_{\text{top}}$, instead of their nominal combination, as described in Section~\ref{sec:background}. Those uncertainties have an impact that ranges from 1\% to 9\% both in the differential and in the production cross-section measurements.
The statistical uncertainty due to the limited statistics in the CRs is also propagated to the fake factor estimate and  has a negligible impact on the final yields. Finally, the uncertainty in the $WZ$ contribution is estimated by varying the fitted normalization within its statistical error and has negligible impact on the estimate.

\subsection{Theoretical uncertainties}
\label{subsec:theorysys}

Uncertainties in the signal prediction due to missing higher-order QCD eﬀects are estimated by using the scheme described in Ref.~\cite{ATL-PHYS-PUB-2023-031}.
Parton shower and multiple-parton interaction modelling uncertainties on the acceptance are estimated by using a prediction based on \Herwig{7} as alternative parton shower generator and using tune eigenvector variations within the nominal parton shower generator tune, A14.
Uncertainties related to the PDF set are evaluated using 40 eigenvector variations of the PDF4LHC21 PDF set.
An additional uncertainty of the \hfourl decay branching ratio of 2.2\%~\cite{Bredenstein:2006rh,Bredenstein:2006ha} is assigned when extrapolating to the total phase-space. These theory-related uncertainties enter the measurements in different ways. %
These uncertainties generally have small effects on both cross-section measurements, mainly through acceptance and migration. In contrast, for the global strength measurement and interpretation of the results, the parameters of interest ($\mu$) are defined relative to the SM predictions. In this case the same modelling uncertainties propagate directly to the predicted signal yields and have a larger impact.

Uncertainties in the prediction of the $\zzstar$ background from missing higher-order QCD effects are taken into account by varying the factorisation and renormalisation scales in the calculation by factors of two and choosing the configuration with the largest impact. PDF-related uncertainties are evaluated using the various PDFs within the \NNPDF[3.0] PDF set using the method outlined in ~\cite{Ball:2014uwa}.
A further modelling uncertainty is assigned by comparing the prediction from the nominal $\Sherpa$ sample to a simulation in \MGNLO matched to \PYTHIA[8] with the \FXFX prescription ~\cite{Frederix:2012ps,Alwall:2014hca}.
As this background contribution is normalised to data, only the change in differential shapes impacts the measurements ($1-3\%$), while the effect in the shapes of the NN discriminants and in the relative contribution of this background between the sidebands and the signal regions for the STXS measurement is negligible. For the tXX process, uncertainties due to PDF and missing higher-order QCD effects are estimated from simulation using a similar procedure as the $\zzstar$ process.

Finally, unfolding-related systematic uncertainties for the inclusive and differential cross-section measurements arise from uncertainties in the production-mode composition that affect the response matrices. Additional uncertainties come from a potential bias introduced by the unfolding method. This bias is evaluated independently in each bin by comparing the unfolded cross-section in simulation with the expected result when the underlying true cross-sections of the simulated sample are varied within their expected statistical uncertainty. The impact of these uncertainties is negligible.


%

%
\section{Results}
\label{sec:results}

\begin{figure}[htbp]
\centering
\includegraphics[width = 0.65 \textwidth]{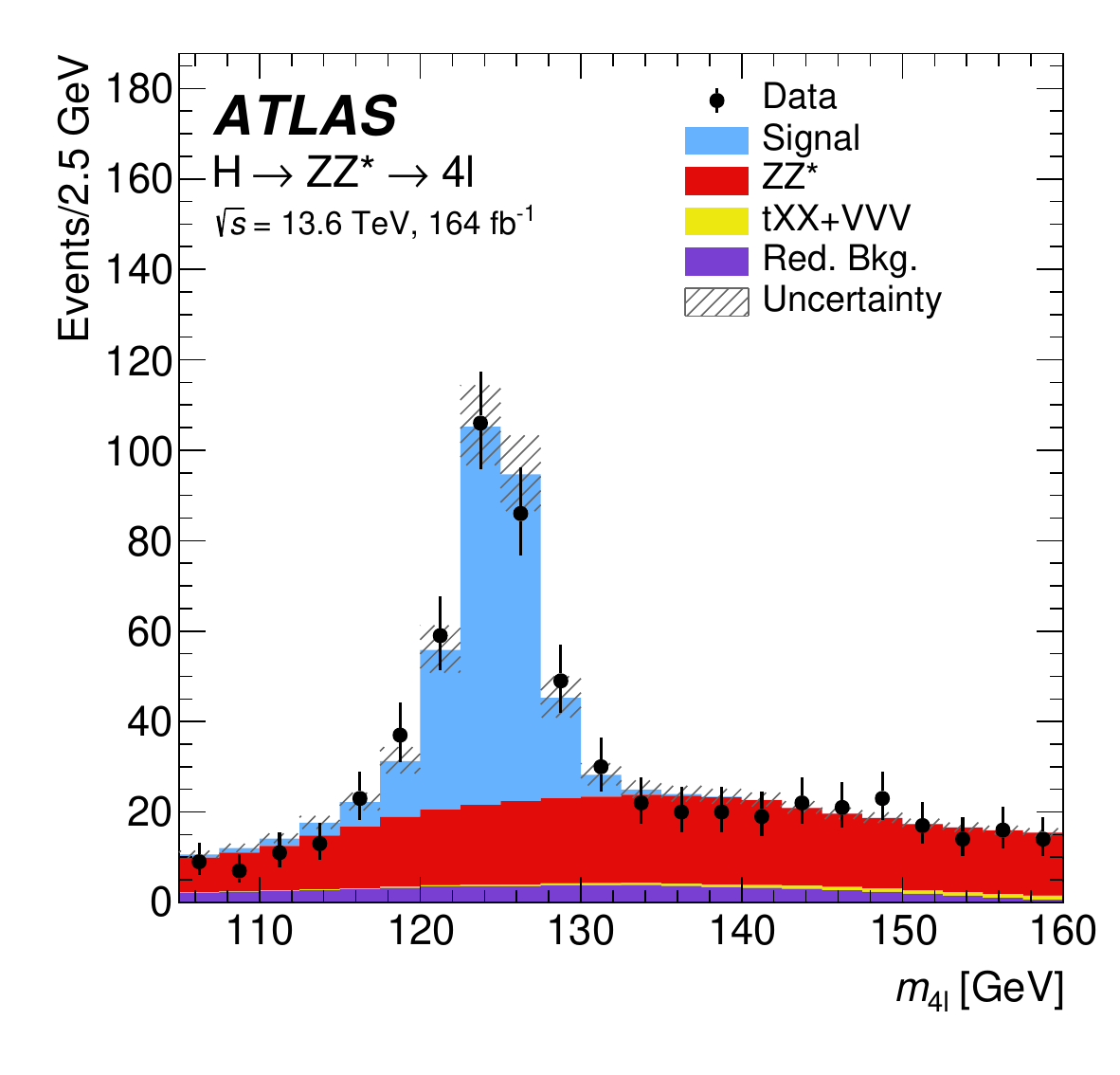}
\caption{The observed and expected (pre-fit) inclusive four-lepton invariant mass distributions for the selected Higgs boson candidates. The uncertainty in the predictions is shown by the hatched band, which includes the theoretical uncertainties of the SM cross-section for the signal and the background processes.  \label{fig:mFourL_SR} }
\end{figure}

Figure~\ref{fig:mFourL_SR} shows the expected and observed four-lepton invariant mass distribution $\mFourL$ for all Higgs boson candidates accepted by the event selection. The observed number of events and the expected signal and background yields in the mass range  \hbox{115~\gev$< \mfourl <$ 130~\gev} are presented in Table~\ref{tab:recoyields}  before fitting to the data (pre-fit). The data agree with predictions.
\begin{table*}[htbb!]
\centering
\caption{Expected (pre-fit) and observed numbers of events in the four decay final states after the event selection, in the mass range  \hbox{115~\gev$< \mfourl <$ 130~\gev}. The column `Other Backgrounds' includes the reducible background and other minor sources discussed in Section~\ref{sec:background}. The sum of the expected number of SM Higgs boson events and the estimated background yield is compared with the data. Combined statistical and systematic uncertainties are included for the predictions (see Section~\ref{sec:systematics}).
\label{tab:recoyields}}
\vspace{0.2cm}

\begin{tabular}{*{7}{c}}
\toprule
Final state  &              Signal &          $\zzstar$ &            Other Backgrounds &               Total SM &  Observed \\
\midrule
$4\mu$      & 90.5 $\pm$ 5.3 & 41.2 $\pm$ 1.6 & 4.5 $\pm$ 0.3 & 136 $\pm$ 6 & 136\\
$2e2\mu$    & 63.4 $\pm$ 4.0 & 30.0 $\pm$ 1.3 & 4.7 $\pm$ 0.5 & 98 $\pm$ 5 & 98\\
$2\mu2e$    & 40.7 $\pm$ 9.1 & 15.6 $\pm$ 4.4 & 5.6 $\pm$ 0.2 & 62 $\pm$ 13 & 67\\
$4e$        & 36.5 $\pm$ 9.2 & 14.1 $\pm$ 4.3 & 5.3 $\pm$ 0.3 & 56 $\pm$ 13 & 59\\
\midrule
$\text{Total}$  & 231 $\pm$ 23   & 101 $\pm$ 10 & 20.1 $\pm$ 1.3  & 352 $\pm$ 32 & 360\\
\bottomrule
\end{tabular}


\end{table*}
\FloatBarrier

\subsection{Inclusive fiducial and total cross-sections}
\begin{figure}[htbp]
\centering
\includegraphics[width = 0.6 \textwidth]{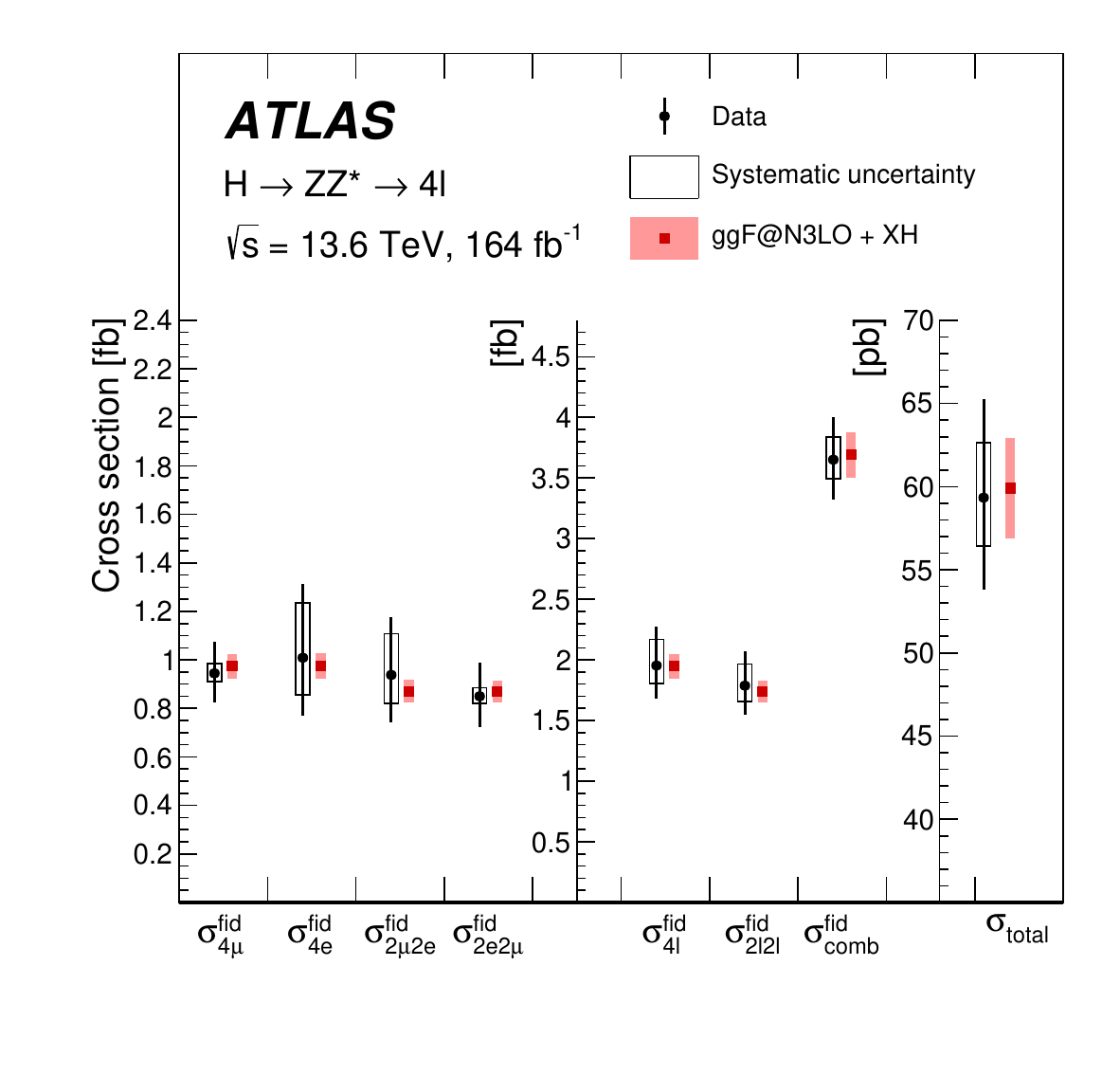}
\caption{Measured fiducial and total cross-sections: (left panel) fiducial cross-sections for the four individual final states; (middle panel) inclusive fiducial cross-section obtained by combining same-flavour, different-flavour, and all final states assuming relative SM branching ratios; (right panel) total cross-section obtained after accounting for the branching ratio of the Higgs boson into four leptons and the fiducial acceptance as determined from the simulated SM Higgs boson samples. The measurement in data is shown as black circular markers with error bands indicating the total error and boxes indicating the systematic uncertainty. The measurements are compared to theoretical predictions with their associated uncertainties.  \label{fig:inclXS}}
\end{figure}

The inclusive fiducial and total cross-sections of the $H\to ZZ^* \to 4\ell$ process are measured by combining the results from the four decay final states assuming the relative SM branching ratio and are summarised in Figure~\ref{fig:inclXS}. A normalisation factor of $N/N_{\mathrm{MC}} = 0.94\pm 0.08$ is obtained for the \zzstar background component. The fiducial cross-section is measured as
\[ \fidObsFull, \]
consistent with the SM prediction of \fidPred. The resulting total cross-section is
\[ \fidTotFull, \]
consistent with the SM value of \fidTotPred. The systematic uncertainty in the measurement is dominated by lepton reconstruction and isolation efficiencies. Larger uncertainties in decay channels with electrons in the subleading lepton pair reflect the conservative estimates of uncertainties on electron reconstruction efficiency. In combination with other final states, the impact of this systematic uncertainty is reduced in the inclusive fiducial measurement.
The probability of compatibility of the measured fiducial cross-section and the SM expectation is 92\% and for the individual channels it ranges from 75\% for $2\mu2e$ to 90\% for $4e$ final states.

\FloatBarrier

\subsection{Differential cross-sections}

\begin{figure}
\centering
\subfloat[]{\includegraphics[width = 0.49 \textwidth]{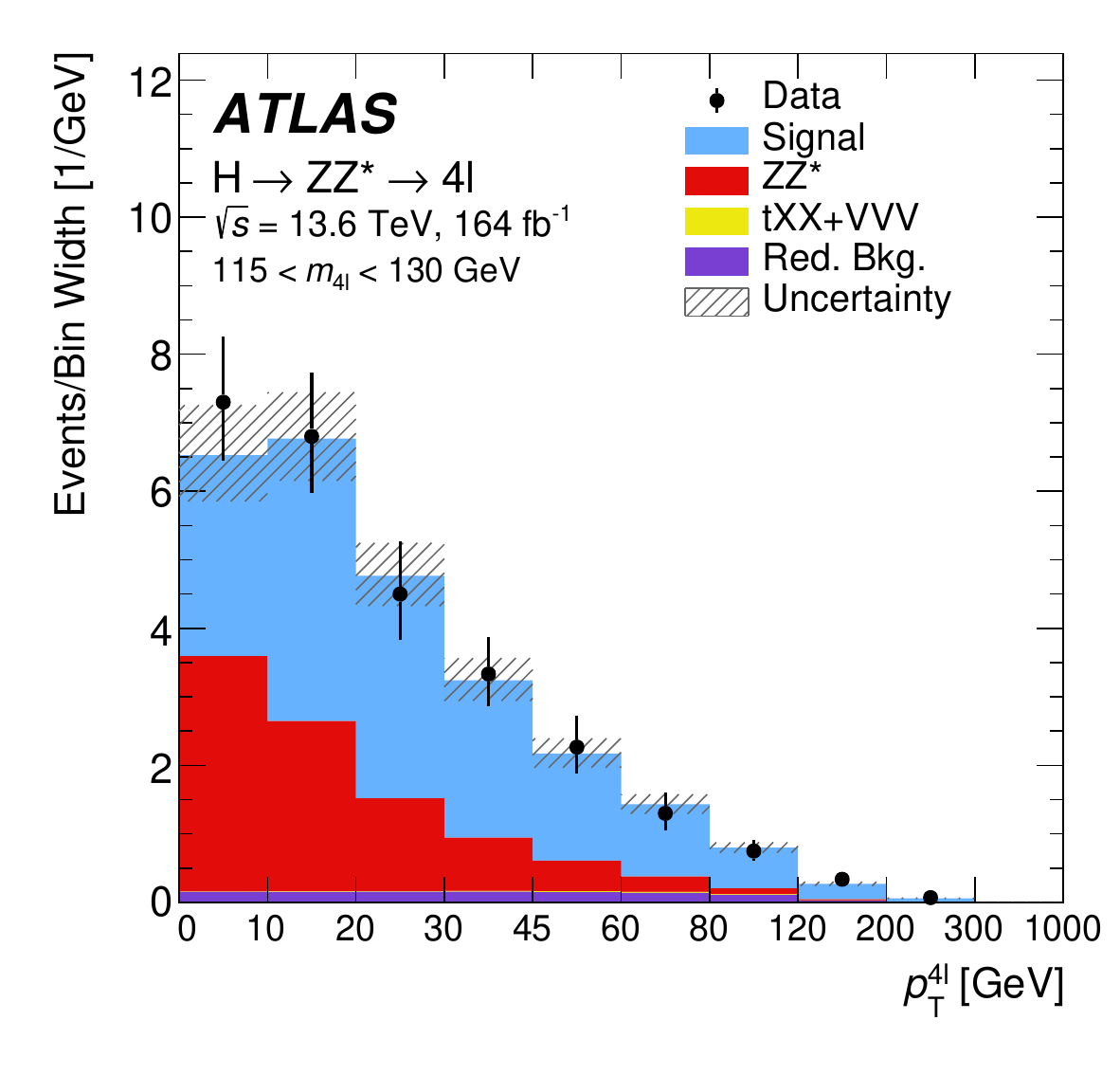}}
\subfloat[]{\includegraphics[width = 0.49 \textwidth]{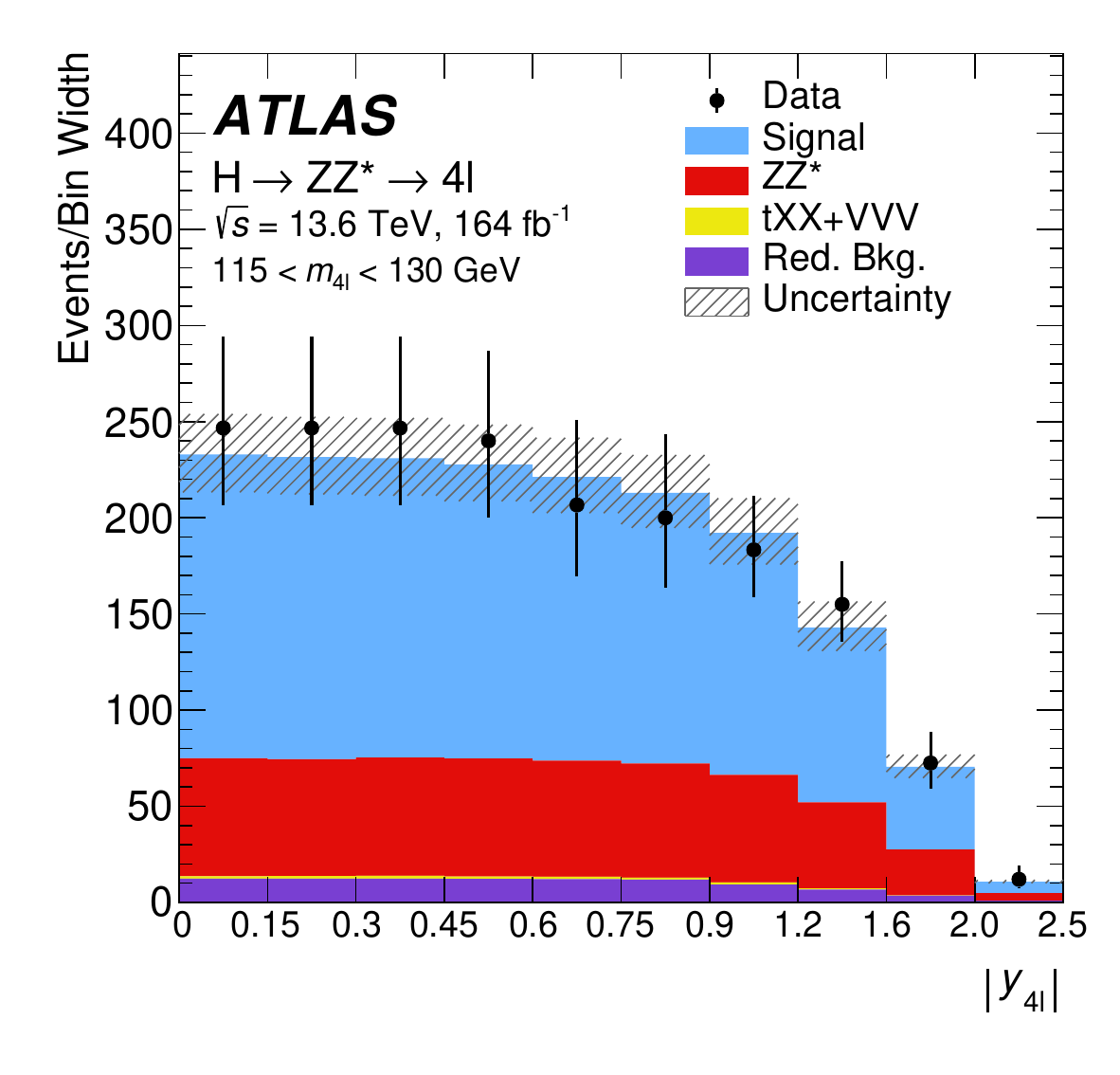}} \\
\subfloat[]{\includegraphics[width = 0.49 \textwidth]{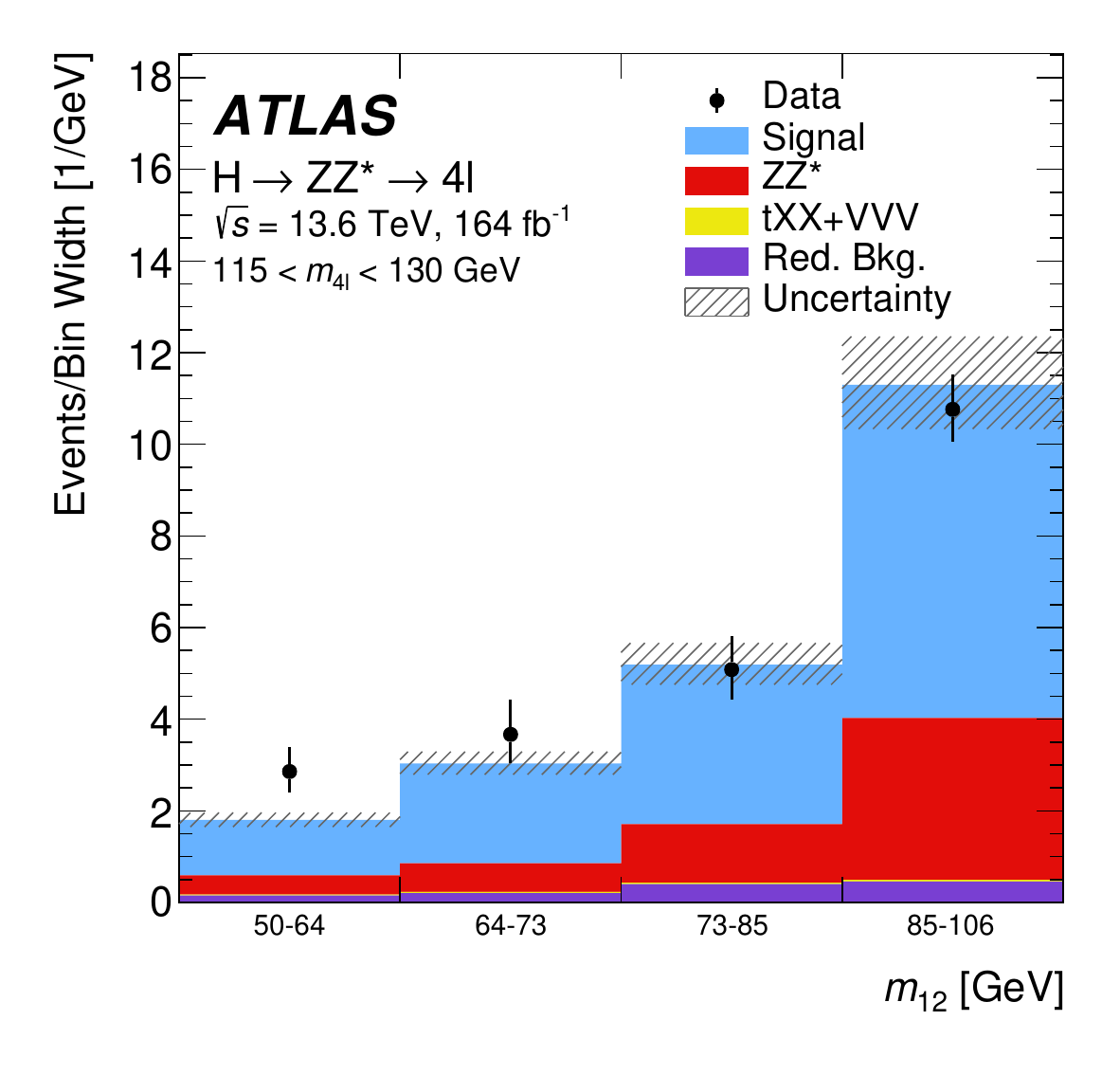}}
\subfloat[]{\includegraphics[width = 0.49 \textwidth]{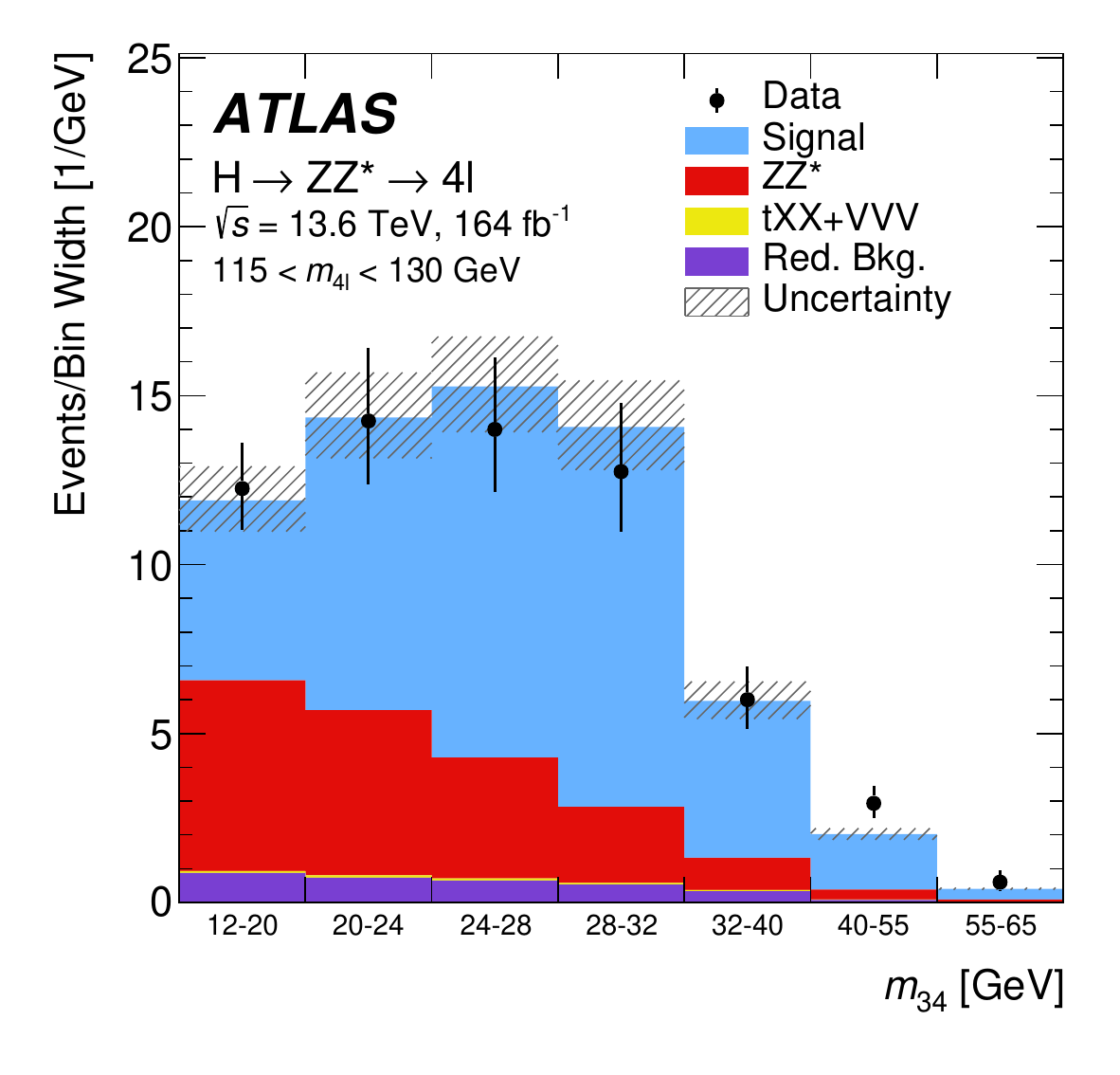}}
\caption{The observed and expected (pre-fit) distributions of (a) $\ptFourL$, (b) $|\yFourL|$, (c) $\mOneTwo$, and (d) $\mThreeFour$ observables, after the event selection in the mass range  \hbox{115~\gev$< \mfourl <$ 130~\gev}. The uncertainty in the predictions is shown by the hatched band, which includes the theoretical uncertainties of the SM cross-section for the signal and the background processes.  \label{fig:differentialPrefit_1} }
\end{figure}

\begin{figure}
\centering
\subfloat[]{\includegraphics[width = 0.49 \textwidth]{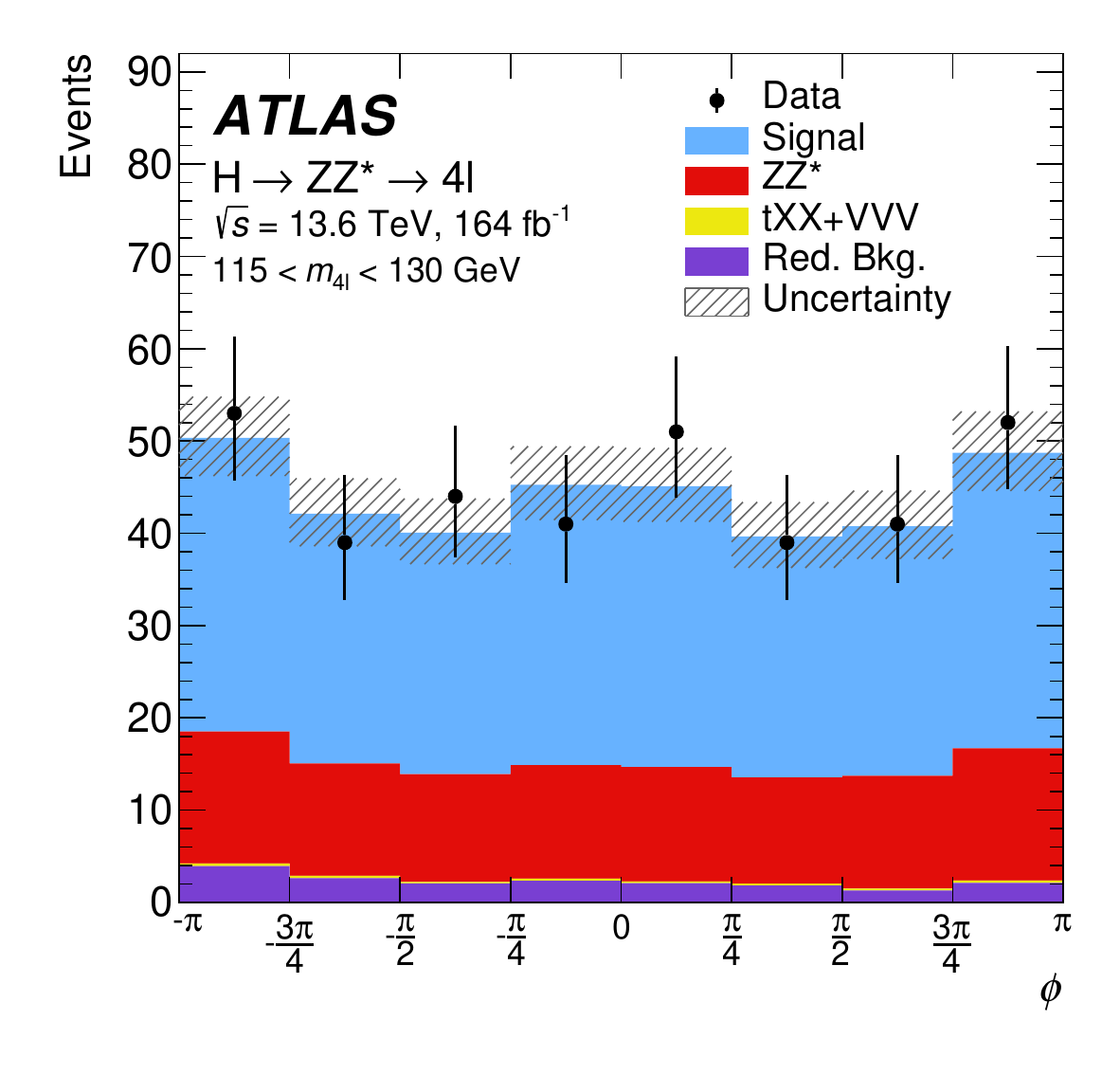}}
\subfloat[]{\includegraphics[width = 0.49 \textwidth]{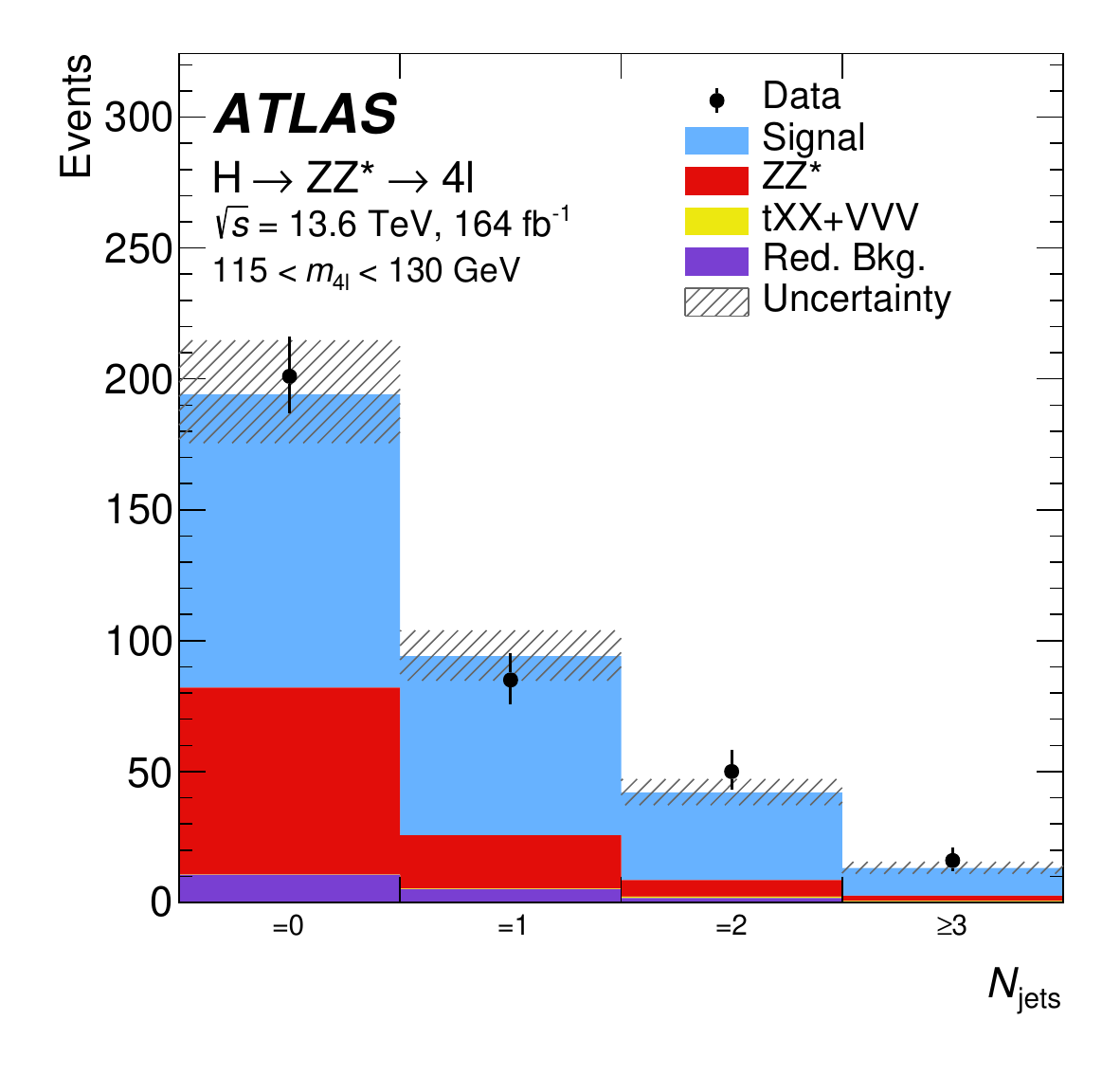}}\\
\subfloat[]{\includegraphics[width = 0.49 \textwidth]{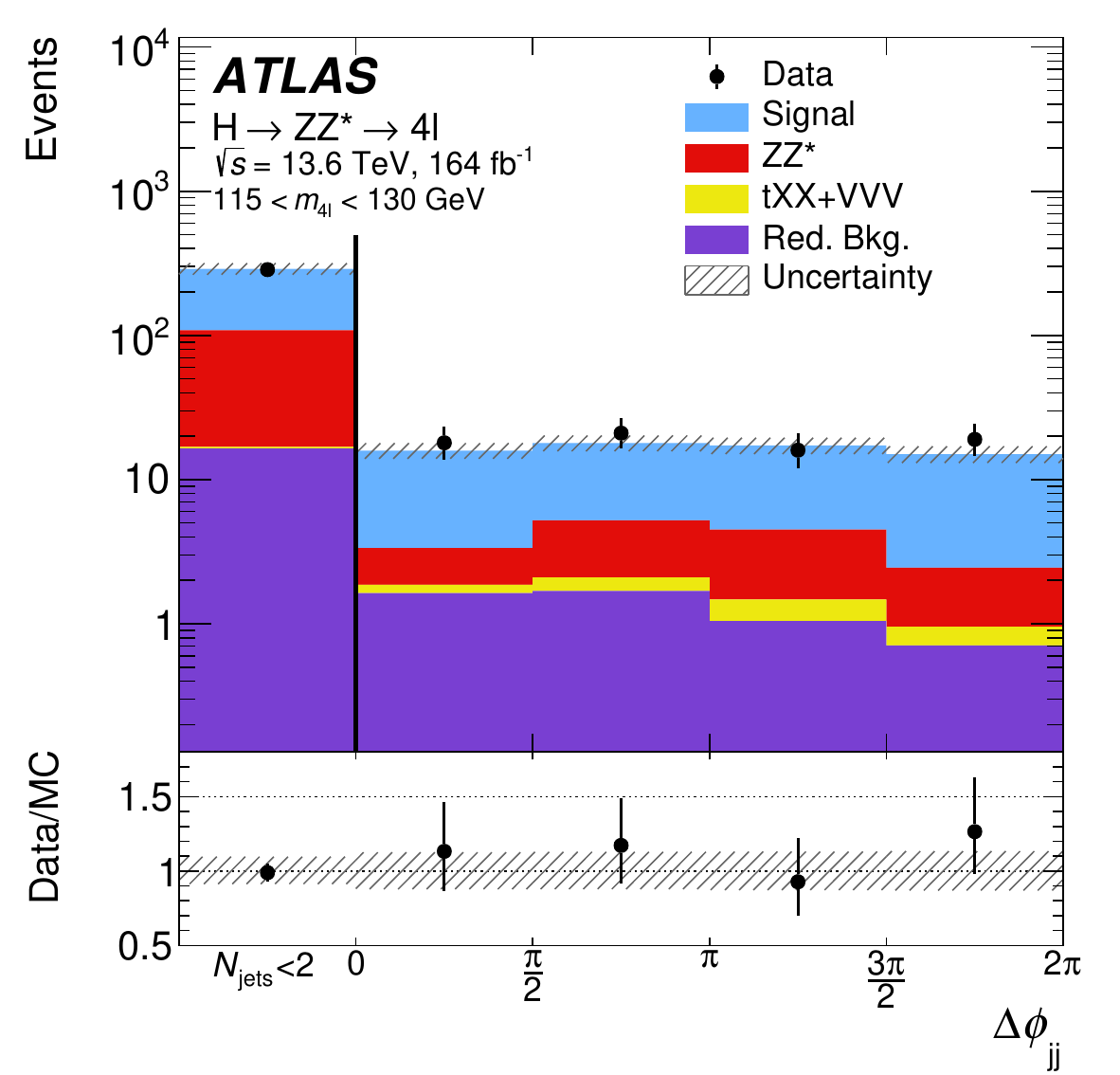}}
\subfloat[]{\includegraphics[width = 0.49 \textwidth]{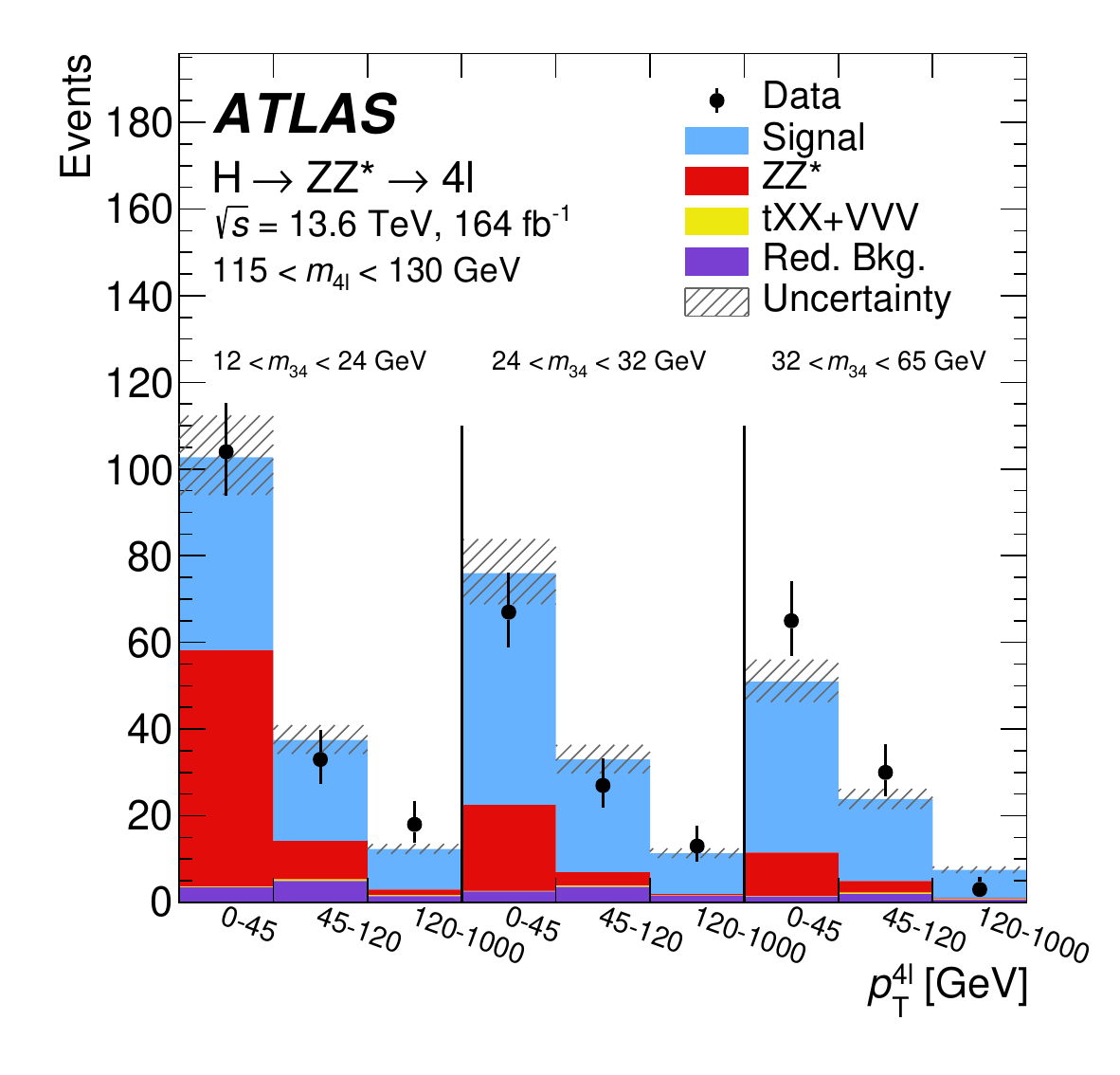}}
\caption{The observed and expected (pre-fit) distributions of (a) $\phi$, (b) $\Njets$, (c) $\dphijj$ and (d) $\pTmthreefour$ observables, after the event selection in the mass range  \hbox{115~\gev$< \mfourl <$ 130~\gev}. The uncertainty in the predictions is shown by the hatched band, which includes the theoretical uncertainties of the SM cross-section for the signal and the background processes. \label{fig:differentialPrefit_2} }
\end{figure}

Figure~\ref{fig:differentialPrefit_1} and Figure~\ref{fig:differentialPrefit_2} show the expected (pre-fit) and observed distributions of the $\ptFourL$, $\yFourL$, $\mOneTwo$, $\mThreeFour$, $\Njets$, $\dphijj$, $\phi$ observables used for the differential cross-section measurements, and the $\pTmthreefour$ distribution used for the double-differential cross-section measurement. The data are found to be largely in agreement with the prediction.

\begin{figure}[!htbp]
\centering
\subfloat[]{\includegraphics[width = 0.45 \textwidth]{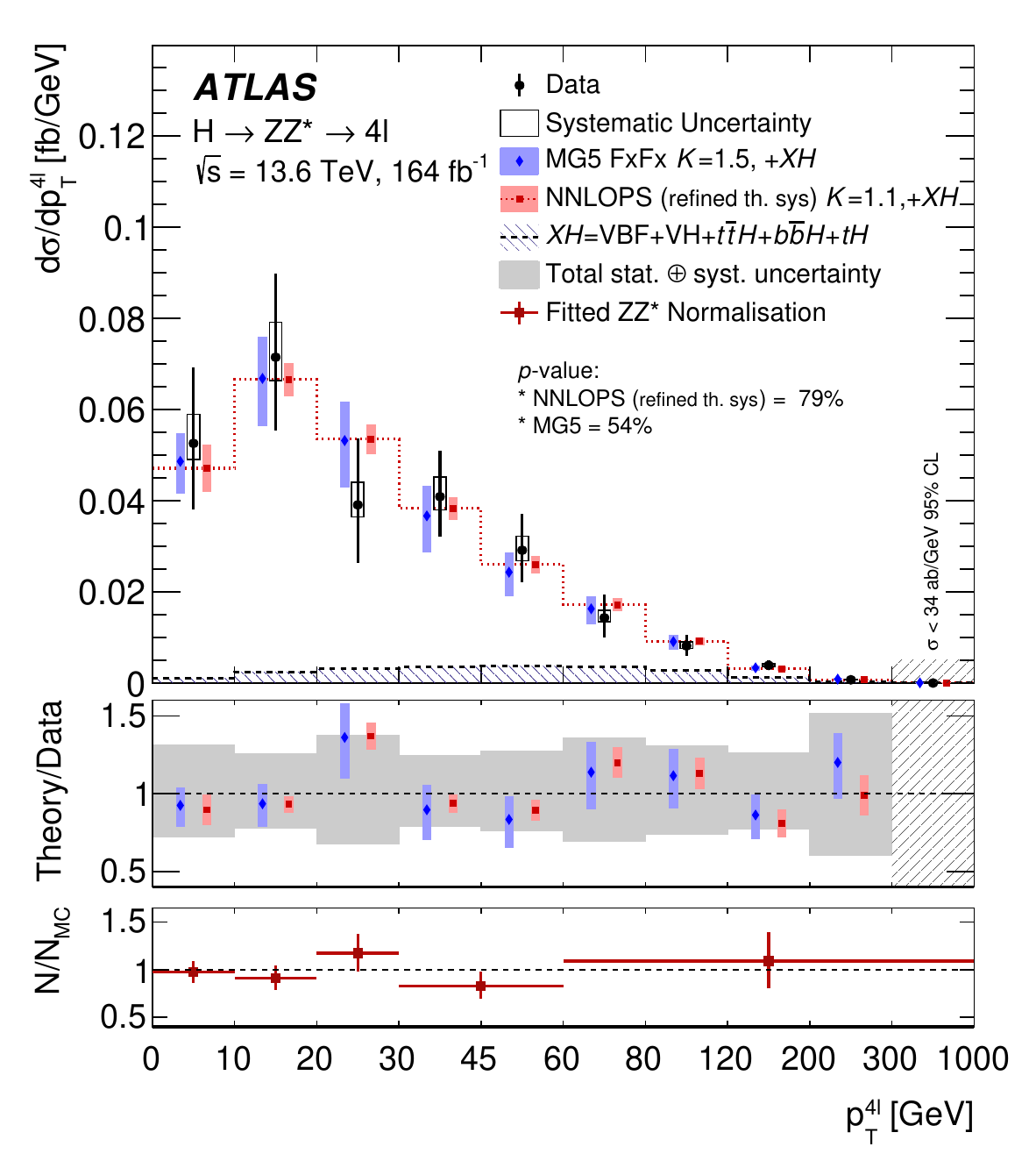}}
\subfloat[]{\includegraphics[width = 0.45 \textwidth]{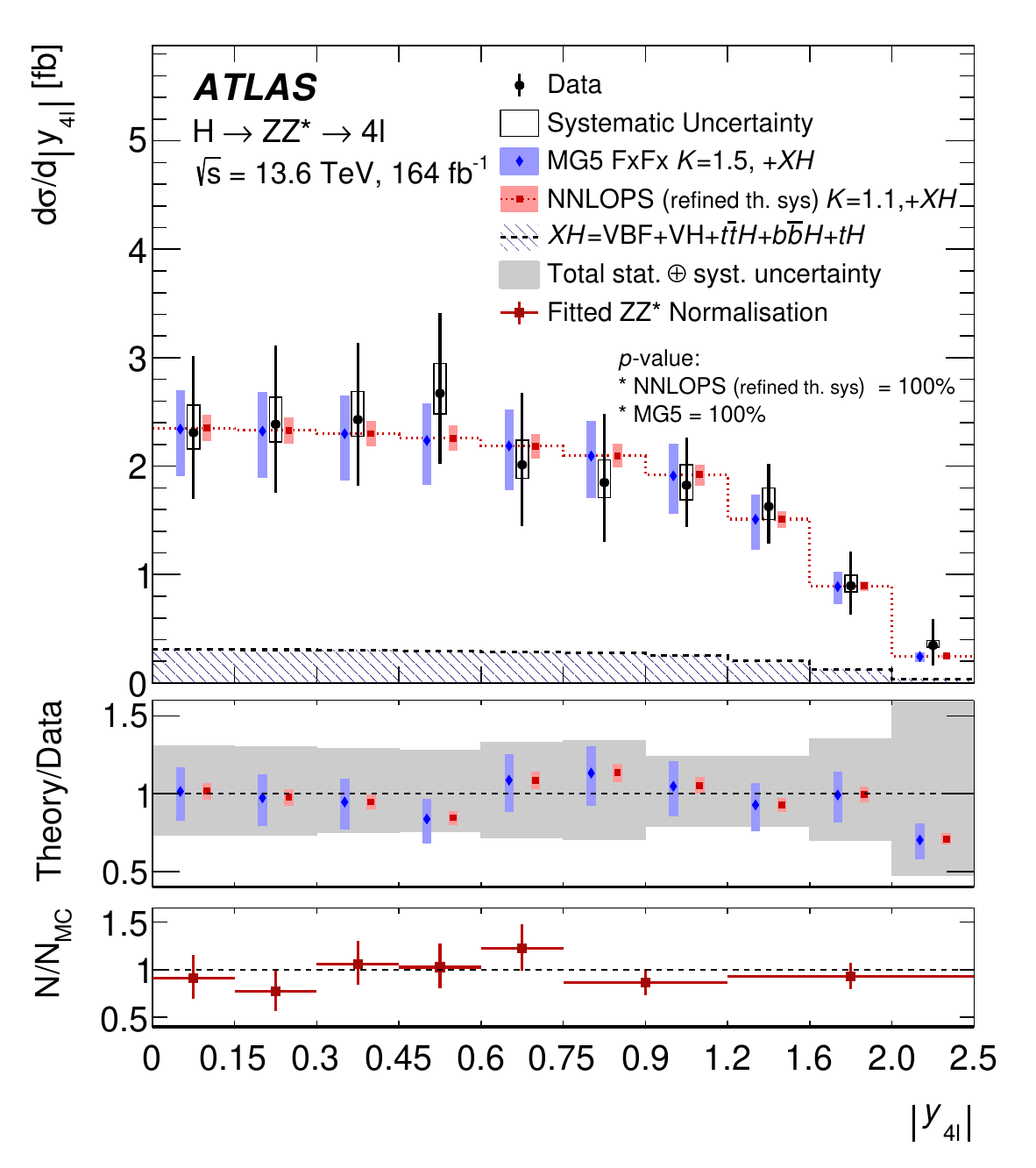}} \\
\subfloat[]{\includegraphics[width = 0.45 \textwidth]{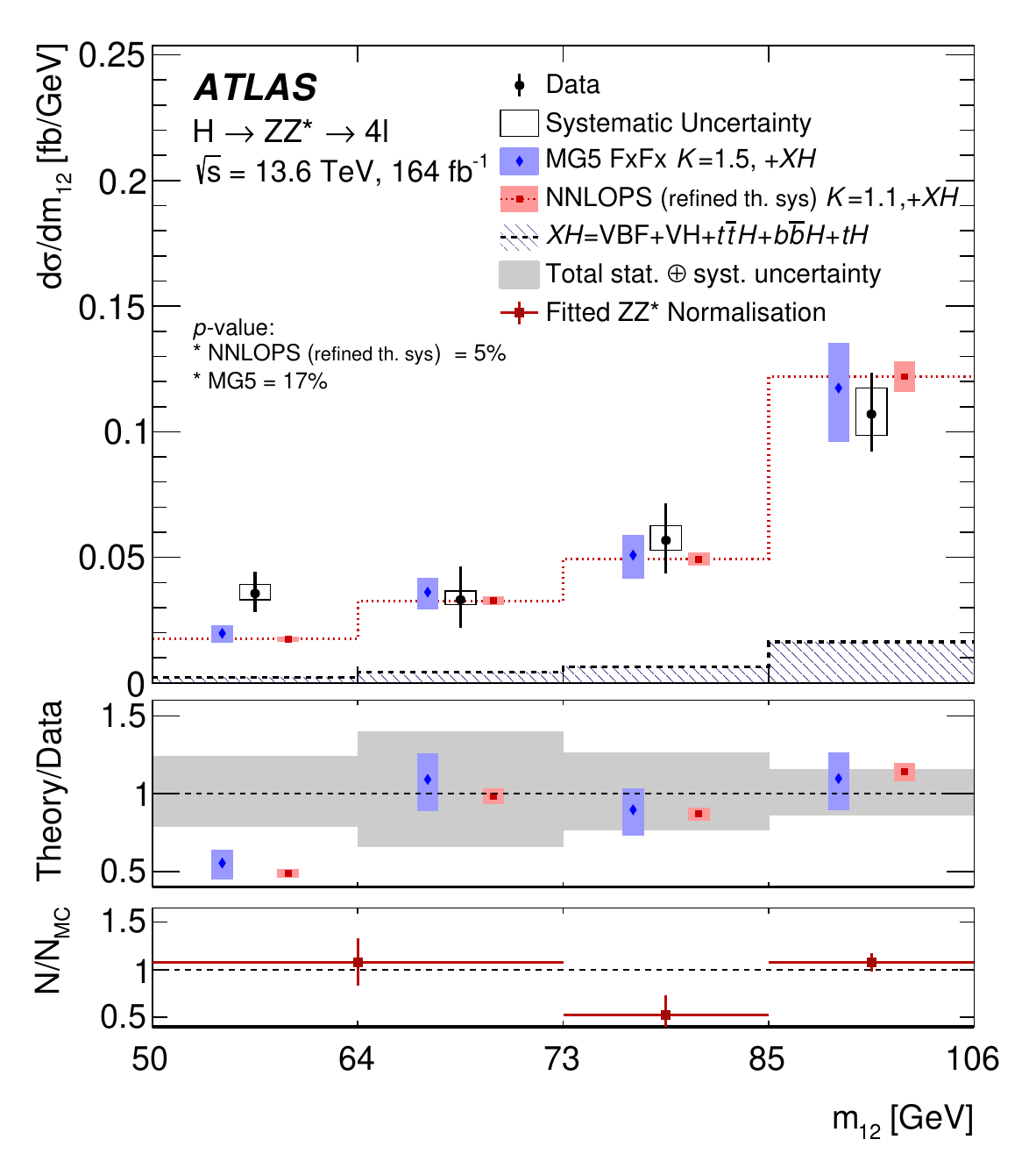}}
\subfloat[]{\includegraphics[width = 0.45 \textwidth]{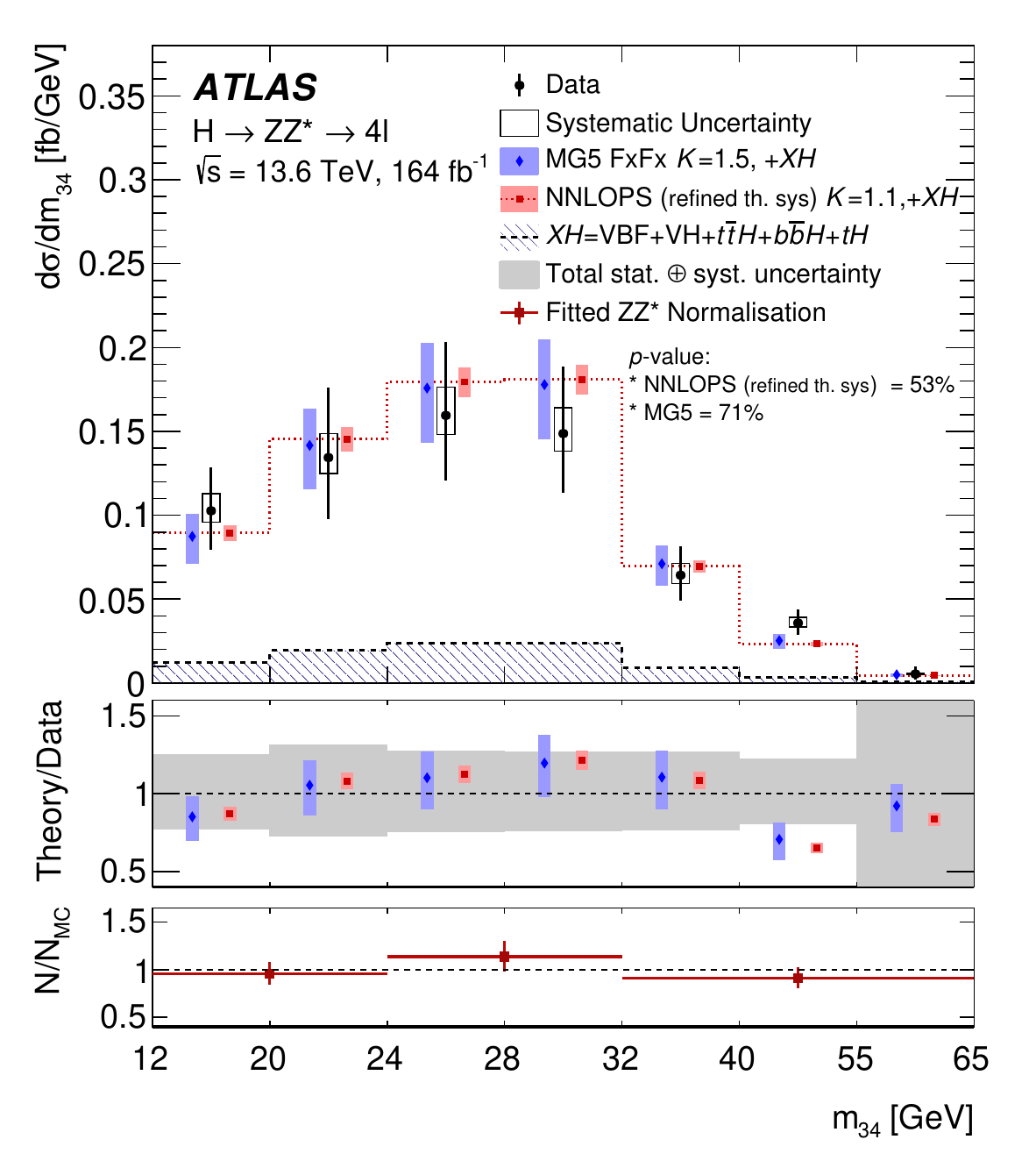}}
\caption{Differential fiducial cross-sections as a function of (a) \ptFourL, (b) $|\yFourL|$, (c) \mOneTwo, and (d) \mThreeFour. The upper panel shows the measured cross-sections, compared with ggF predictions by \POWHEG (NNLOPS) and \MGNLO (MG5), which are normalised to the N$^{3}$LO total cross-section with the listed $K$-factors. The shaded bands on the expected cross-sections indicate the PDF and scale systematic uncertainties, in particular the ones for NNLOPS are calculated as described in Ref.~\cite{ATL-PHYS-PUB-2023-031}, while for MG5 are estimated from varying factorisation and renormalisation scales. The markers denoting the simulation are horizontally displaced relative to the measurement in data to improve readability. MC-based predictions for all other Higgs boson production-modes $XH$ are normalised to the SM predictions. The error bars on the data points show the total uncertainties, while the systematic uncertainties are indicated by the boxes. The middle panel shows the ratio of different predictions to the data, and the grey shaded area represents the total uncertainty of the measurement. The bottom panel shows the ratios of the fitted values of the $ZZ^*$ normalisation factors to the predictions from MC simulation. No data candidates are observed in the final bin of the \ptFourL distribution in (a), and a 95\% CL limit on the cross-section is set as indicated by the hatched lines. \label{fig:diffXS_1} }
\end{figure}

\begin{figure}[!htbp]
\centering
\subfloat[]{\includegraphics[width = 0.45 \textwidth]{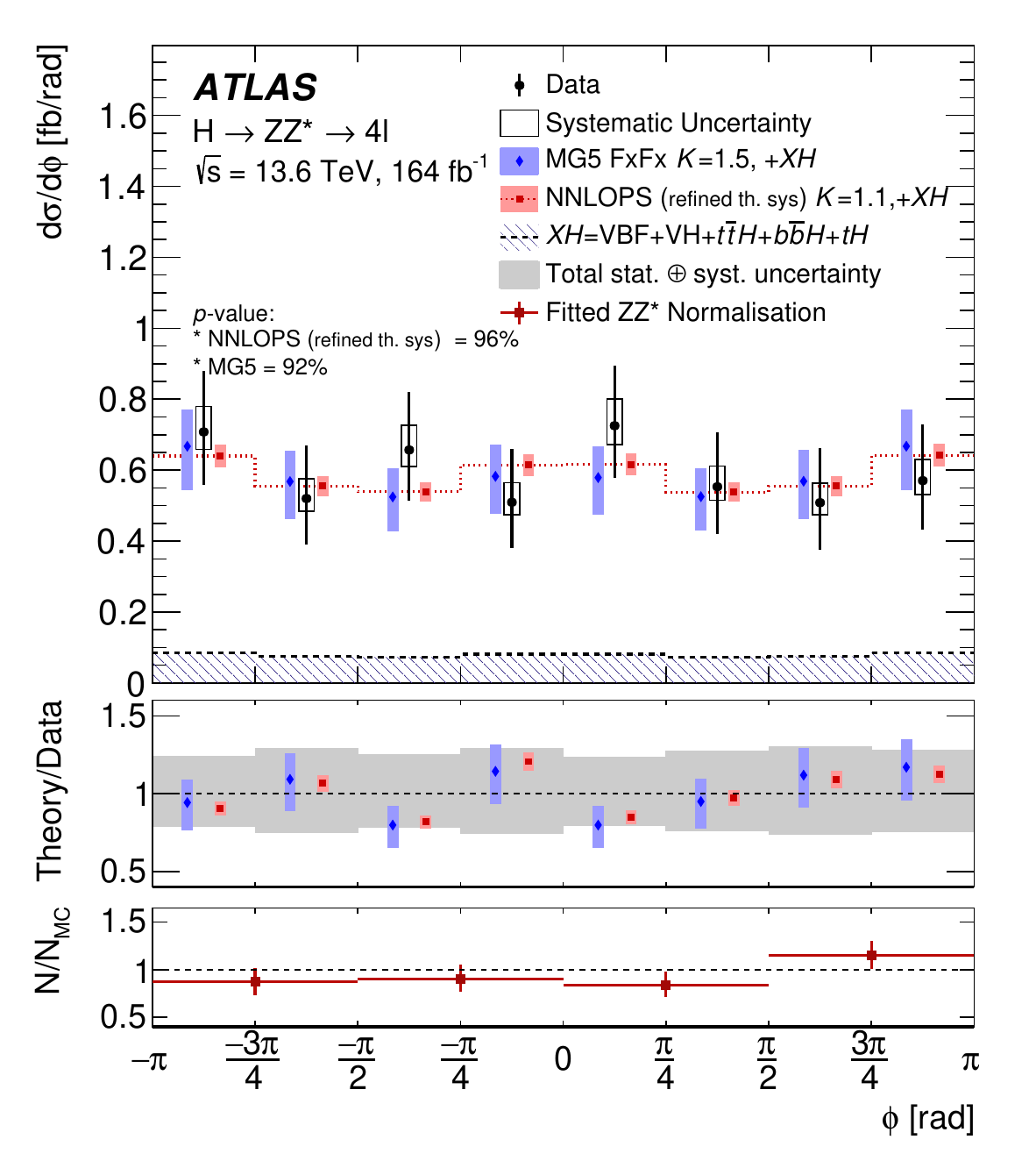}}
\subfloat[]{\includegraphics[width = 0.45 \textwidth]{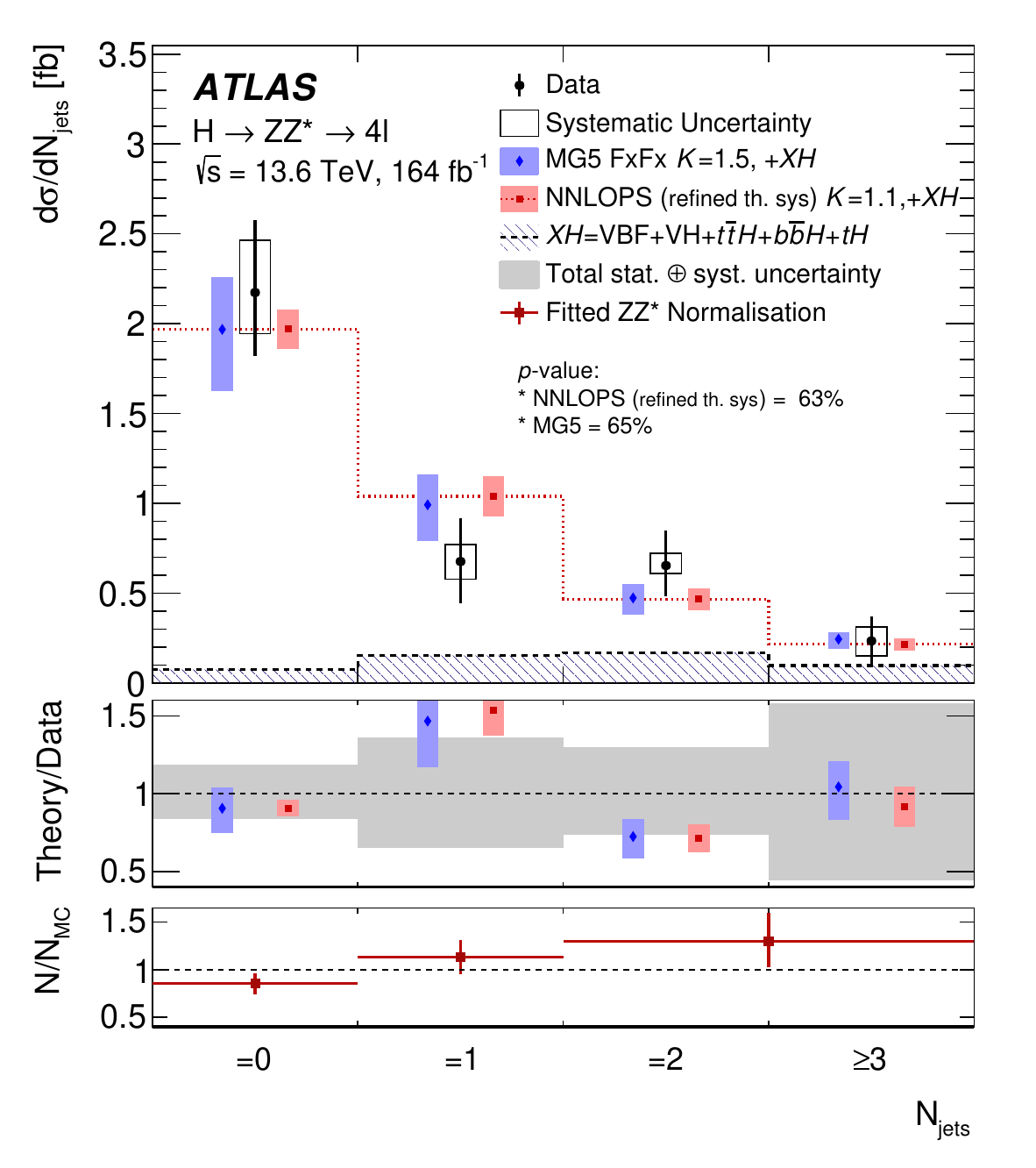}} \\
\subfloat[]{\includegraphics[width = 0.45 \textwidth]{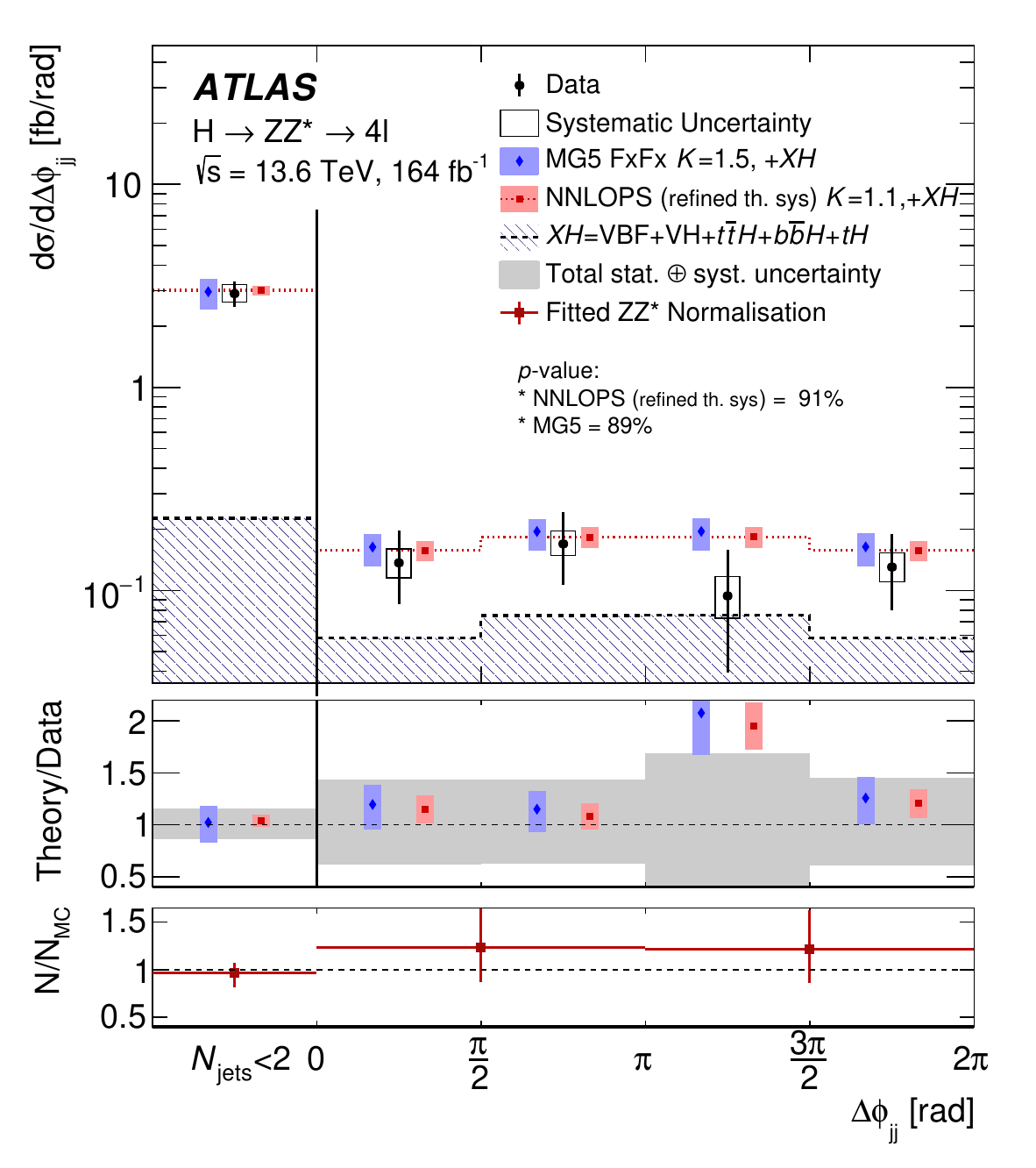}}
\subfloat[]{\includegraphics[width = 0.45 \textwidth]{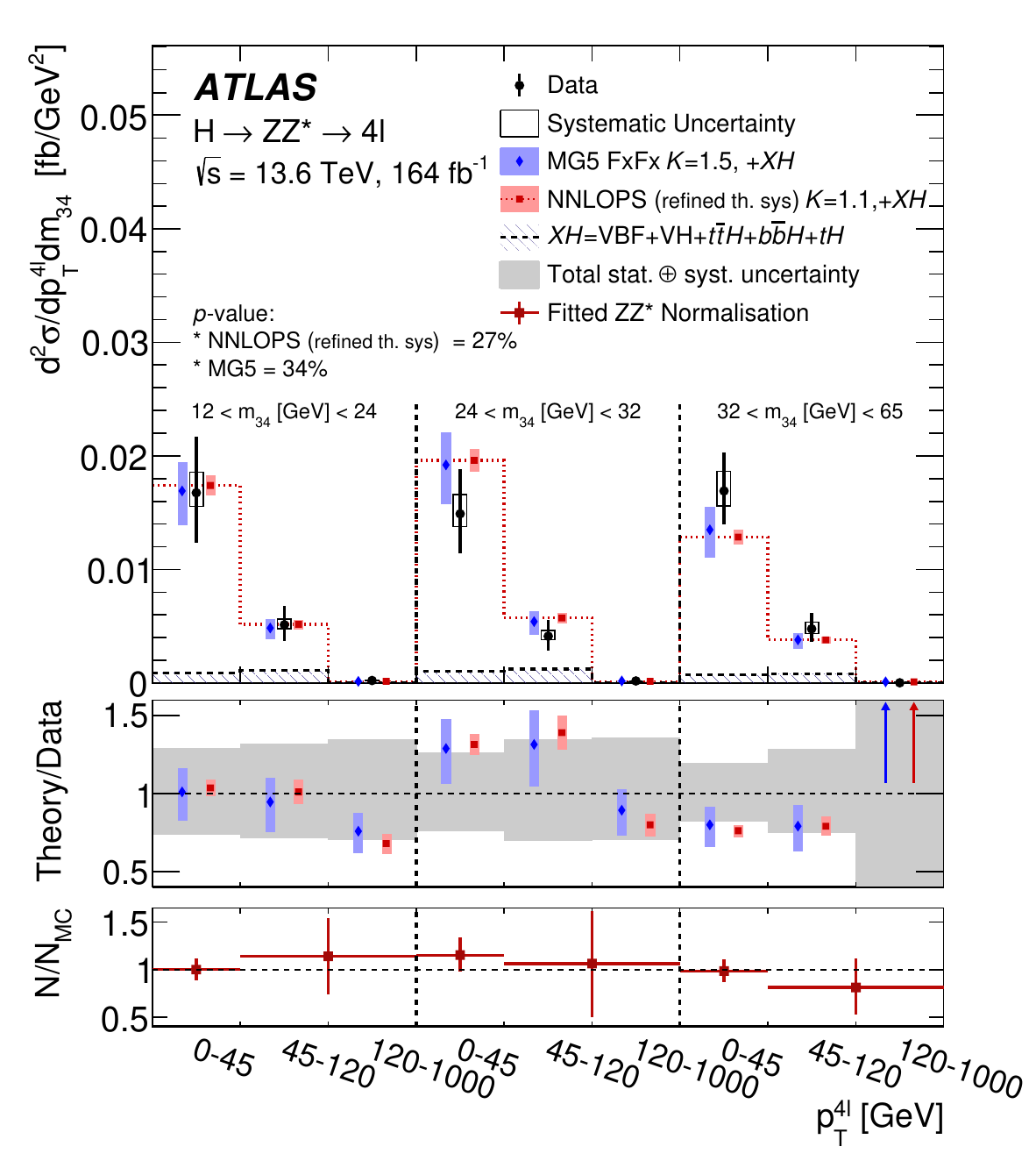}}
\caption{Differential fiducial cross-sections as a function of (a) $\phi$, (b) $\Njets$, (c) $\dphijj$ and (d) $\pTmthreefour$. The upper panel shows the measured cross-sections, compared with ggF predictions by \POWHEG (NNLOPS) and \MGNLO (MG5), which are normalised to the N$^{3}$LO total cross-section with the listed $K$-factors. The shaded bands on the expected cross-sections indicate the PDF and scale systematic uncertainties, in particular the ones for NNLOPS are calculated as described in Ref.~\cite{ATL-PHYS-PUB-2023-031}, while for MG5 are estimated from varying factorisation and renormalisation scales. The markers denoting the simulation are horizontally displaced relative to the measurement in data to improve readability. MC-based predictions for all other Higgs boson production-modes $XH$ are normalised to the SM predictions. The error bars on the data points show the total uncertainties, while the systematic uncertainties are indicated by the boxes. The middle panel shows the ratio of different predictions to the data, and the grey shaded area represents the total uncertainty of the measurement. The bottom panel shows the ratios of the fitted values of the $ZZ^*$ normalisation factors to the predictions from MC simulation. \label{fig:diffXS_2}} %
\end{figure}

The corresponding measured differential cross-sections obtained from these distributions using the procedure discussed in Section~\ref{subsec:diffxs} and the corresponding normalisation factors for the background, $N/N_{\mathrm{MC}}$, are depicted in Figure ~\ref{fig:diffXS_1} and Figure ~\ref{fig:diffXS_2}. Uncertainties are statistically dominated, and the subdominant systematic component arises primarily from lepton efficiency and (for jet-related observables) jet energy calibration uncertainties. The data are compared with SM expectations constructed from the ggF predictions provided by \POWHEG and \MGNLO, which are normalised to N$^3$LO cross-section as described in Section~\ref{sec:simulation:signal:norm}, with k-factors of 1.1 and 1.5 respectively. The uncertainty for \POWHEG predictions is estimated from the procedure described in Section~\ref{sec:systematics}, while those for \MGNLO are estimated from varying factorisation and renormalisation scales. The probability of compatibility of the measured differential cross-section and the SM expectation ranges from 5\% to approximately 100\%, with the $\mOneTwo$ being the least compatible with SM predictions due to a fluctuation in the first bin of the distribution.

\FloatBarrier
\subsection{Cross-sections per production-mode}

\begin{figure}[!htbp]
\begin{center}
\subfloat[]{
\includegraphics[width=0.45\linewidth]{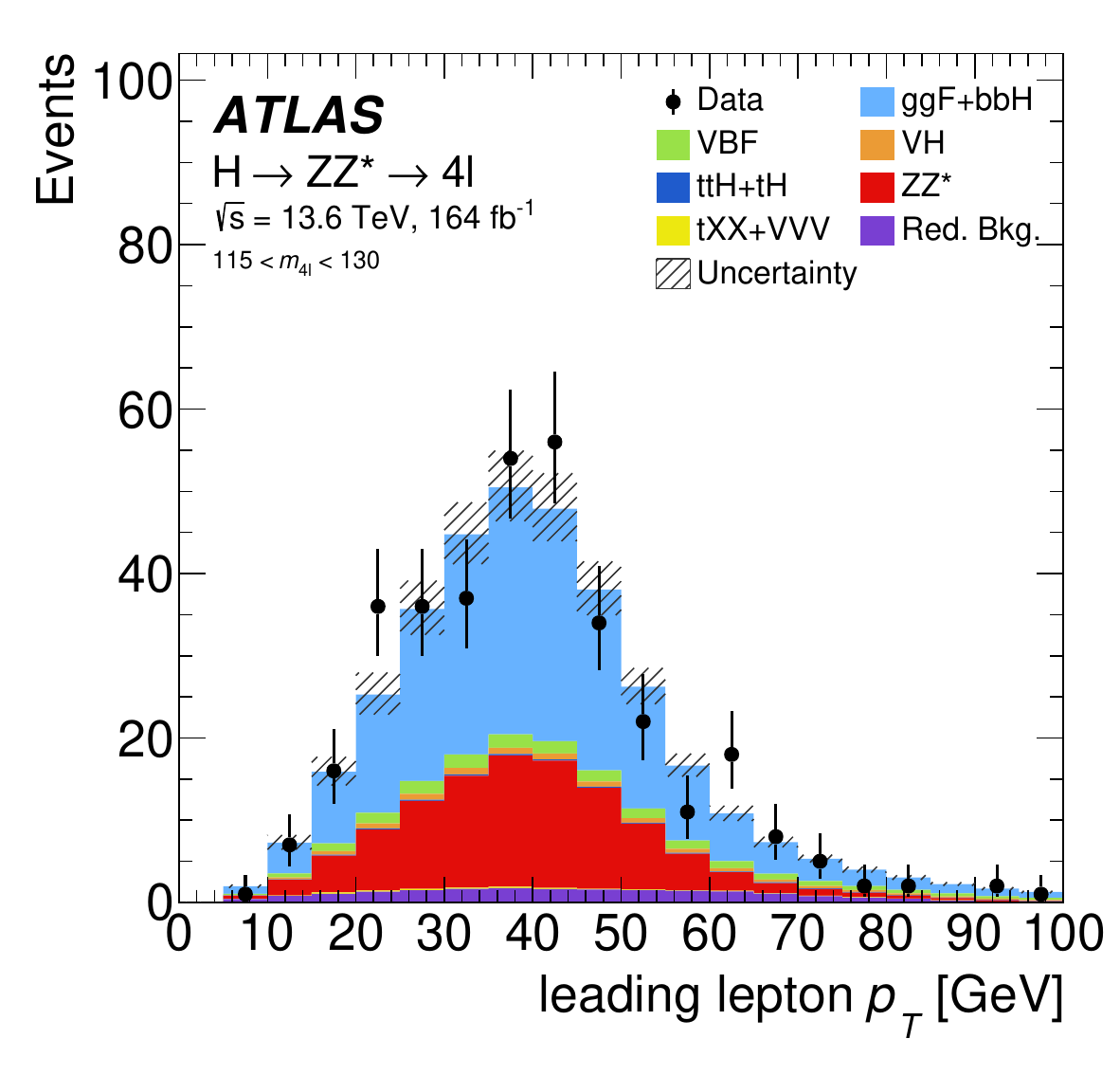}}
\subfloat[]{
\includegraphics[width=0.45\linewidth]{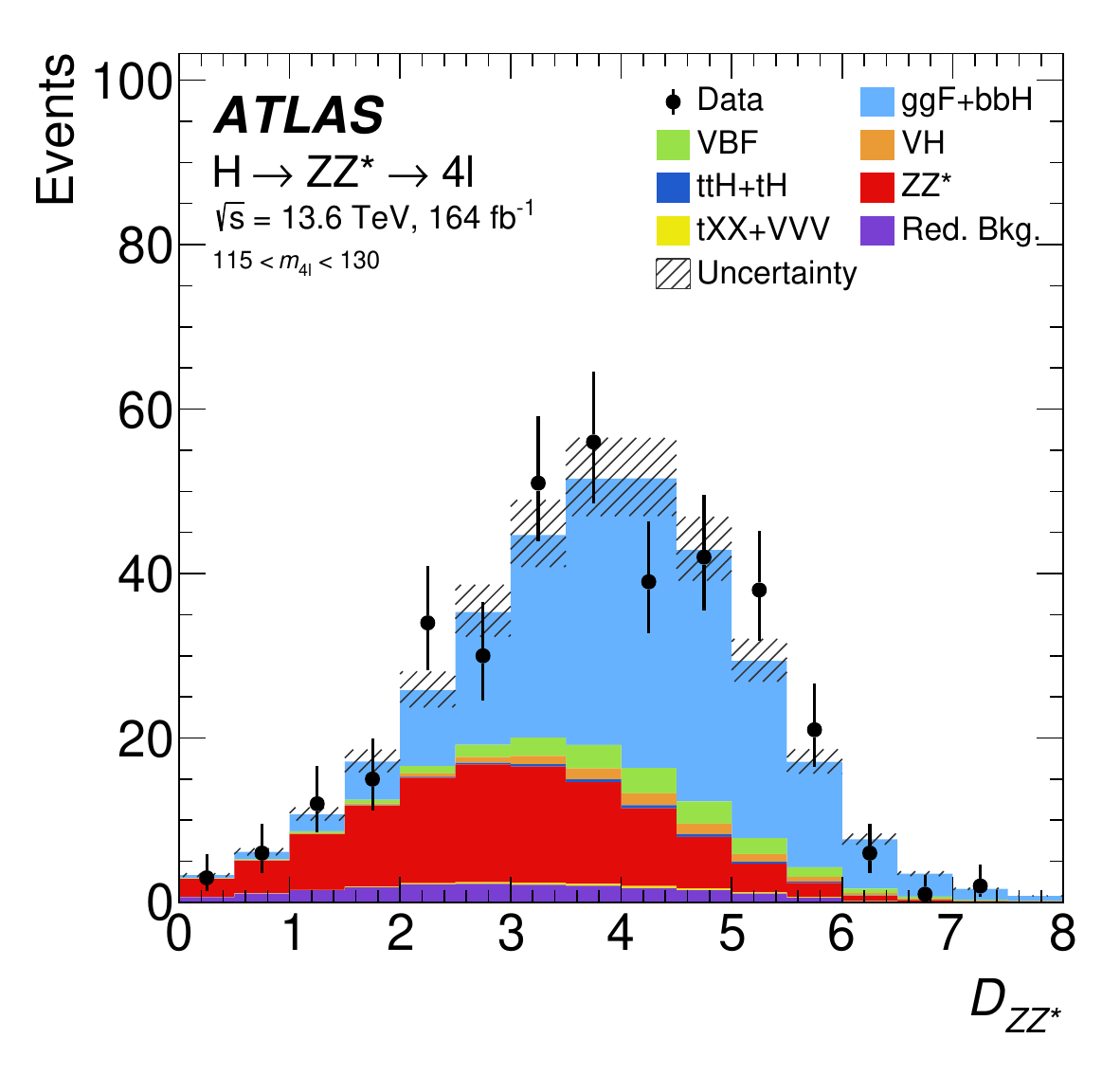}}\\
\subfloat[]{
\includegraphics[width=0.45\linewidth]{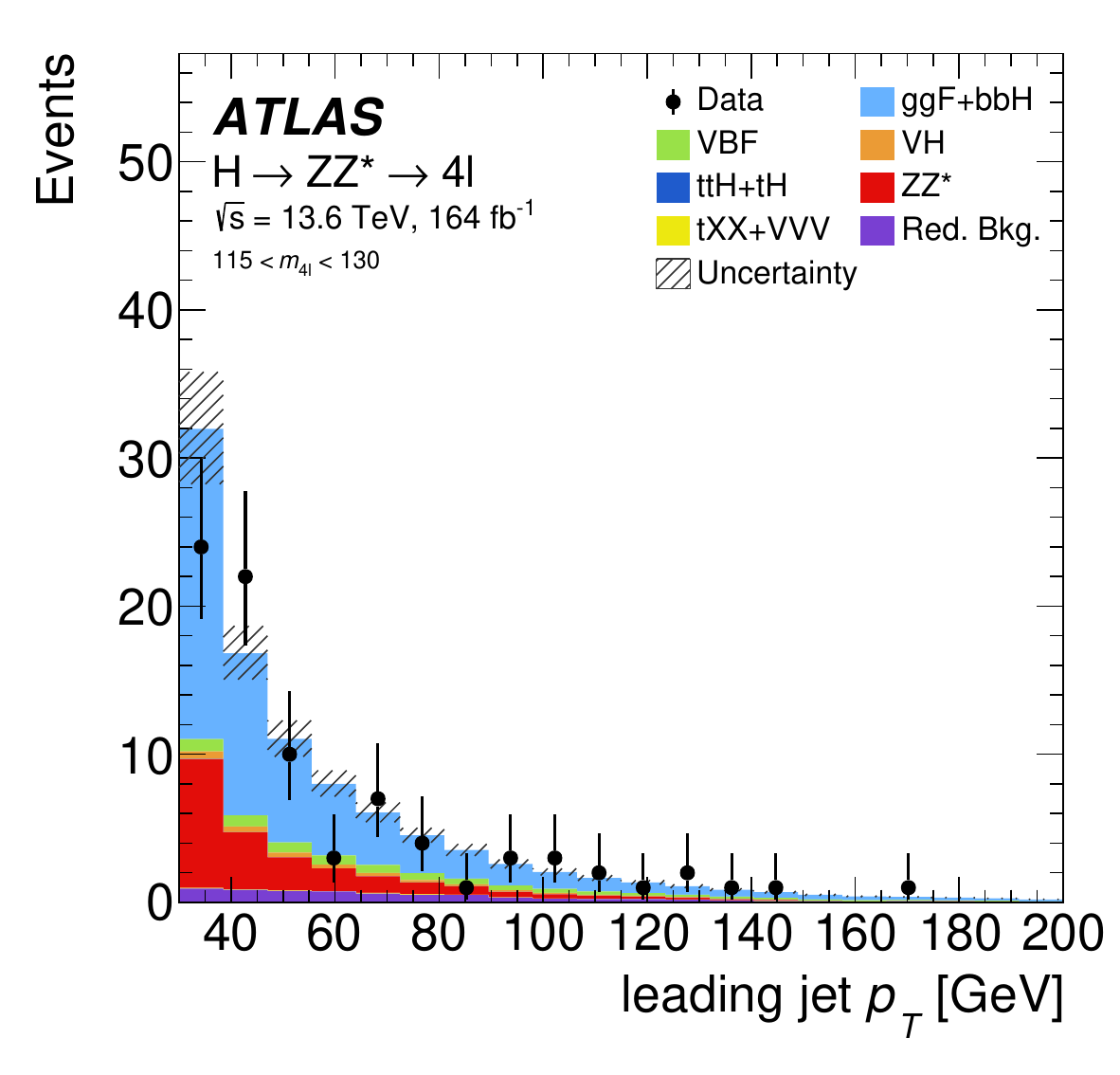}}
\subfloat[]{
\includegraphics[width=0.45\linewidth]{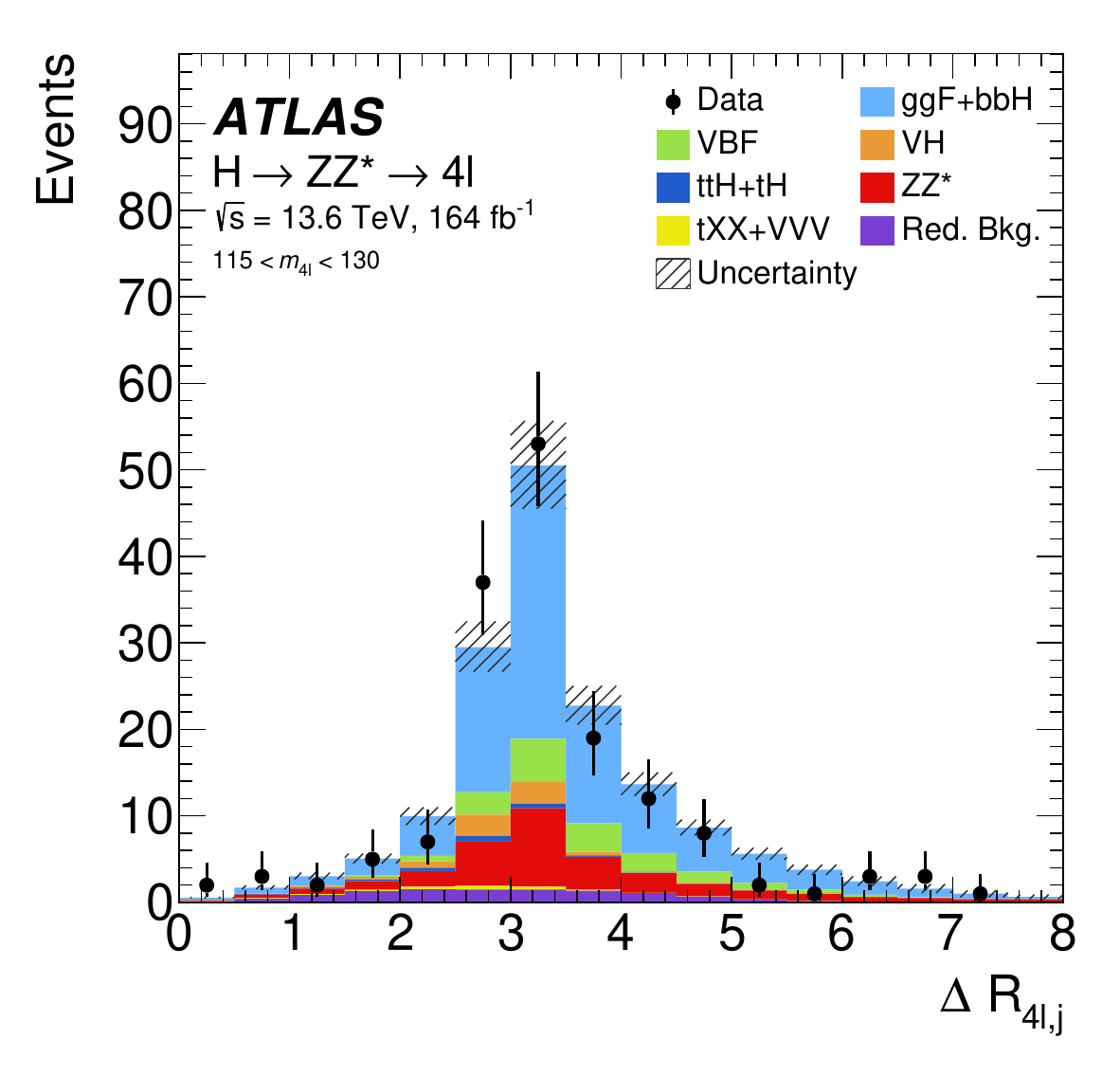}}
\end{center}
\caption{The observed and expected distributions (pre-fit) of (a) the transverse momenta of the leading lepton  (b) the kinematic discriminant $D_{ZZ^{*}}$  (c) the transverse momenta of the leading jet and (d) angular separation between the four lepton system and the leading jet $\Delta R_{4\ell,j}$. The SM Higgs boson signal is assumed to have a mass $m_{H}$ = 125~\gev. The uncertainty in the prediction is shown by the hatched band, calculated as described in Section~\ref{sec:systematics}. }
\label{fig:Cat_observables}
\end{figure}

For the production cross-section measurement, the expected numbers of signal and background events in each reconstructed event category are shown in Table~\ref{tab:event_yields_all} together with the corresponding observed number of events. They are in good agreement. The observed and expected (pre-fit) distributions of the transverse momentum of the leading lepton, the kinematic discriminant $D_{ZZ^{*}}$, the transverse momentum of the leading jet and the angular separation between the four lepton system and the leading jet $\Delta R_{4\ell,j}$, which are used as inputs variables for the NN training in the reconstructed categories, are shown in Figure \ref{fig:Cat_observables} and are in good agreement as well.
\begin{table*}[!htbp]
\centering
\caption{The expected (pre-fit) and observed numbers of events are shown for an integrated luminosity of \intLum\ at \sqrtS. Results are presented for the signal region \mbox{$115<m_{4\ell}<130$~\GeV\ } and the sideband regions \mbox{$105<m_{4\ell}<115$~\GeV\ } and \mbox{$130<m_{4\ell}<160$~\GeV\ (350~\GeV\ for \tXX-enriched)} in each reconstructed event category, assuming a SM Higgs boson signal with a mass $m_{H}=125$~\GeV. The sum of the expected SM Higgs boson signal and the estimated background yields is compared with the data. Combined statistical and systematic uncertainties are included for the predictions. Expected contributions below 0.2\% of the total yield in each reconstructed event category are not shown and are replaced by `-'.}
\label{tab:event_yields_all}
\vspace{0.1cm}
\vspace{0.1cm}
{\renewcommand{\arraystretch}{1.1}
\resizebox{\linewidth}{!}{
\begin{tabular}{l ccccccc}
\hline\hline
Reconstructed&
\multicolumn{1}{c}{Signal}&
\multicolumn{1}{c}{\zzstar}&
\multicolumn{1}{c}{$tXX$}&
\multicolumn{1}{c}{Other}&
\multicolumn{1}{c}{Total}&
\multicolumn{1}{c}{Observed}\\

event category&
&
\multicolumn{1}{c}{background}&
\multicolumn{1}{c}{background}&
\multicolumn{1}{c}{backgrounds}&
\multicolumn{1}{c}{expected}&
\\

\hline

Signal&
\multicolumn{6}{c}{$115<m_{4\ell}<130$ \GeV}\\  \hline
\CatZeroJL  & 26 $\pm$ 5        & 31.4 $\pm$ 3.2        & -                     & 1.750 $\pm$ 0.035     & 59 $\pm$ 7        & 64\\
\CatZeroJM  & 85 $\pm$ 11       & 40 $\pm$ 4            & -                     & 5.86 $\pm$ 0.26       & 131 $\pm$ 15      & 135\\
\CatZeroJH  & 0.38 $\pm$ 0.05   & 0.041 $\pm$ 0.007     & 0.0243 $\pm$ 0.0019   & 0.149 $\pm$ 0.014     & 0.60 $\pm$ 0.06   & 1\\
\CatOneJL   & 38 $\pm$ 5        & 15.3 $\pm$ 2.1        & 0.112 $\pm$ 0.015     & 3.21 $\pm$ 0.33       & 57 $\pm$ 7        & 54\\
\CatOneJM   & 23.2 $\pm$ 3.4    & 4.3 $\pm$ 0.4         & 0.138 $\pm$ 0.019     & 1.69 $\pm$ 0.18       & 29 $\pm$ 4        & 24\\
\CatOneJH   & 5.4 $\pm$ 0.9     & 0.56 $\pm$ 0.05       & 0.047 $\pm$ 0.007     & 0.2861 $\pm$ 0.0013   & 6.3 $\pm$ 0.9     & 7\\
\CatOneJBSM & 1.4 $\pm$ 0.4     & 0.083 $\pm$ 0.007     & 0.0076 $\pm$ 0.0010   & 0.030 $\pm$ 0.009     & 1.5 $\pm$ 0.4     & 0\\
\CatTwoJ    & 44 $\pm$ 7        & 8.9 $\pm$ 1.4         & 0.96 $\pm$ 0.14       & 4.07 $\pm$ 0.33       & 58 $\pm$ 8        & 67\\
\CatTwoJBSM & 4.1 $\pm$ 0.8     & 0.197 $\pm$ 0.013     & 0.033 $\pm$ 0.004     & 0.58 $\pm$ 0.06       & 4.9 $\pm$ 0.8     & 4\\
\CatVHLep   & 1.42 $\pm$ 0.12   & 0.256 $\pm$ 0.029     & 0.059 $\pm$ 0.006     & 0.365 $\pm$ 0.018     & 2.10 $\pm$ 0.15   & 2\\
\CatttH     & 1.68 $\pm$ 0.20   & 0.037 $\pm$ 0.008     & 0.29  $\pm$ 0.04      & 0.301 $\pm$ 0.027     & 2.31 $\pm$ 0.21   & 2\\

\hline
Sideband &  \multicolumn{5}{c}{$105<m_{4\ell}<115$ \GeV~or $130<m_{4\ell}<160$ \GeV}\\
\hline
\CatSBZeroJ   & 5.9 $\pm$ 1.0 & 161 $\pm$ 15    & -                 & 21.06 $\pm$ 0.35  & 189 $\pm$ 16  & 153\\
\CatSBOneJ    & 3.8 $\pm$ 0.6 & 53 $\pm$ 5      & 1.64 $\pm$ 0.021  & 12.9 $\pm$ 0.4    & 71 $\pm$ 6    & 74\\
\CatSBTwoJ    & 3.0 $\pm$ 0.5 & 26 $\pm$ 4      & 4.4 $\pm$ 0.6     & 7.237 $\pm$ 0.020 & 40 $\pm$ 4    & 48\\

\hline
& \multicolumn{5}{c}{$105<m_{4\ell}<115$ \GeV~or $130<m_{4\ell}<350$ \GeV}\\
\hline
\CatSBtXX & 0.096 $\pm$ 0.012 & 0.27 $\pm$ 0.07 & 5.3 $\pm$ 0.8 & 2.29 $\pm$ 0.14 & 7.9 $\pm$ 0.8 & 15\\

\hline \hline
\noalign{\vspace{0.05cm}}
\label{tab:yields_obs_sum}
\end{tabular}
}}
\end{table*}

\FloatBarrier

\begin{figure}[htbp]
\centering
\subfloat[]{
\includegraphics[width=0.33\linewidth]{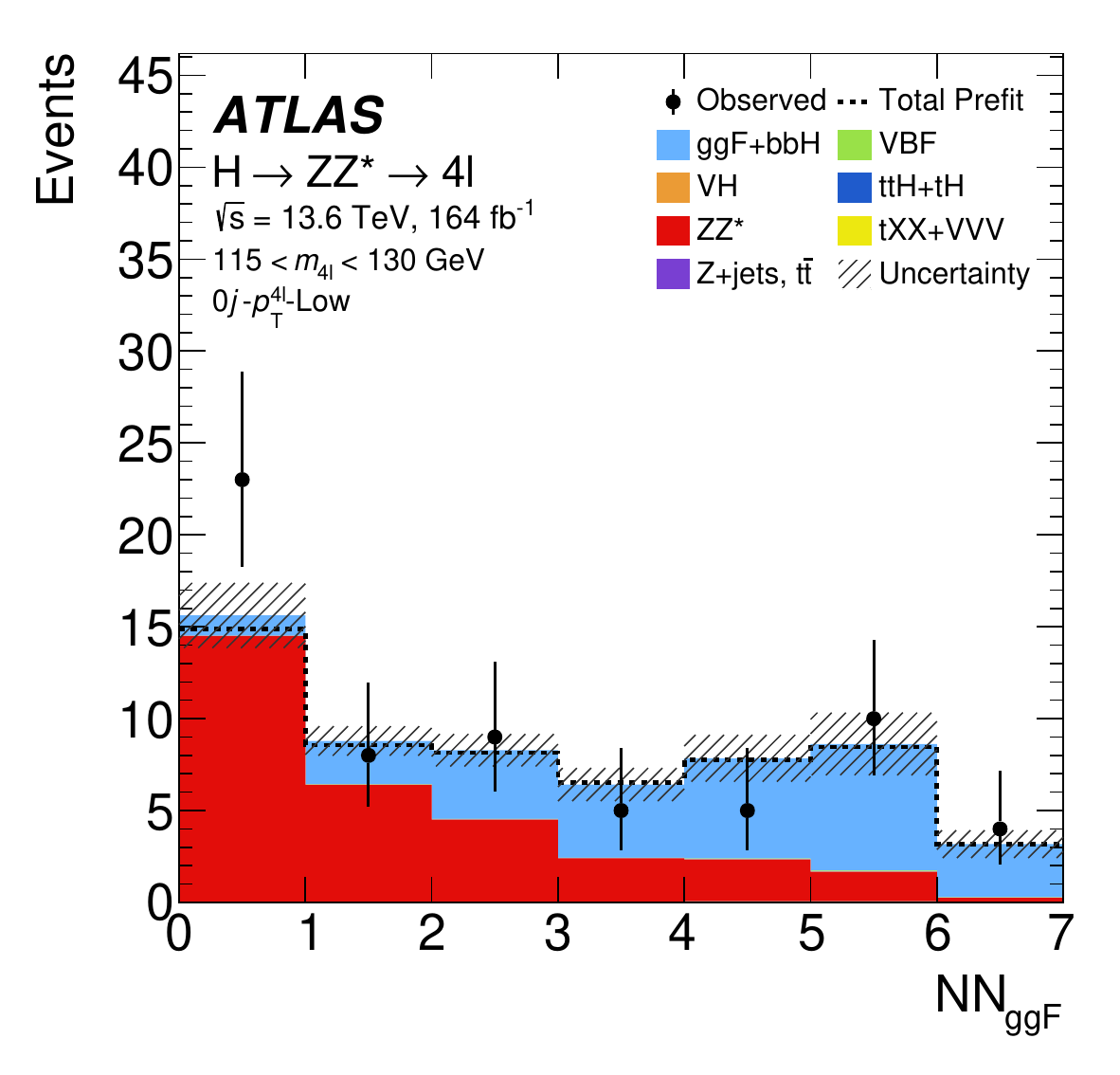}\label{fig:oNN0jptl}}
\subfloat[]{
\includegraphics[width=0.33\linewidth]{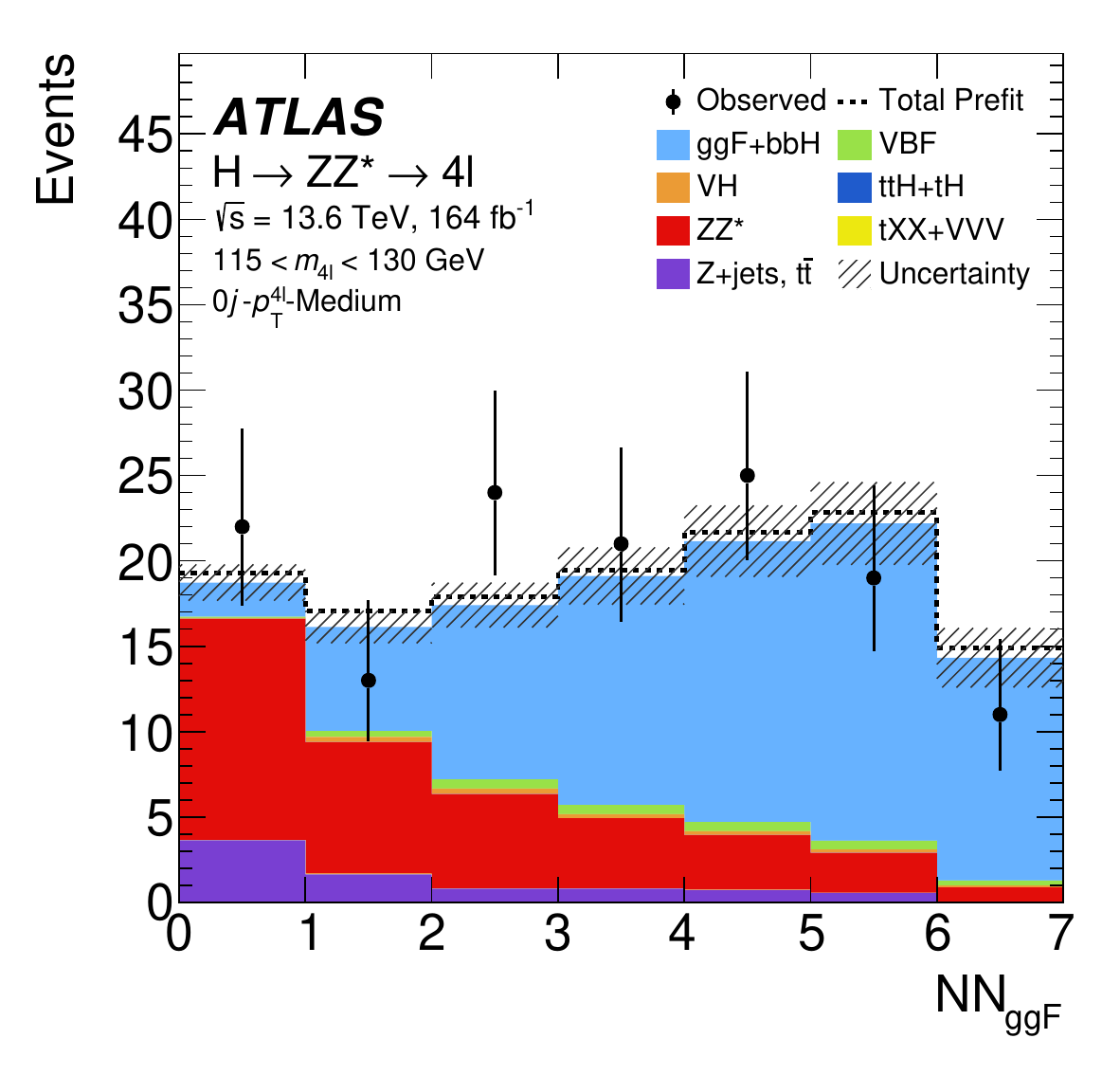}\label{fig:oNN0jpthm}}\\
\subfloat[]{
\includegraphics[width=0.33\linewidth]{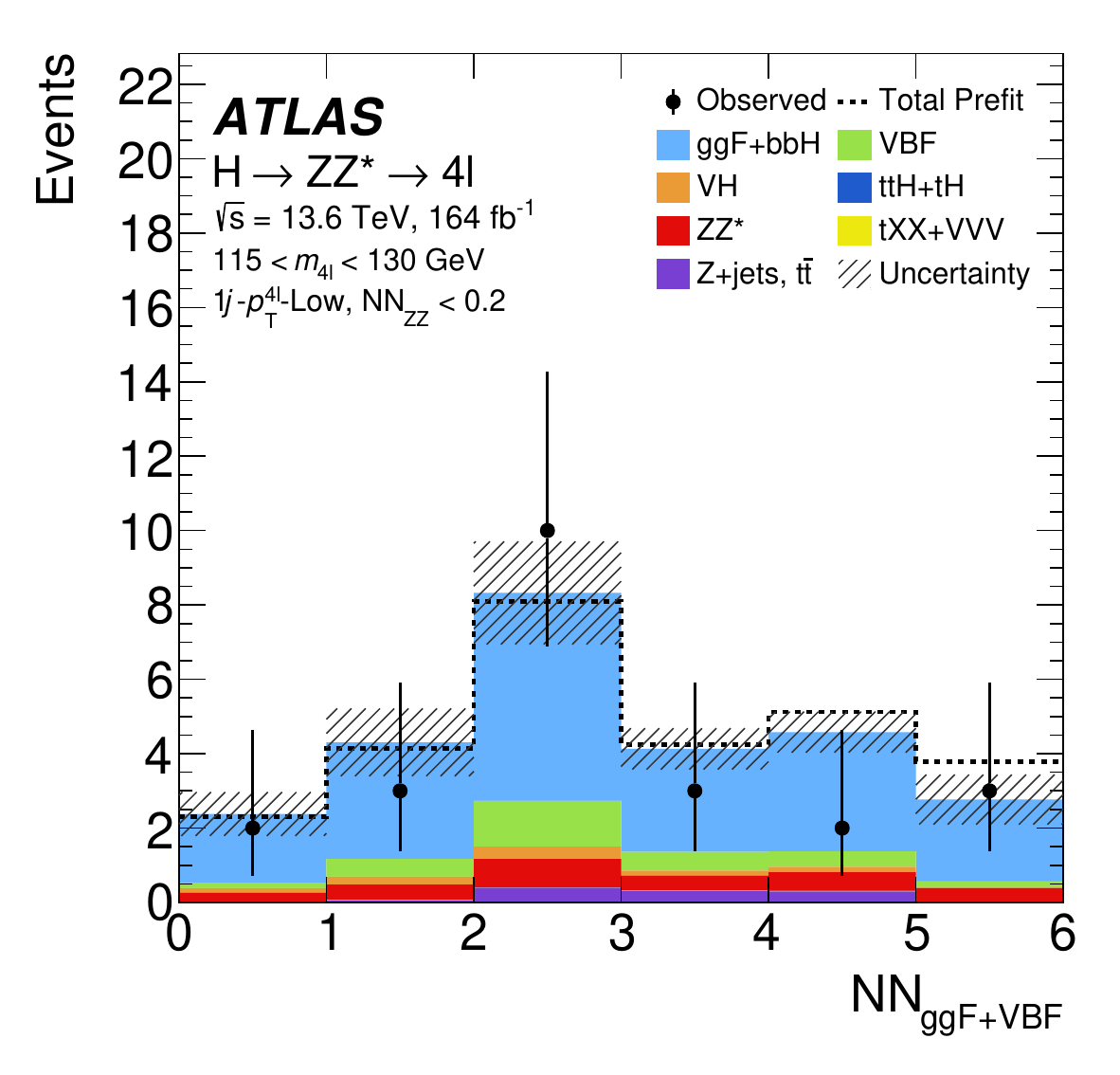}\label{fig:oNN1jpthlvbf}}
\subfloat[]{
\includegraphics[width=0.33\linewidth]{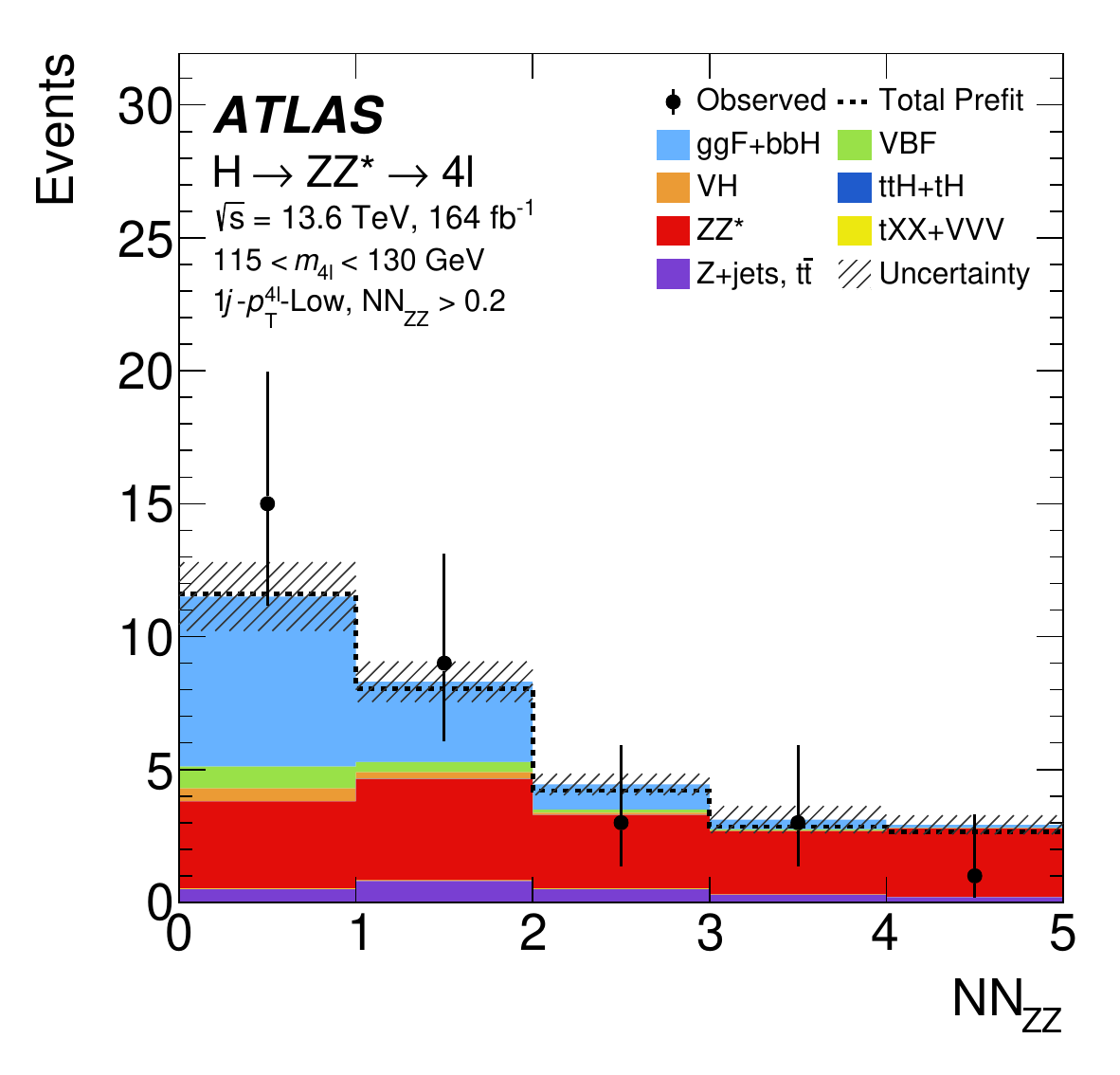}\label{fig:oNN1jptlzz}}\\
\subfloat[]{
\includegraphics[width=0.33\linewidth]{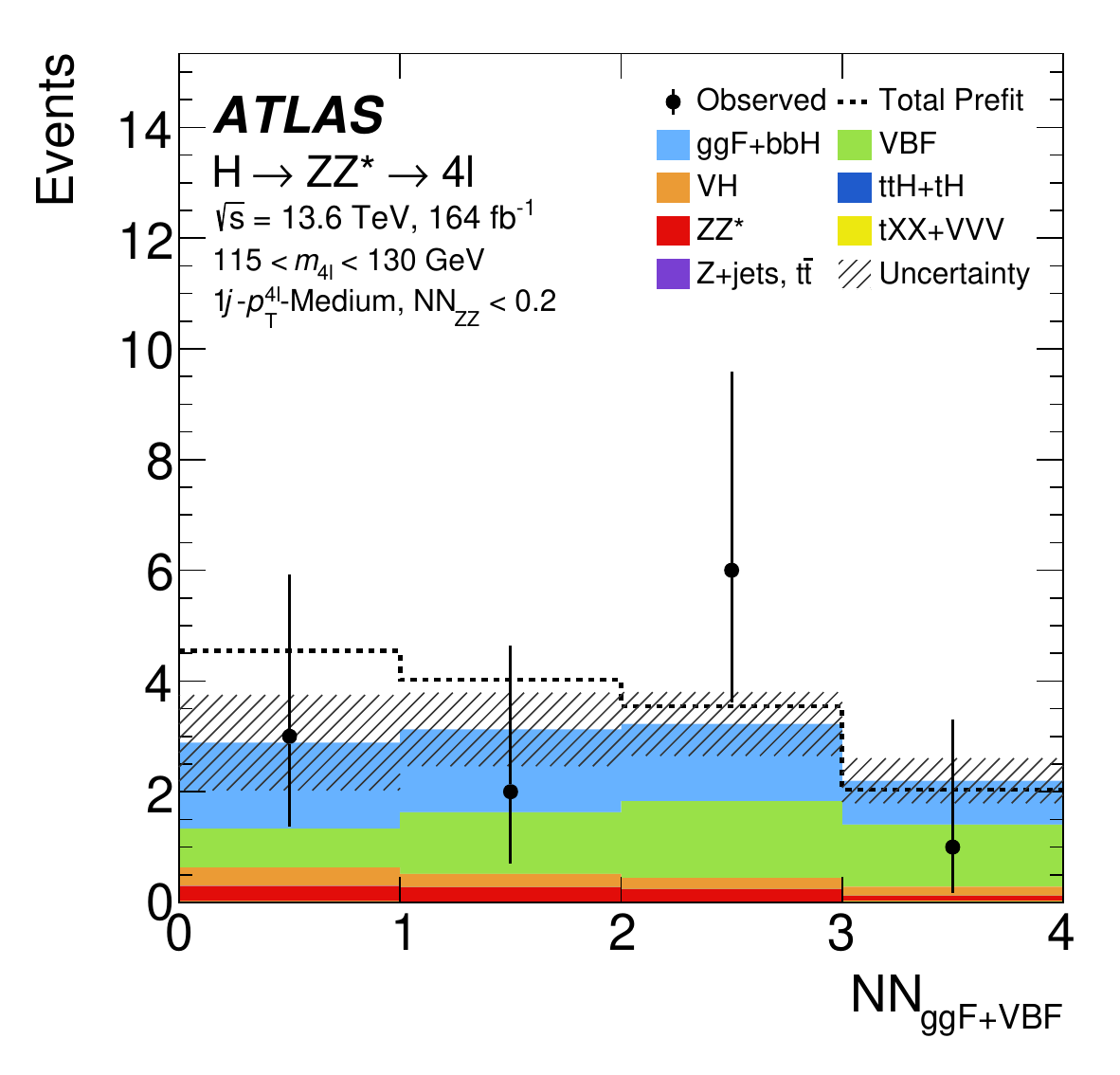}\label{fig:oNN1jpthmvbf}}
\subfloat[]{
\includegraphics[width=0.33\linewidth]{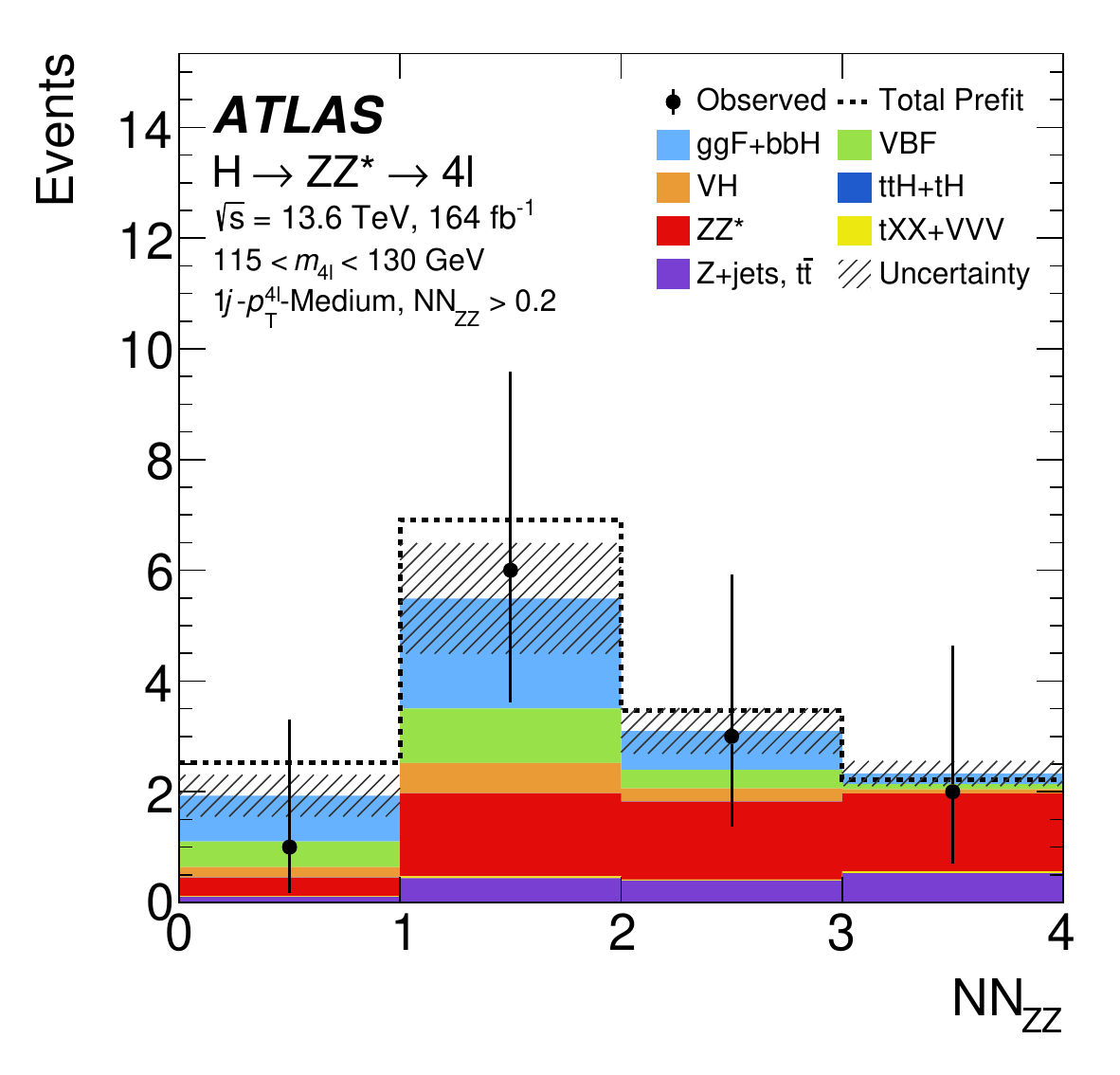}\label{fig:oNN1jpthmzz}}
\subfloat[]{
\includegraphics[width=0.33\linewidth]{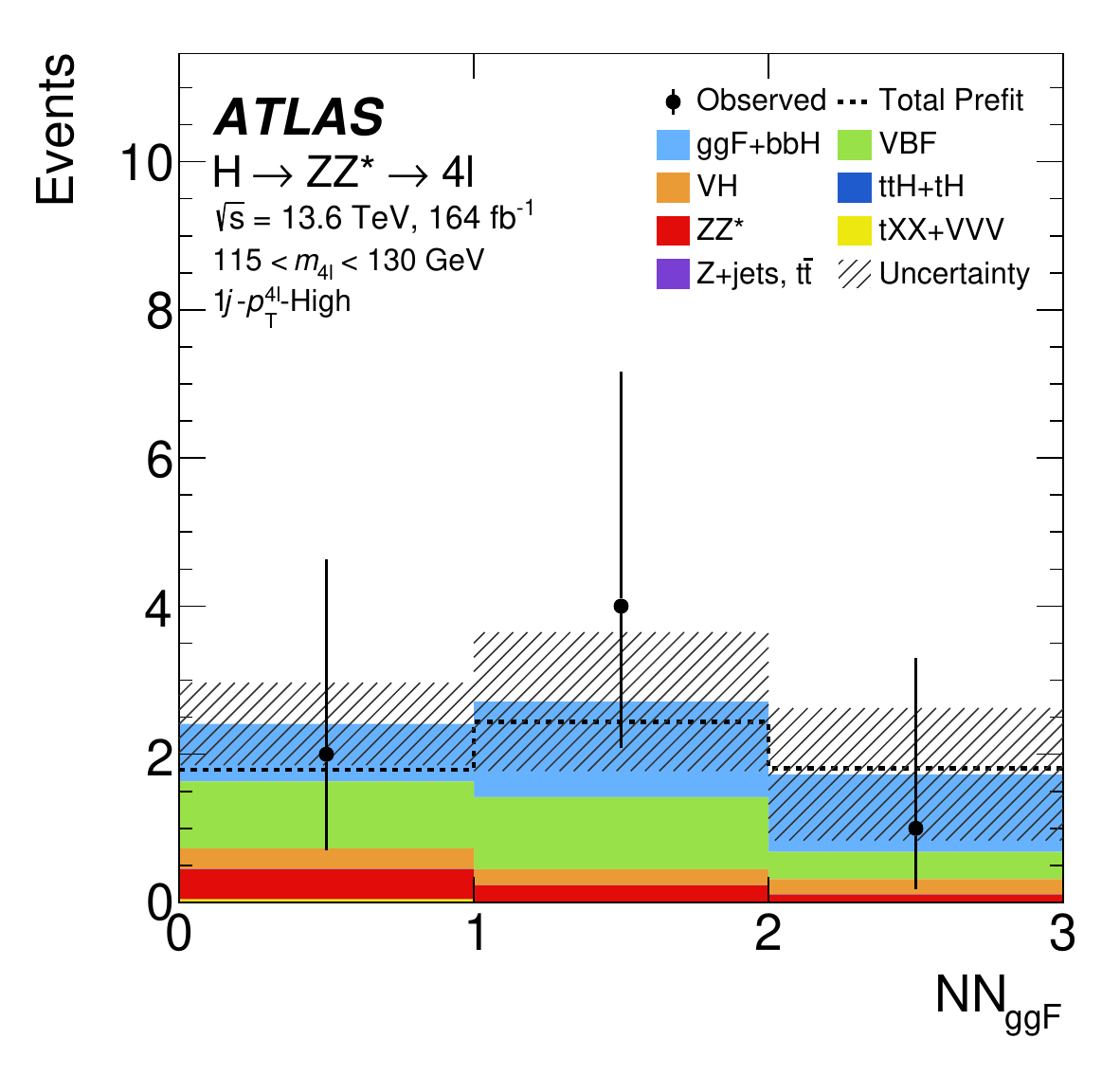}\label{fig:oNN1jpthhzz}}

\caption{The observed and expected neural network output (post-fit) distributions for an integrated luminosity of \intLum at \sqrtS in the different reconstructed event categories:
(a) $\mathrm{NN_{ggF}}$\ in \CatZeroJL,
(b) $\mathrm{NN_{ggF}}$\ in \CatZeroJM,
(c) $\mathrm{NN_{ggF+VBF}}$\ and (d) $\mathrm{NN}_{ZZ}$\ in \CatOneJL\ for low and high $\mathrm{NN}_{ZZ}$, respectively,
(e) $\mathrm{NN_{ggF+VBF}}$\ and (f) $\mathrm{NN}_{ZZ}$\ in \CatOneJM\ for low and high $\mathrm{NN}_{ZZ}$, respectively and
(g) $\mathrm{NN_{ggF}}$\ in \CatOneJH. The total post-fit distribution is shown as a stacked histogram, while the total pre-fit distribution is shown with dotted black line for comparison. The uncertainty in the prediction, corresponding to the total error associated with the post-fit signal cross-sections and background normalization factors, is shown by the hatched band. The bin boundaries are chosen to maximise the significance of the targeted signal in each category. The individual bins do not map linearly to discriminant values. \label{fig:STXS_NNs_1}}
\end{figure}

\begin{figure}[htbp]
\centering
\subfloat[]{
\includegraphics[width=0.33\linewidth]{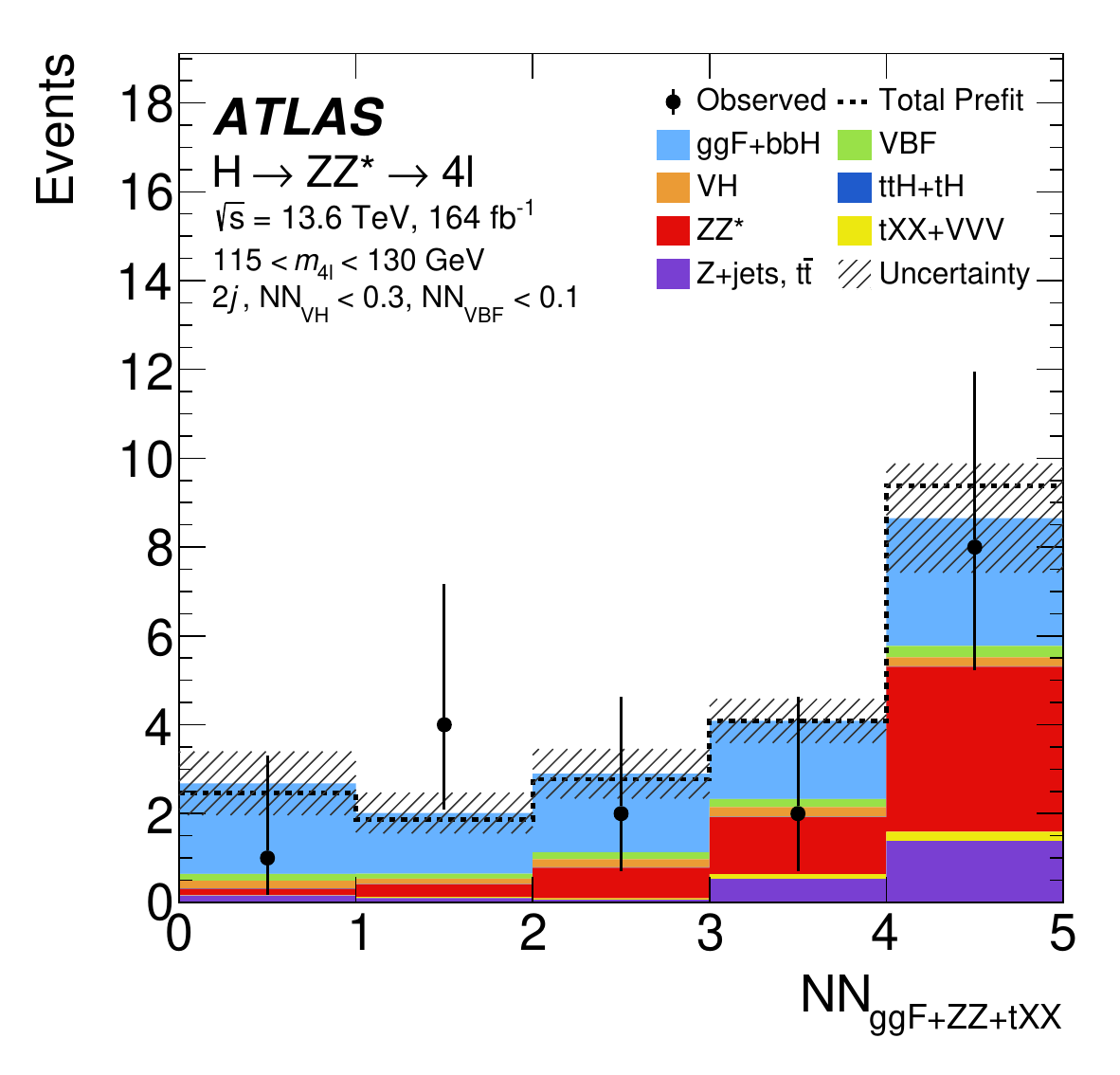}\label{fig:oNN2jggH}}
\subfloat[]{
\includegraphics[width=0.33\linewidth]{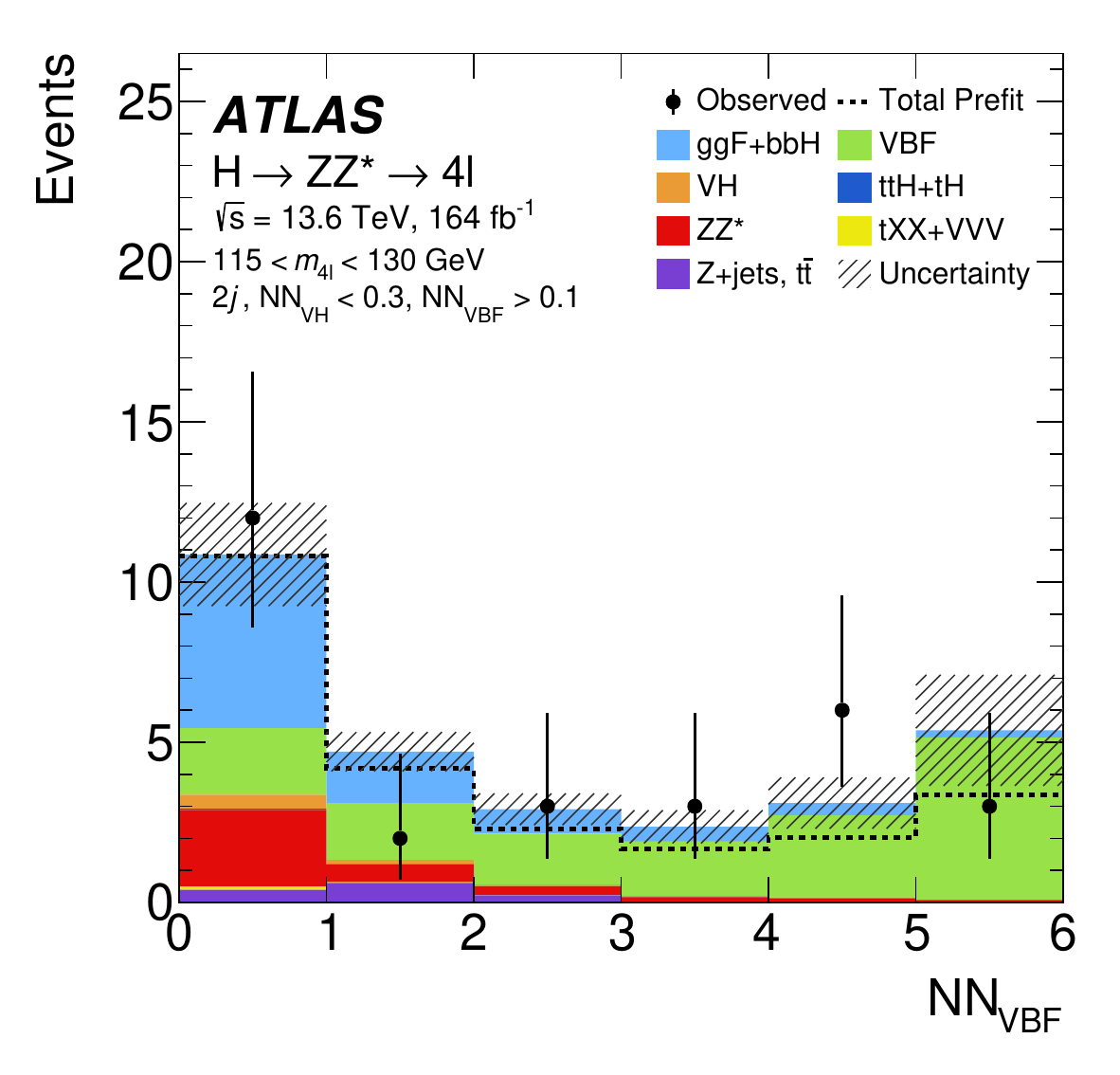}\label{fig:oNN2jvbf}}
\subfloat[]{
\includegraphics[width=0.33\linewidth]{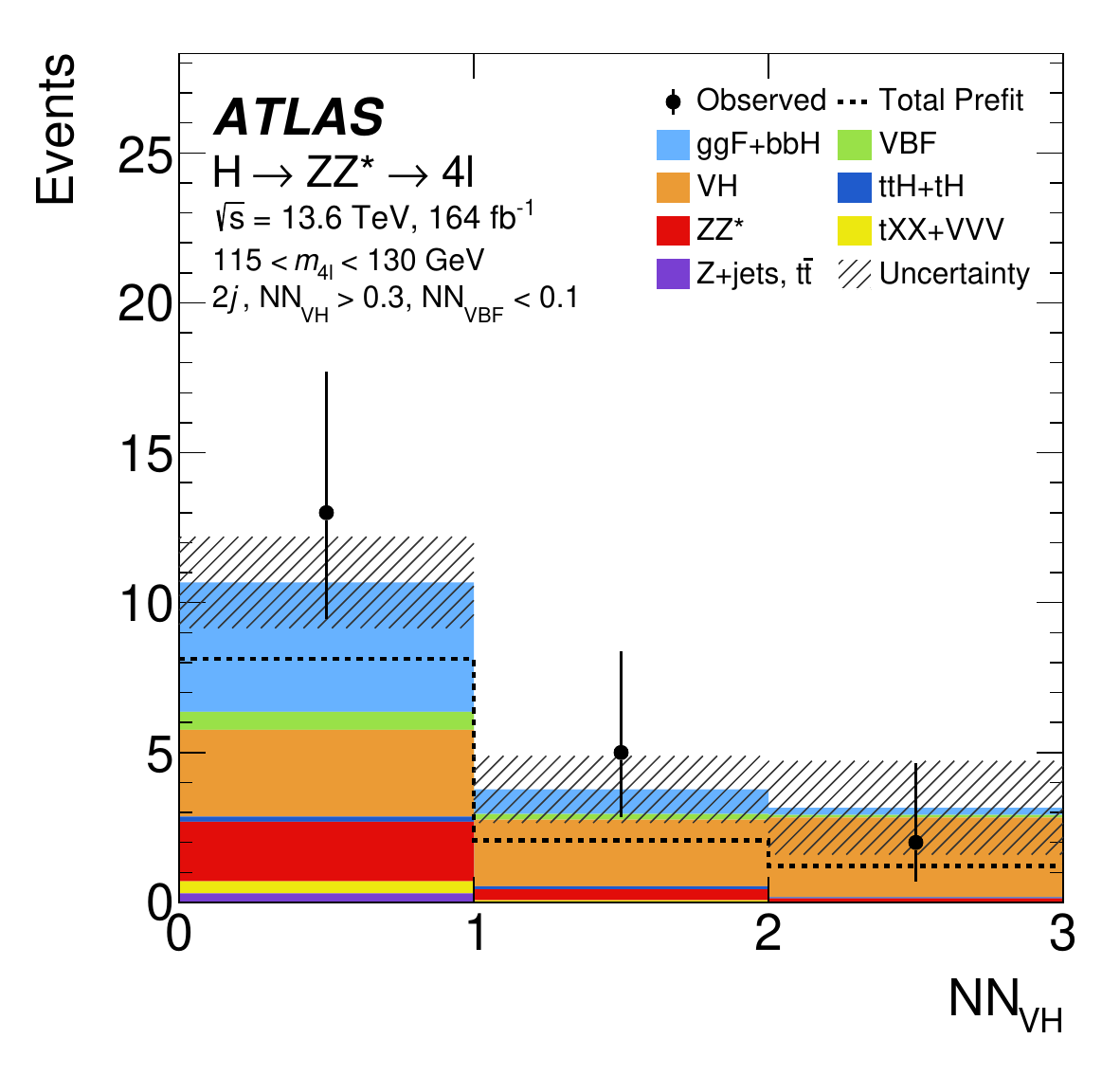}\label{fig:oNN2jvh}}\\
\subfloat[]{
\includegraphics[width=0.33\linewidth]{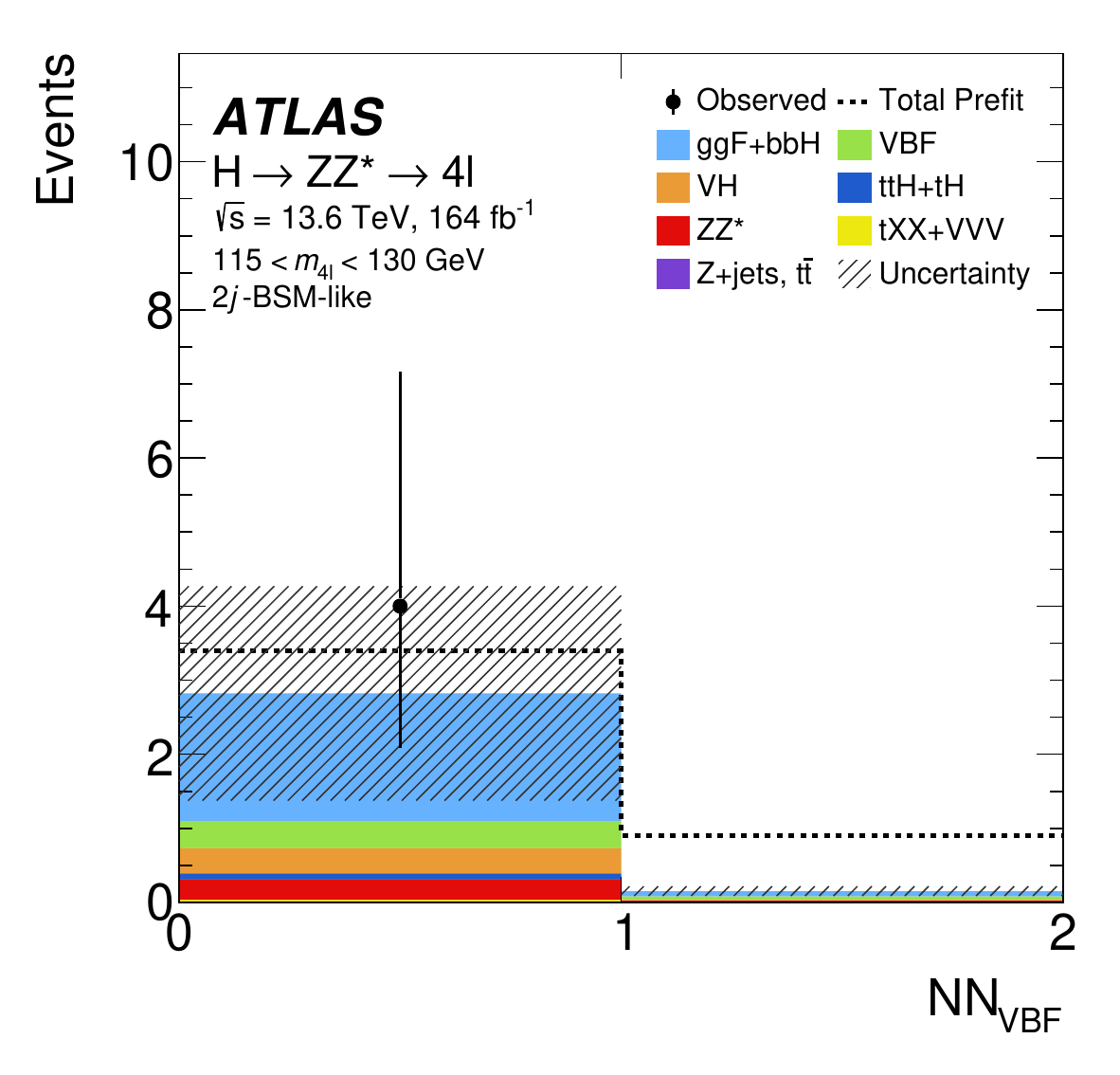}\label{fig:oNNbsmvbf}}
\subfloat[]{
\includegraphics[width=0.33\linewidth]{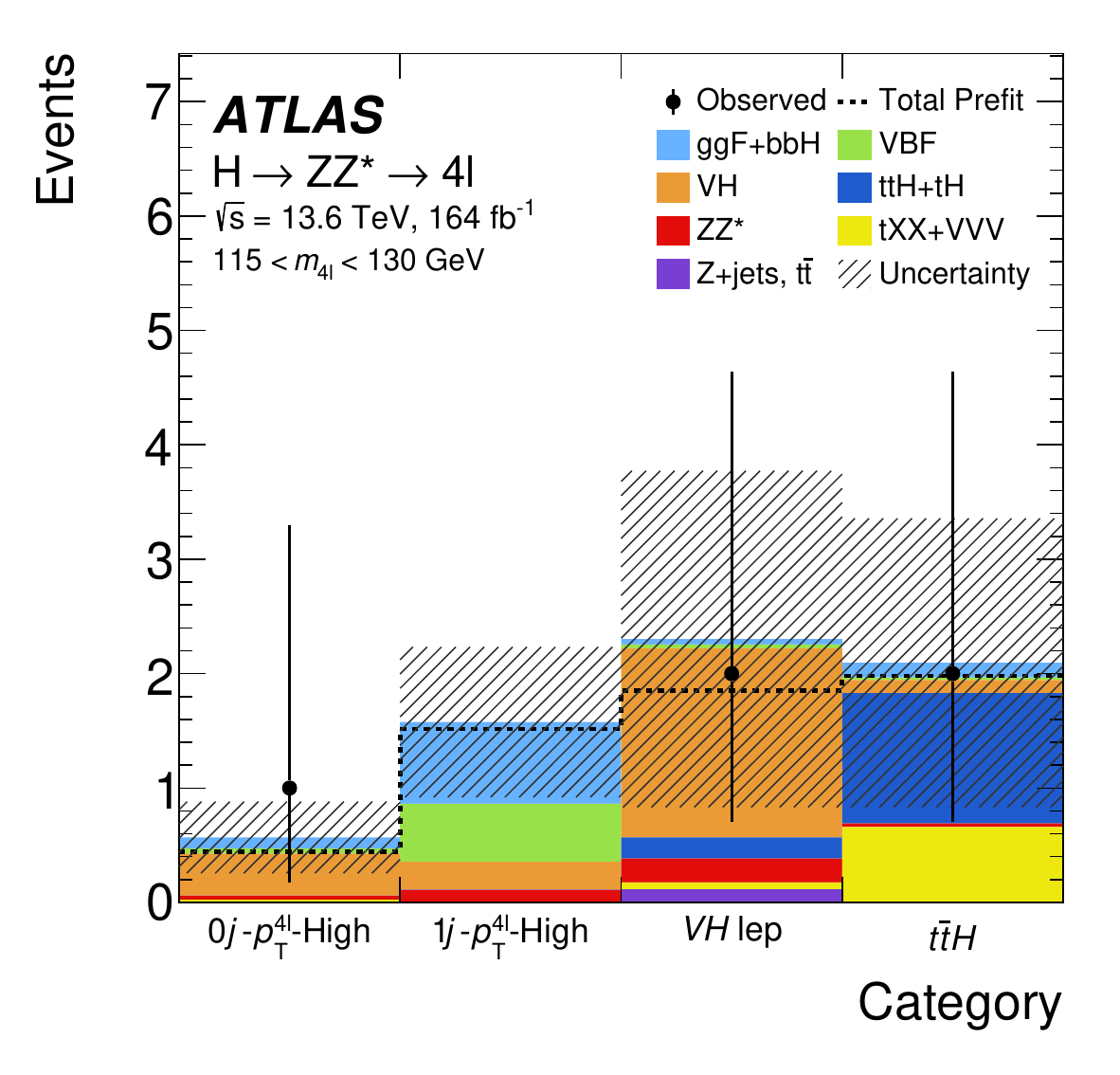}\label{fig:SRcount}}
\subfloat[]{
\includegraphics[width=0.33\linewidth]{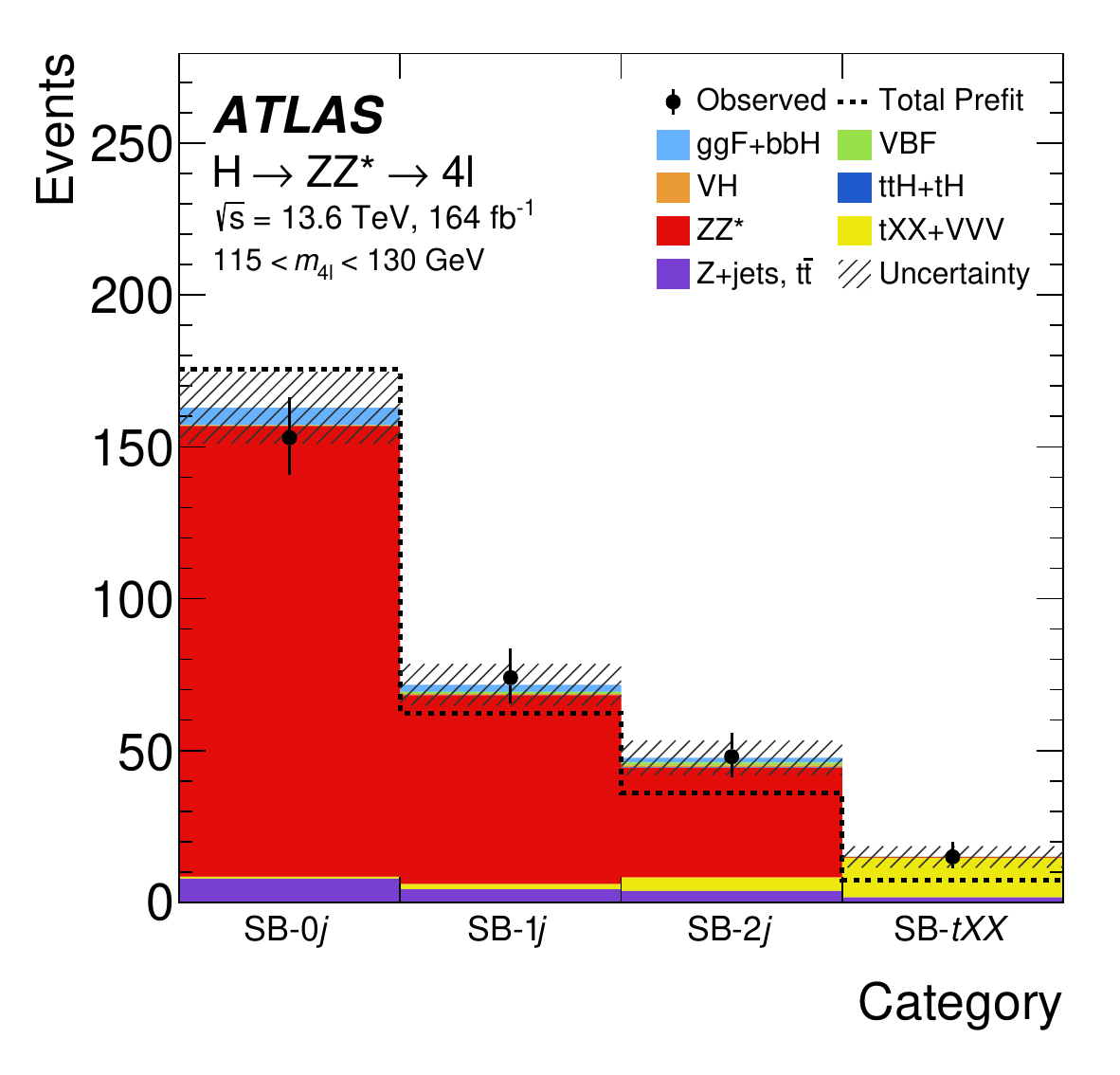}\label{fig:SBcount}}
\caption{The observed and expected neural network output (post-fit) distributions for an integrated luminosity of \intLum at \sqrtS in the different reconstructed event categories:
(a) $\mathrm{NN_{ggF+ZZ+tXX}}$\, (b) $\mathrm{NN_{VBF}}$\ and (c) $\mathrm{NN_{VH}}$\ with the different selection on NN$_{\textrm{VBF}}$\ and NN$_{\textrm{VH}}$\ in \CatTwoJ\ categories,
(d) $\mathrm{NN_{VBF}}$\ in \CatTwoJBSM. (e) Shows the categories where no NN discriminant is used while (f) shows the sidebands used to constrain the \zzstar and $tXX$ backgrounds. The total post-fit distribution is shown as a stacked histogram, while the total pre-fit distribution is shown with dotted black line for comparison. The uncertainty in the prediction, corresponding to the total error associated with the post-fit signal cross-sections and background normalization factors, is shown by the hatched band. The bin boundaries are chosen to maximise the significance of the targeted signal in each category. The individual bins do not map linearly to discriminant values. \label{fig:STXS_NNs_2}}
\end{figure}

The observed and expected (post-fit) distributions of the NN discriminants are shown in  Figure~\ref{fig:STXS_NNs_1} and Figure~\ref{fig:STXS_NNs_2}. The \CatOneJL~ category is additionally split into two sub-categories with $\mathrm{NN_{ZZ}}$ greater or smaller than 0.2 to target different sub-processes as defined in Table \ref{tab:stxs_nn_training_summary}. Similarly, the \CatTwoJ is divided into three sub-categories with ($\mathrm{NN}_{\textrm{VH}} < 0.3,\ \mathrm{NN}_{\textrm{VBF}} < 0.1)$, ($\mathrm{NN}_{\textrm{VH}} < 0.3,\ \mathrm{NN}_{\textrm{VBF}} > 0.1)$ and ($\mathrm{NN}_{\textrm{VH}} > 0.3,\ \mathrm{NN}_{\textrm{VBF}} < 0.1)$ to target the \ggF, \VBF and \VH processes.
In addition, Figure \ref{fig:SRcount} and Figure \ref{fig:SBcount} show the observed and expected yields in the categories where no NN discriminant is used and in the mass sidebands used to constrain the \zzstar\ and $tXX$ background, respectively. All distributions are in good agreement with the data.

\begin{figure}[htbp]
\centering
\includegraphics[width = 0.6 \textwidth]{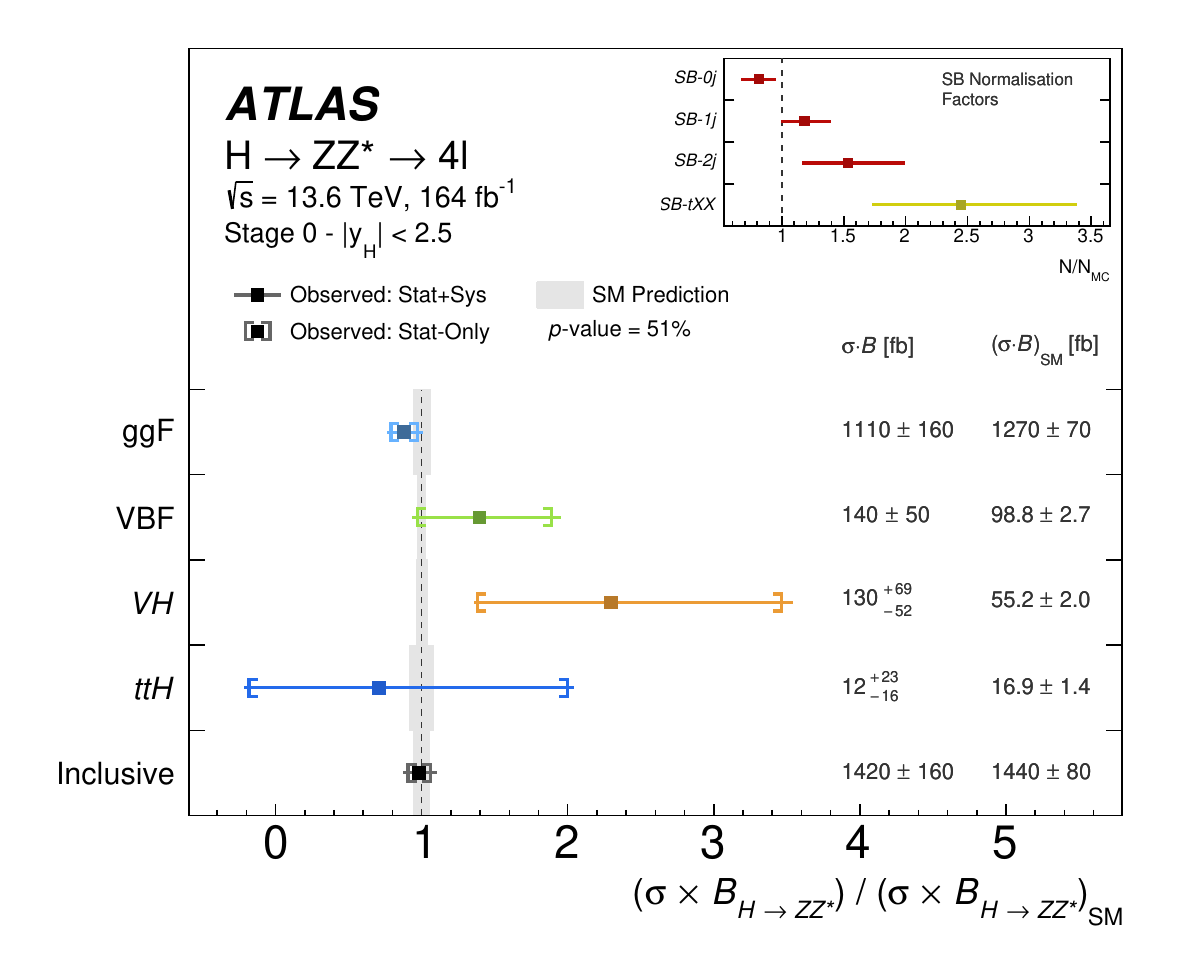}
\caption{The observed and expected SM values of the inclusive and individual production-mode cross-sections $\sigma_{|y_H|<2.5} \cdot \mathcal{B}\left(H\rightarrow ZZ^\ast\right)$ normalised by the SM expectation for an integrated luminosity of \intLum at \sqrtS. The branching ratio for the $\zzstar \rightarrow4\ell$ decay is assumed to be given by the SM prediction.
The horizontal bands around the expected SM result indicate the theoretical uncertainty in the prediction.
\label{fig:prodModes_S0} }
\end{figure}

\begin{figure}[htbp]
\centering
\includegraphics[width = 0.6 \textwidth]{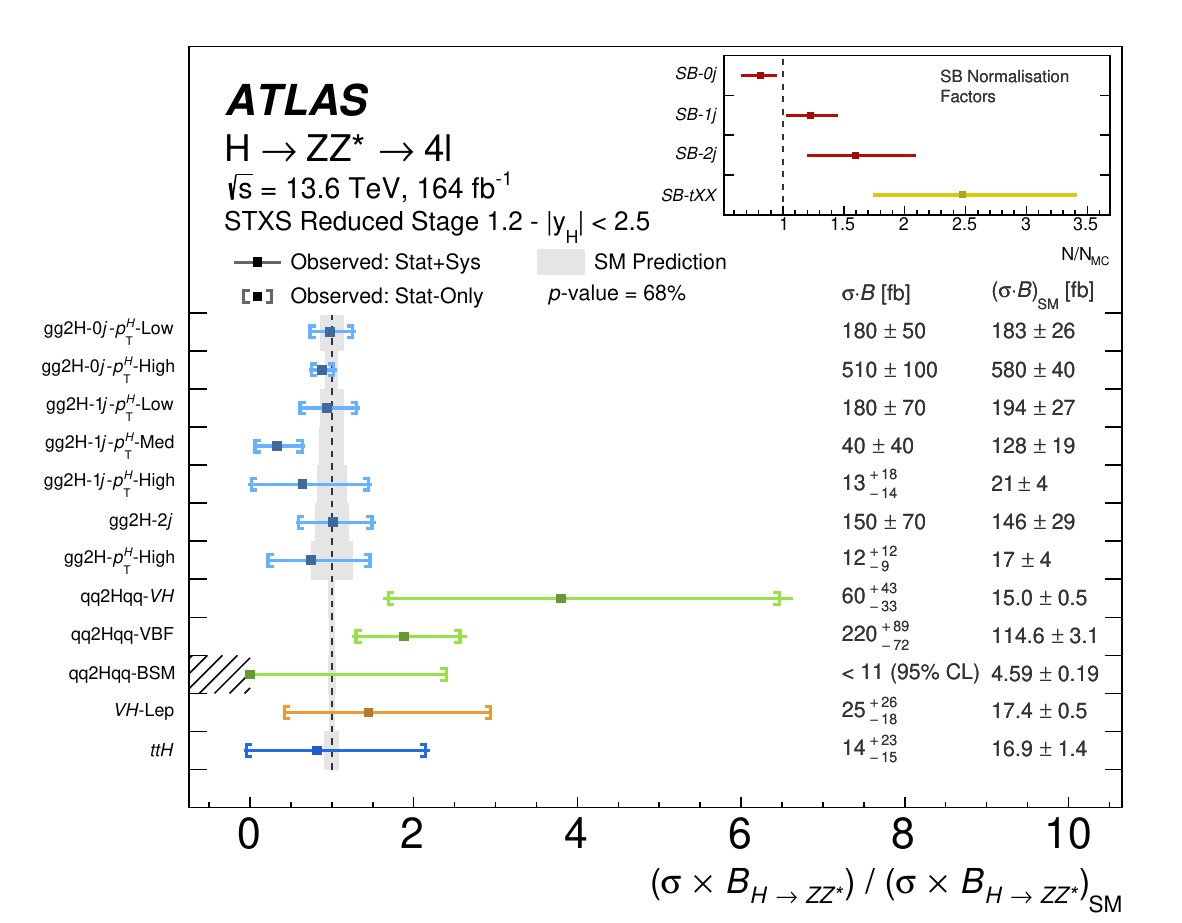}
\caption{The observed and expected SM values of the inclusive and individual production-mode cross-sections $\sigma_{|y_H|<2.5} \cdot \mathcal{B}\left(H\rightarrow ZZ^\ast\right)$ normalised by the SM expectation for an integrated luminosity of \intLum at \sqrtS. The branching ratio for the $\zzstar \rightarrow4\ell$ decay is assumed to be given by the SM prediction.
The horizontal bands around the expected SM result indicate the theoretical uncertainty in the prediction.
Due to zero observed events in the bins with highest sensitivity, an upper limit at 95\%CL for the \STXSqqtoHqqBSM bin is set. The hatched area indicates that the \STXSqqtoHqqBSM parameter of interest is constrained to positive values in the fit.
\label{fig:prodModes_S1} }
\end{figure}

The observed values of  $\sigma_{|y_H|<2.5} \cdot \mathcal{B}\left(H\rightarrow ZZ^\ast\right)$ in the Higgs boson rapidity range of $|y_H| < 2.5$, the corresponding expected SM cross-sections, and their ratios are shown in Figure~\ref{fig:prodModes_S0} for the inclusive production and for the individual production-modes as defined in Section~\ref{subsec:stxs}. Similarly, the results for the Reduced Stage 1.2 bins are shown in Figure~\ref{fig:prodModes_S1}. The branching ratio for the $ZZ^{\ast}\rightarrow4\ell$ decay is fixed to the SM prediction, such that $\mathcal{B}\left(H\rightarrow ZZ^\ast\right)=2.62\times10^{-2}$ \cite{YR4}. In the ratio calculation, uncertainties on the SM expectation are presented separately and not included in the systematics. An inclusive cross-section of
\[ \stxsIncObsFull \]
is measured, in agreement with the SM prediction of \stxsIncPred. Normalisation factors for the $\zzstar$ continuum background of $N_{ZZ}^{\textrm{SB}-0j}/N_{\textrm{MC}} = 0.82 \pm 0.14$, $N_{ZZ}^{\textrm{SB}-1j}/N_{\textrm{MC}} = 1.18^{+0.22}_{-0.19} $, and $N_{ZZ}^{\textrm{SB}-2j}/N_{\textrm{MC}} = 1.53^{+0.46}_{-0.37} $ are obtained from the sidebands, separately for the three event categories with different jet multiplicities, while the one for the $tXX$ backgrounds from the specific category $N_{tXX}^{\textrm{SB}-tXX}/N_{\textrm{MC}} = 2.5^{+0.9}_{-0.7}$.

The data are also interpreted in terms of the global signal strength, where the theoretical uncertainties in the SM cross-section predictions are included in the measurement, yielding
\[ \muIncObsFull \]
The dominant systematic uncertainty in inclusive cross-section measurement is the experimental uncertainty in the lepton efficiency and the integrated luminosity measurements. For the global signal strength measurement theoretical uncertainties in the \ggF prediction are dominant.

The dominant contribution to the measurement uncertainty in the \ggF production-mode originates from the same sources as in the inclusive measurement and in the Reduced Stage 1.2 bins with zero jets. For the \VBF production bin, the theoretical systematic uncertainties related to missing higher-order QCD corrections for both \VBF and \ggF predictions
and experimental uncertainties related to the jet energy scale and resolution become relevant. For \VH and \ttH production-modes, theoretical uncertainties in the \ggF prediction and those experiment ones on jet flavour tagging are dominant. For the Reduced Stage 1.2 bins with higher jet multiplicity, the dominant uncertainties on the cross-sections are the jet energy scale and resolution, and missing higher-order corrections.  The probability of compatibility of the measured production-mode (Reduced Stage 1.2) measurement and the SM Higgs boson expectation is 51\% (68\%).

\FloatBarrier


%

\section{Interpretations of the differential distributions and production cross-sections}
\label{sec:interpretations}

\subsection{Constraints on Higgs boson couplings within the $\kappa$-framework}

To probe physics beyond the SM, the measured production-mode cross-sections are interpreted in the $\kappa$ framework~\cite{Heinemeyer:2013tqa}, which parameterises possible deviations of Higgs boson couplings from their SM values using coupling-strength modifiers $\vec{\kappa}$. Under the narrow-width approximation, Higgs production and decay factorise, and $\sigma\cdot\mathcal{B}$ can be expressed in terms of modified Higgs production cross-sections and partial widths. The $\vec{\kappa}$ dependence at $\sqrt{s}=13.6\,\TeV$ is obtained by linear interpolation between the corresponding terms at 13 and 14~\TeV~\cite{Hamilton:2013fea}.

The interpretation is performed in a benchmark scenario with two universal modifiers: $\kappa_V=\kappa_W=\kappa_Z$ for Higgs-to-vector-boson couplings and $\kappa_f=\kappa_t=\kappa_b=\kappa_c=\kappa_\tau=\kappa_\mu$ for Higgs-to-fermion couplings, assuming a single Higgs boson state that is even under the combined charge-parity (CP) transformation, with the SM tensor structure and no undetected or invisible decays. For loop-induced processes, the resolved treatment is adopted, in which the effective loop couplings are expressed in terms of the modified couplings to the massive SM particles.

The results are statistically combined with the corresponding measurements performed with the full Run~2 data set~\cite{HIGG-2018-28}, accounting for the systematic uncertainties described in Section~\ref{sec:systematics}. The following experimental uncertainties are treated as correlated between Run~2 and Run~3 results: the muon momentum scale and reconstruction efficiency; the electron energy scale and energy resolution; and the jet energy scale and resolution, except for Run~3-specific components. Other experimental uncertainties are treated as uncorrelated. Theoretical uncertainties from missing higher-order corrections affecting the Higgs boson signal cross-sections in exclusive kinematic regions, the Higgs boson mass uncertainty, and uncertainties in background estimates are treated as correlated. All other theoretical uncertainties, including PDF and parton shower uncertainties, are treated as uncorrelated.

\begin{figure}[htbp]
\centering
\includegraphics[width=0.50\linewidth]{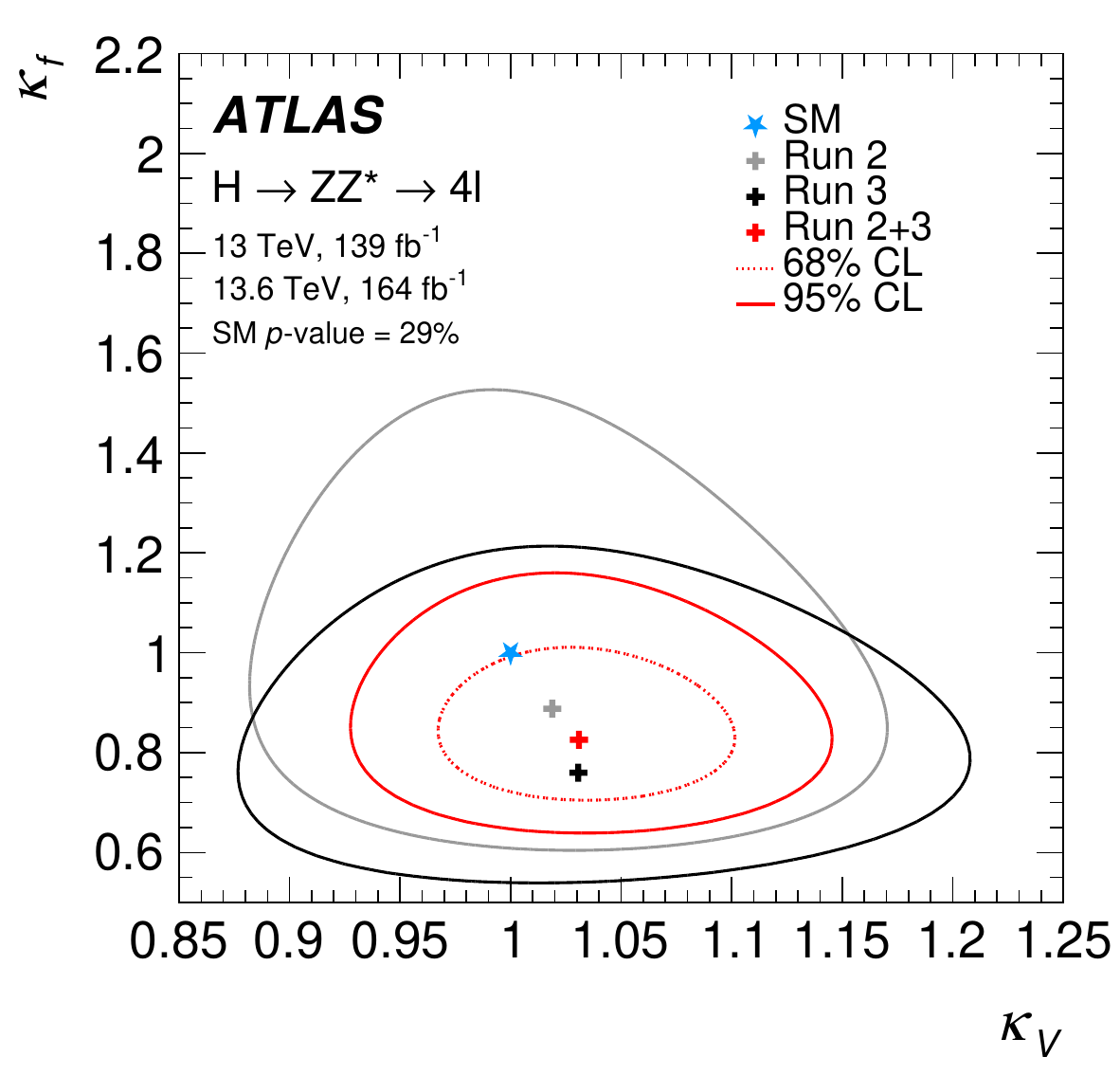}
\caption{Likelihood contours in the $(\kappa_{f}, \kappa_{V})$ plane comparing the current results with those obtained in Run~2. The statistical combination of the two measurements is also presented. The area outside the solid (dashed) contours is excluded at 95\% (68\%) confidence level. The best-fit point (solid cross) is shown together with the SM prediction (star).}

\label{fig:kappas}
\end{figure}

The likelihood contours and best-fit values in the $(\kappa_f,\kappa_V)$ plane are shown in Figure~\ref{fig:kappas}. Only the quadrant with $\kappa_f>0$ and $\kappa_V>0$ is displayed, as the $H\to ZZ^*\to 4\ell$ channel has limited sensitivity to the relative sign of the two modifiers. The combined measurements improves on the previous result \cite{HIGG-2018-28} by about 40\% (25\%)  on $\kappa_V$ ($\kappa_f$). The compatibility of the combined measurement with the SM expectation corresponds to a $p$-value of 29\%.

\FloatBarrier
\subsection{Constraints on CP-even and CP-odd effective field theory operators}
\label{subsec:eft}

The $\kappa$-framework assumes that the tensor structure of the Higgs boson couplings is the same as in the SM. To probe possible non-SM tensor structures, selected unfolded differential measurements are interpreted in an effective field theory (EFT) approach. In this approach, the BSM interactions are introduced via additional higher-dimensional operators
$\mathcal{O}_i^{(d)}$ of dimension $d$, supplementing the SM Lagrangian $\cal{L}_{\textrm{SM}}$,
\begin{equation*}
\mathcal L_{\textrm{EFT}} = \mathcal L_{\textrm{SM}} +  \sum_{i}^{} \frac{C_i^{(d)}}{\Lambda^{(d-4)}} \mathcal O_i^{(d)} \ \ \ \text{for} \ d>4,
\end{equation*}
where $C_i^{(d)}$ are the Wilson coefficients and $\Lambda$ denotes the characteristic scale of new physics. In this analysis, only dimension-six operators are considered. Dimension-five and dimension-seven operators are not included, as they violate lepton or baryon number conservation and are therefore not relevant in the present context. Although dimension-eight operators can interfere with the SM amplitudes at a same order as the quadratic contributions from dimension-six operators, their effect is not considered. The data are sensitive only to the combinations $c_i = C_i^{(d=6)}/\Lambda^2$, which are the quantities constrained in this analysis.

Constraints are derived within the Standard Model EFT (SMEFT) formalism~\cite{SMEFT} in the Warsaw basis~\cite{wbasis}. Given the sensitivity in the $H\to ZZ^*\to 4\ell$ channel, only three CP-even and three CP-odd operators modifying Higgs boson interactions with electroweak gauge bosons are probed as summarised in Table~\ref{tab:OpSMEFT}. Operators affecting only double-Higgs boson production, as well as operators modifying Higgs couplings to quarks and leptons, are not considered. The impact of these operators on the total Higgs boson width is neglected.

\begin{table}[!htbp]
\renewcommand*{\arraystretch}{1.0}
\caption{Summary of CP-even and CP-odd EFT operators in the SMEFT formalism probed in the $H\to ZZ^* \to 4\ell$ channel. The corresponding tensor structure in terms of the SM fields from Ref.~\cite{SMEFT} is shown together with the associated Wilson coefficients and the affected interaction vertices. $V_{\mu\nu}$ is the field-strength tensor for a given gauge field $V=W,B$.}
\begin{center}
\begin{tabular}{lll}
\toprule
Operator & Structure & Coeff.\\
\midrule
$\mathcal O_{HW}$  & $H H^{\dagger} W^{l}_{\mu\nu} W^{\mu\nu l}$        &  $c_{HW}$ \\
$\mathcal O_{HB}$  & $H H^{\dagger} B_{\mu\nu} B^{\mu\nu}$              & $c_{HB}$ \\
$\mathcal O_{HWB}$ & $H H^{\dagger} \tau^l W^{l}_{\mu\nu} B^{\mu\nu}$   & $c_{HWB}$  \\
\midrule
$\mathcal{O}_{H\tilde{W}}$ & $H H^\dagger \tilde{W}_{\mu\nu}^l W^{\mu\nu l}$ & $c_{H\tilde{W}}$ \\
$\mathcal{O}_{H\tilde{B}}$ & $H H^\dagger \tilde{B}_{\mu\nu} B^{\mu\nu}$ & $c_{H\tilde{B}}$ \\
$\mathcal{O}_{H\tilde{W}B}$ & $H H^\dagger \tau^l \tilde{W}_{\mu\nu}^l B^{\mu\nu}$ & $c_{H\tilde{W}B}$ \\
\bottomrule
\end{tabular}
\end{center}
\label{tab:OpSMEFT}
\end{table}

The EFT interpretation is performed using observables selected to enhance sensitivity to the targeted operators and to retain discrimination between the interference (linear) and BSM-squared (quadratic) contributions. The two-dimensional $(\ptFourL, m_{34})$ measurement is used to constrain $c_{HW}$, since it simultaneously probes production- and decay-side effects: $\ptFourL$ is sensitive to modifications of the production kinematics, while $m_{34}$ provides sensitivity to changes in the decay matrix element in $H\to ZZ^*\to4\ell$. The sensitivity of this approach depends on the chosen binning in $m_{34}$ as the balance between statistical precision and shape sensitivity is required to be optimised.

For the remaining CP-even operators, the angular observable $\phi$ provides improved sensitivity compared with $(\ptFourL, m_{34})$. This improvement arises as the linear SMEFT contribution exhibits sign changes as a function of $\phi$, leading to a periodic modulation that cancels when integrated over the full phase space. Consequently, the SM-BSM interference is suppressed in $\phi$-inclusive observables, similarly to those observed in non-Higgs boson decays~\cite{Azatov_2016, Panico_2018}. Although $\phi$ is a CP-even observable, it also exhibits sensitivity to CP-odd operators, with the interference term showing an analogous behaviour to the CP-even case but with different periodicity. The $\dphijj$ observable is also studied for CP-odd operators, but it does not provide discrimination between CP-even and CP-odd operators in the present dataset due to the dominant dependence on quadratic EFT terms.

The Wilson coefficients are constrained by first parametrizing the unfolded signal cross-section in each bin of the chosen distribution to the SMEFT expectation, with the expected cross-section and acceptance using a matrix-element-based procedure~\cite{HIGG-2022-17} and refitting the data. For a bin $j$ of a given observable, the expected cross-section is written as
\begin{equation}
\resizebox{0.9\linewidth}{!}{$%
\sigma_{pp\to H \to 4\ell, j}^{\mathrm{SMEFT}}\!\left( c_i \right) =
\sigma_{pp\to H \to 4\ell, j}^{\mathrm{SM}} \cdot
\sum_{p}{ \left( 1 + \sum_{i}{\alpha_{i}^{(p,j)}c_i} + \sum_{i,k\ge i}{\beta_{ik}^{(p,j)}c_i c_k} \right)} \cdot
\frac{\left( 1 + \sum_{i}{\alpha_{i}^{(\Gamma_{4\ell}, j)}c_i} + \sum_{i,k\ge i}{\beta_{ik}^{(\Gamma_{4\ell}, j)}c_i c_k} \right)}{\left( 1 + \sum_{i}{\alpha_{i}^{(\Gamma_{tot}, j)}c_i} + \sum_{i,k\ge i}{\beta_{ik}^{(\Gamma_{tot}, j)}c_ic_k} \right)} \,,
$%
}
\label{eq:smeft_parametrisation}
\end{equation}
where $p$ labels Higgs boson production processes, and $\Gamma_{4\ell}$ and $\Gamma_{tot}$ denote the partial width to four leptons and the total width, respectively. The terms linear in $c_i$ encode SM--EFT interference, while the quadratic terms represent the pure dimension-six contribution. The coefficients $\alpha$ and $\beta$ are obtained from matrix elements computed using dedicated simulated samples generated at LO in QCD with \progname{MadGraph5\_aMC@NLO} and the NNPDF23lo PDF set. The SMEFT contributions to \VBF production and the $HZZ$ decay vertex are modelled using \progname{SMEFTsim}~\cite{SMEFT,Aebischer_2018}. The SMEFT impact on $\VH$ production is neglected, as it has a negligible effect on the considered distributions.

\begin{figure}[htbp]
\centering
\subfloat[]{\includegraphics[width = 0.315 \textwidth]{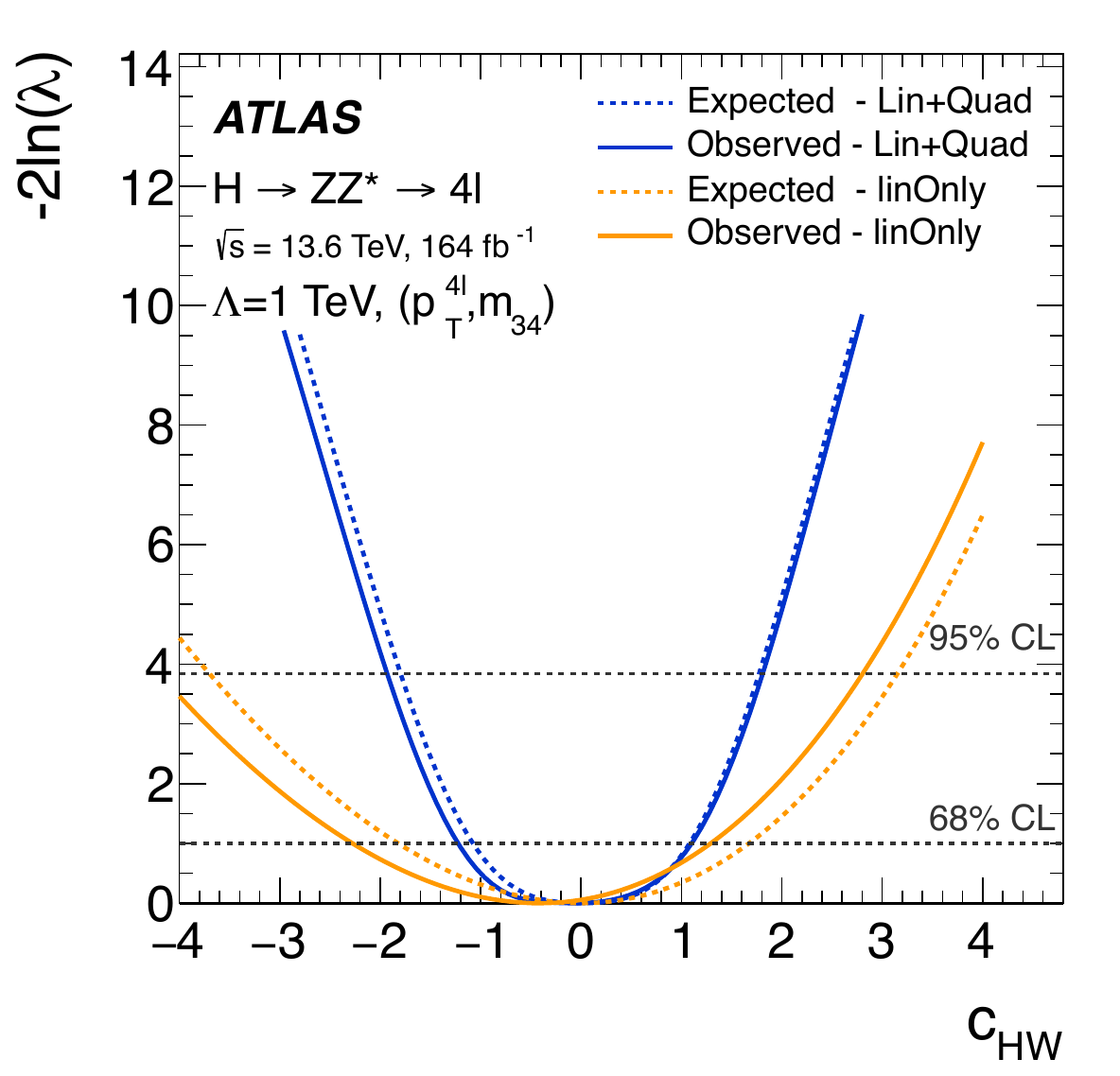}}
\subfloat[]{\includegraphics[width = 0.315 \textwidth]{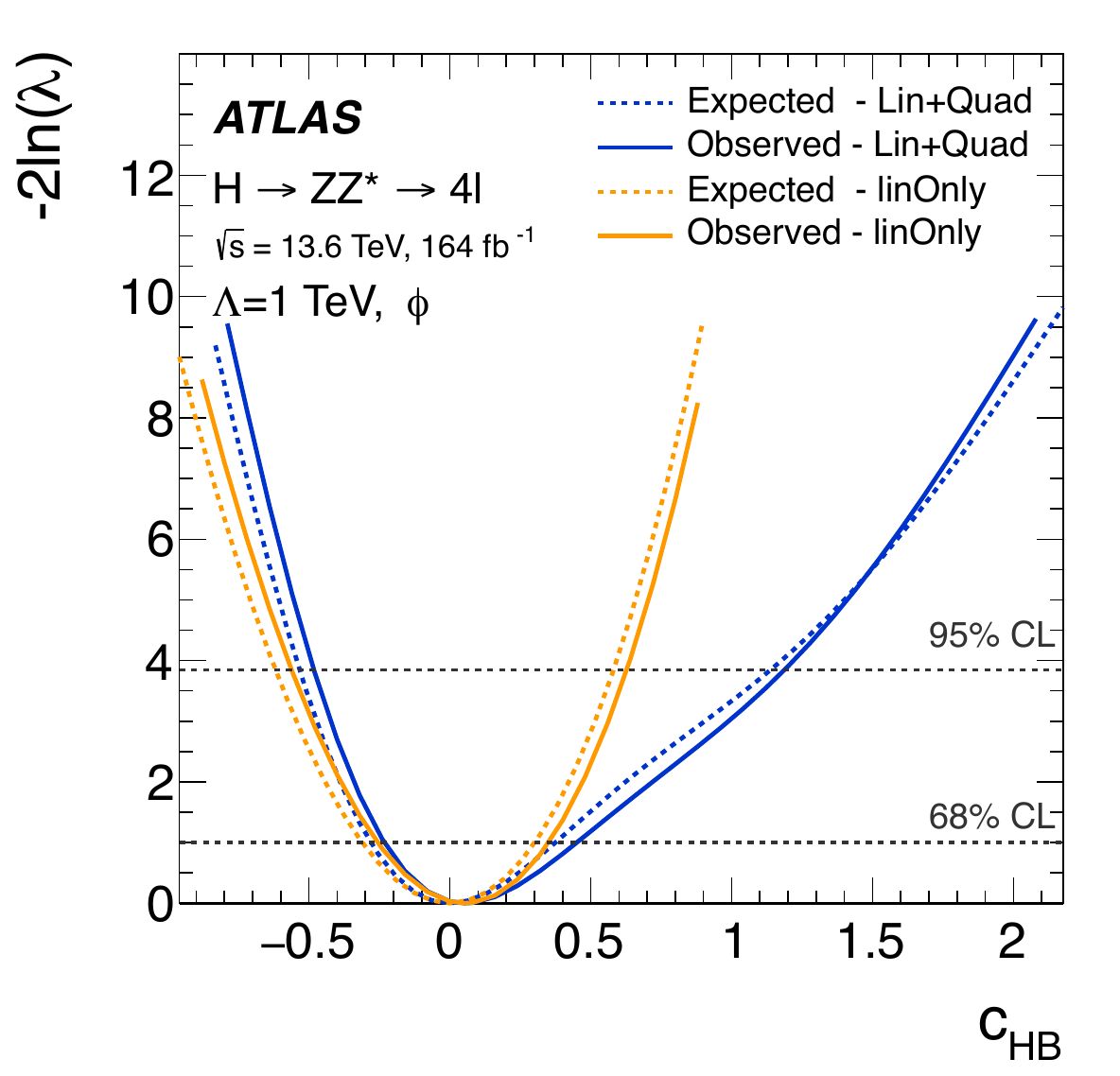}}
\subfloat[]{\includegraphics[width = 0.315 \textwidth]{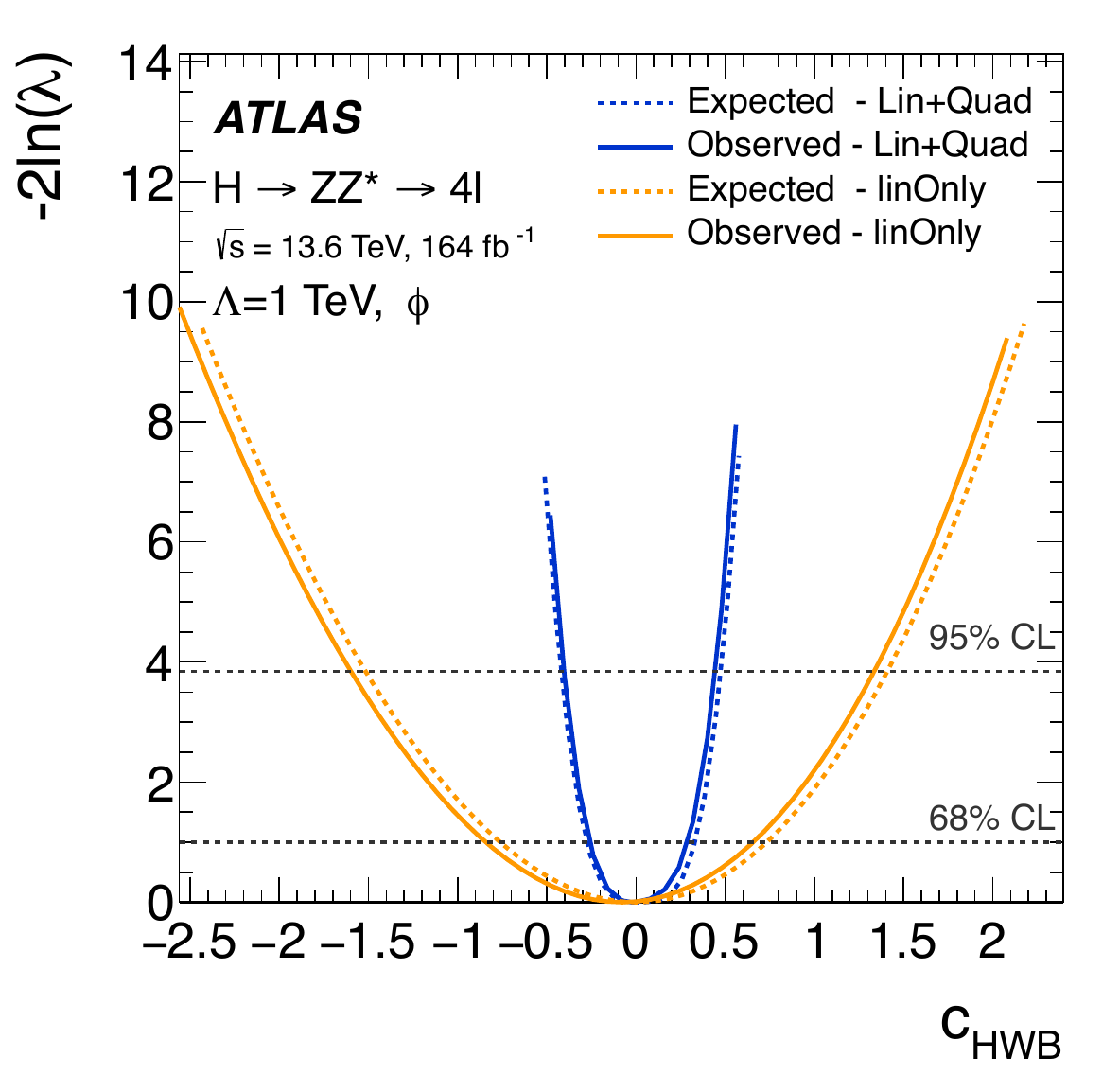}}
\caption{
One-dimensional likelihood scans for the CP-even SMEFT Wilson coefficients, comparing the linear-only and linear+quadratic SMEFT treatments. The $\cHW$ scan uses the unfolded $(\ptFourL,m_{34})$ measurement, while the other scans use the unfolded $\phi$ measurement. A new physics scale of $\Lambda = 1~\TeV$ is assumed.}
\label{fig:eft_scans_1D}
\end{figure}

\begin{figure}[htbp]
\centering
\subfloat[]{\includegraphics[width = 0.315 \textwidth]{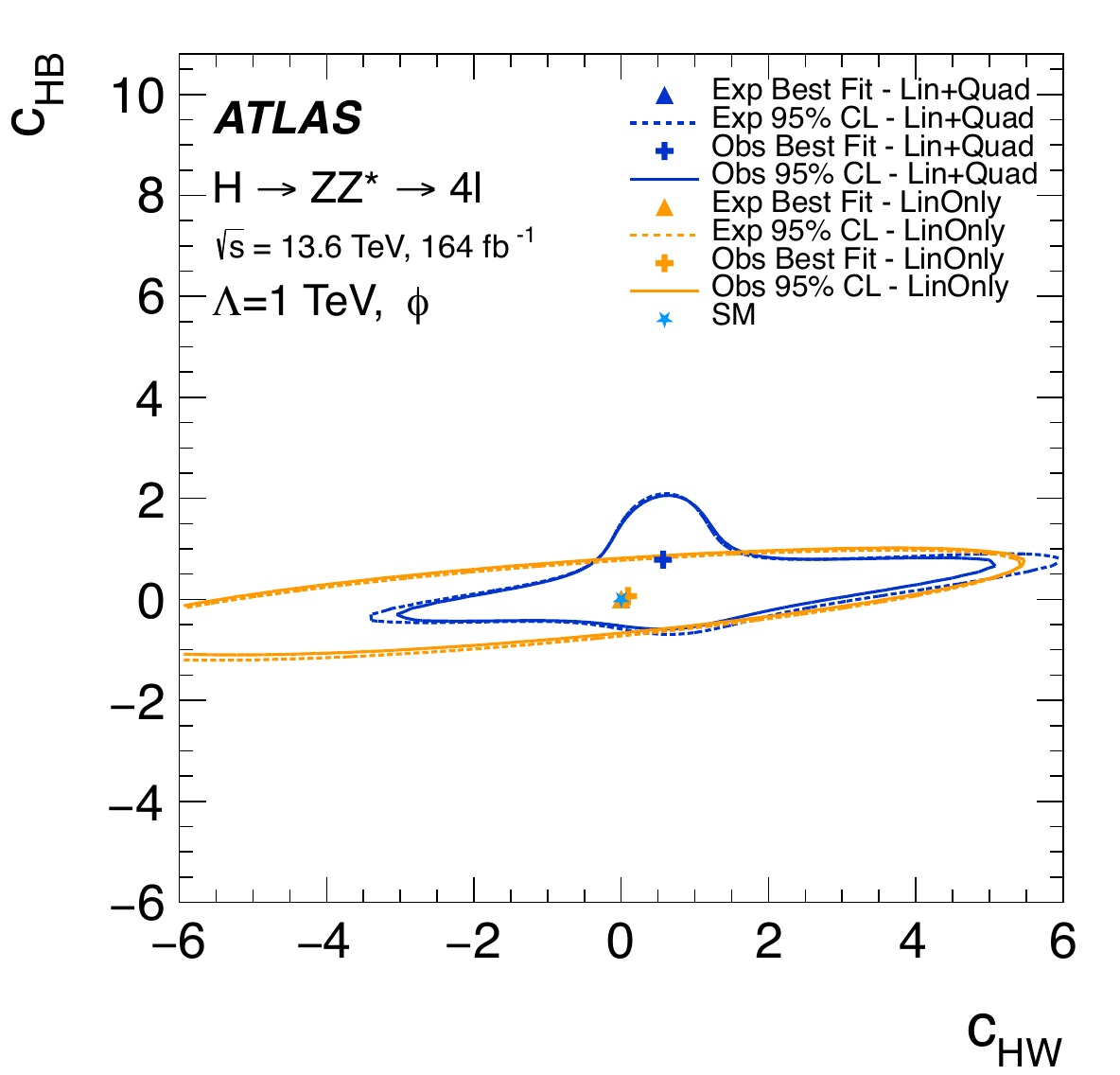}}
\subfloat[]{\includegraphics[width = 0.315 \textwidth]{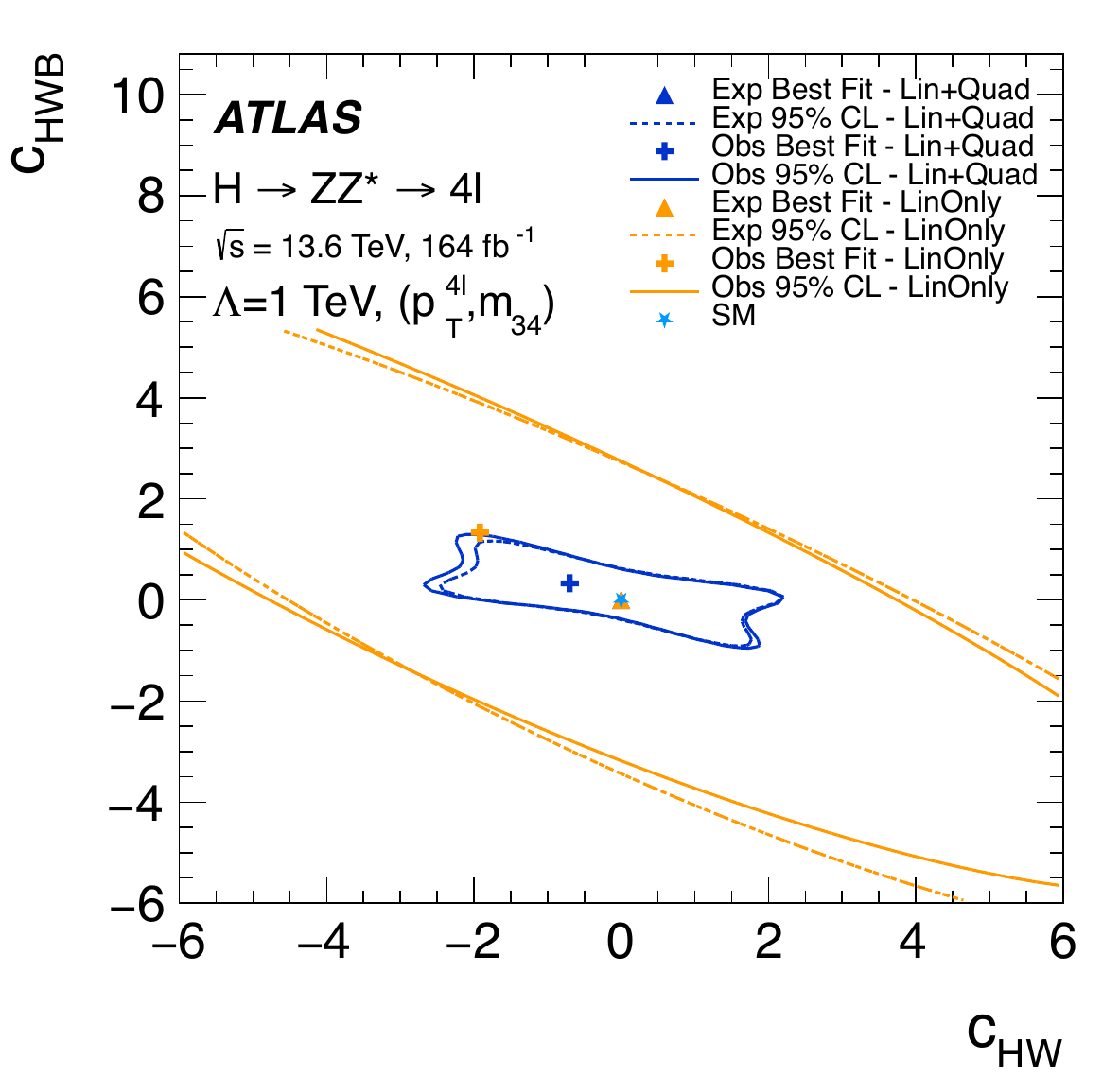}}
\subfloat[]{\includegraphics[width = 0.315 \textwidth]{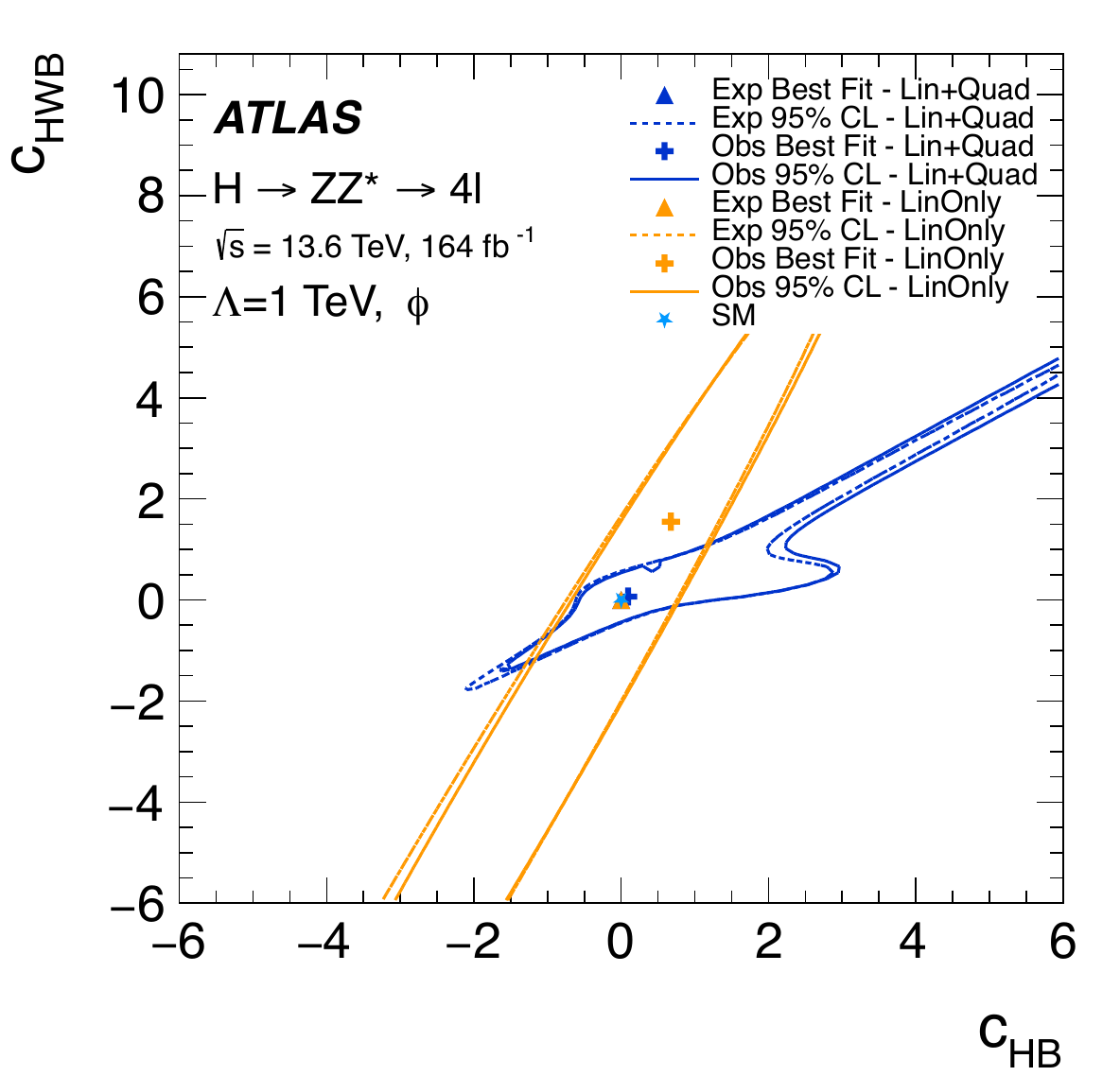}} \\
\subfloat[]{\includegraphics[width = 0.315 \textwidth]{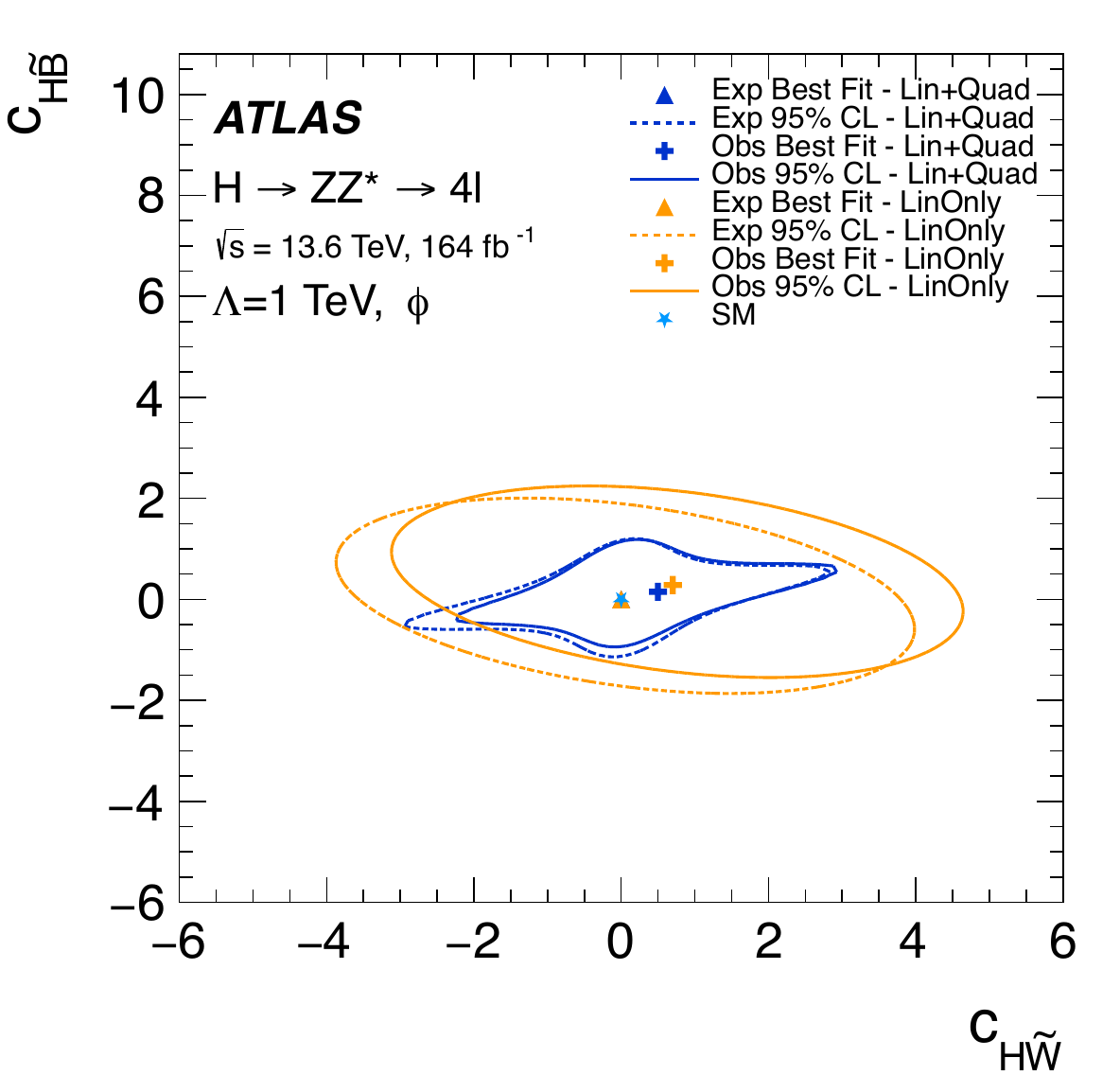}}
\subfloat[]{\includegraphics[width = 0.315 \textwidth]{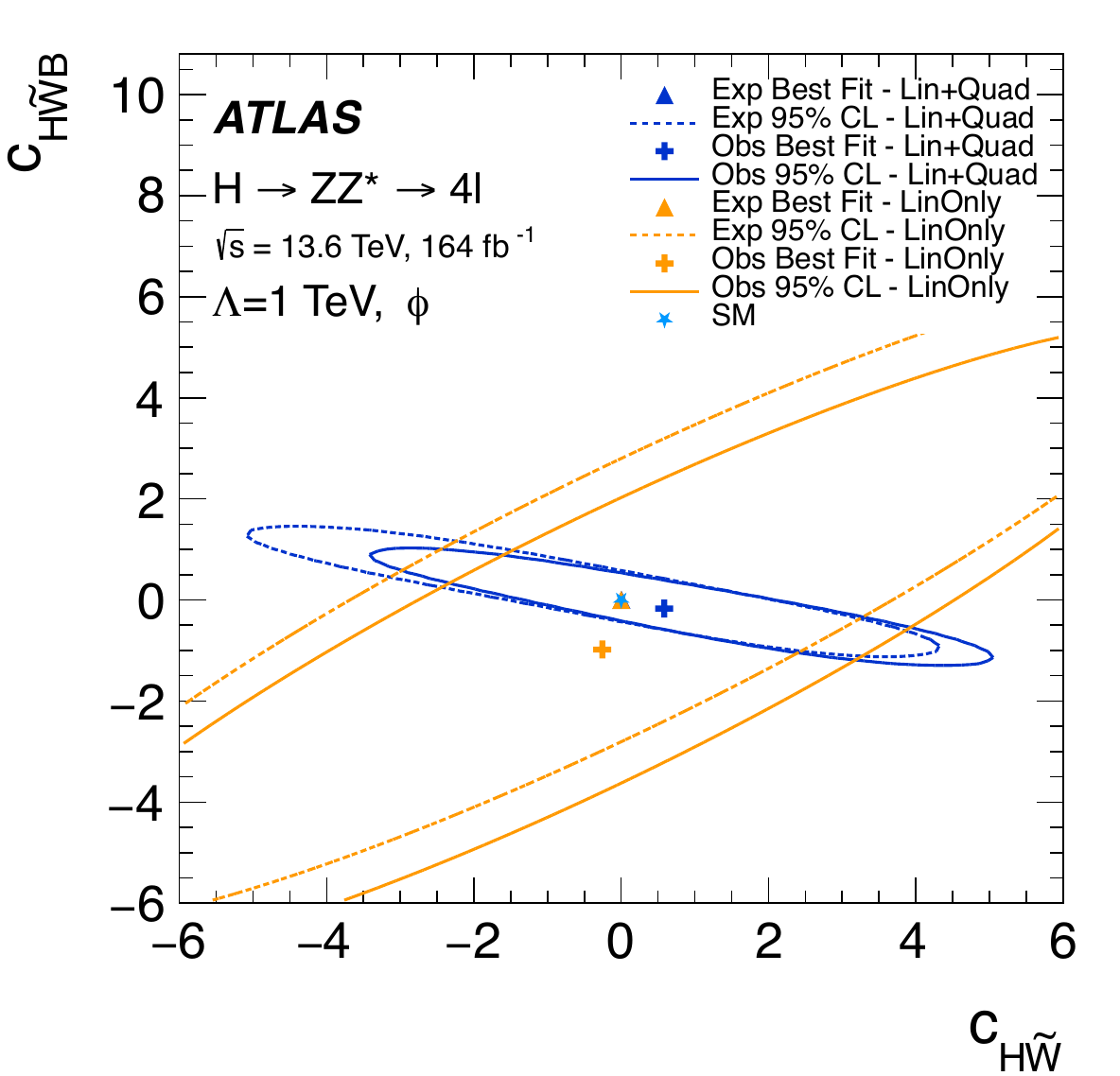}}
\subfloat[]{\includegraphics[width = 0.315 \textwidth]{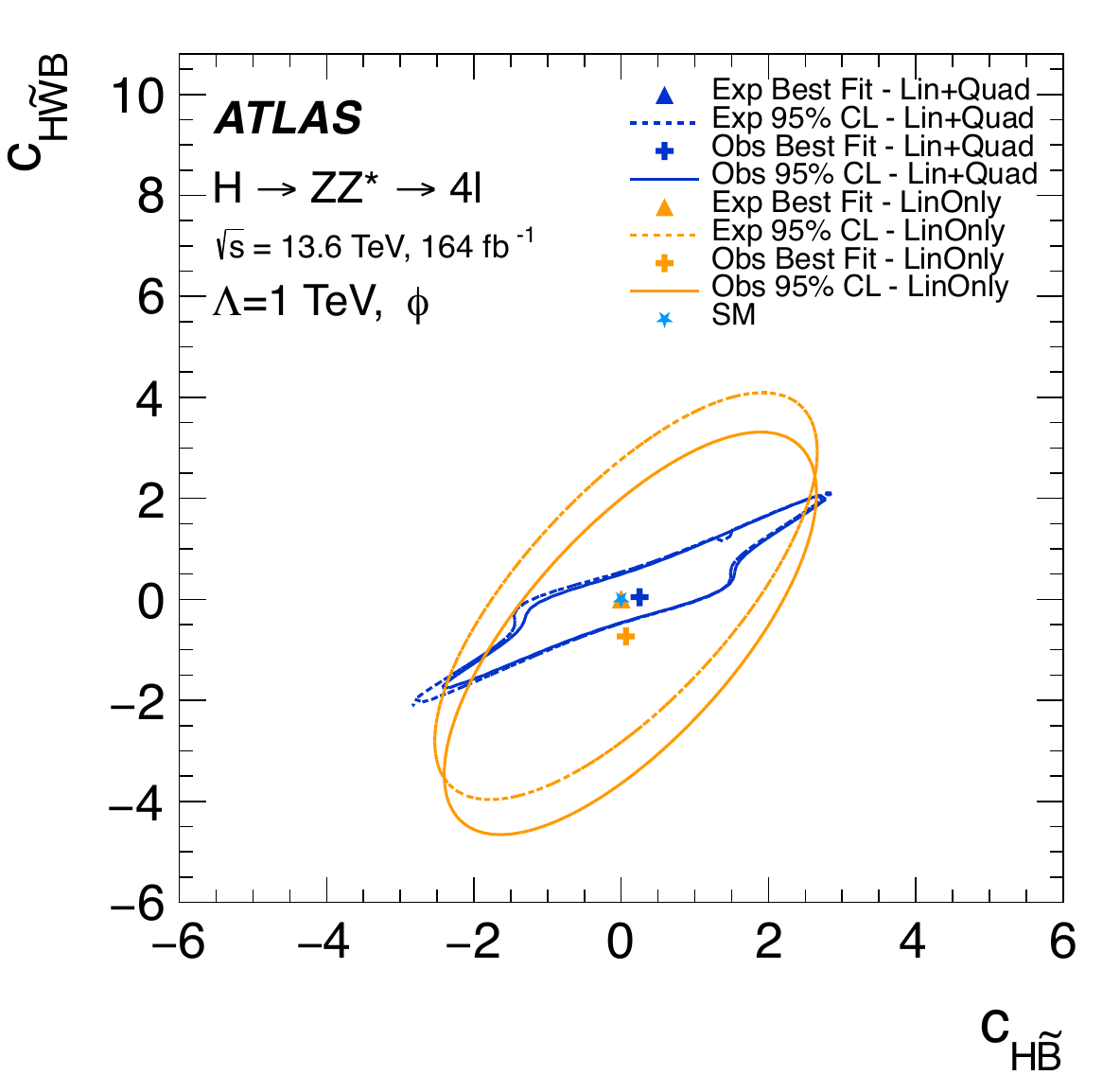}}
\caption{Expected (dashed line) and observed (full line) two-dimensional likelihood scans for the SMEFT Wilson coefficients, comparing the linear-only and linear+quadratic SMEFT treatments. Coupling values outside the solid lines are excluded at $95\%$ confidence level (CL). The best fit to the data (solid cross) and the SM prediction (triangle) are also indicated. A new physics scale of $\Lambda = 1~\TeV$ is assumed. Except for the two fitted Wilson coefficients, all others are set to zero. The CP-even $\cHW~ \text{vs} ~\cHWB$ scan uses the unfolded $(\ptFourL,m_{34})$ measurement, while the other scans use the unfolded $\phi$ measurement.}
\label{fig:eft_scans_2D}
\end{figure}

Since the measurements do not provide sufficient sensitivity to constrain the full set of coefficients simultaneously, fits are performed for one coefficient at a time, with all others set to zero. Two treatments of the SMEFT expansion are presented: a \emph{linear-only} interpretation retaining only the interference term, and a \emph{linear+quadratic} interpretation including both interference and BSM-squared terms. Quadratic terms typically increase sensitivity through additional changes in total cross-section, and a comparison of the two treatments provides information on the impact of higher-order contributions and the possible onset of perturbative breakdown as quadratic effects become dominant.
For each probed coefficient, one-dimensional likelihood scans are performed using the observable providing the best expected sensitivity: $(\ptFourL,m_{34})$ for $c_{HW}$ and $\phi$ for the other Wilson coefficients. The observed and expected confidence intervals (CL) are summarised in Table~\ref{tab:SMEFT_conf_intervals1D}, and the corresponding likelihood scans comparing the linear-only and linear+quadratic treatments are shown for the CP-even Wilson Coefficients in Figure~\ref{fig:eft_scans_1D}. Additional results where combinations of two EFT operators are simultaneously probed are in shown in Figure \ref{fig:eft_scans_2D}. For each combination, the scans are obtained from the observable that provides the best expected sensitivity. Two-dimensional EFT scans are not used to derive limits, as the likelihood exhibits poorly constrained directions in parameter space, leading to degeneracies and flat directions.

\begin{figure}[!htbp]
\centering
\includegraphics[width = 0.75\textwidth]{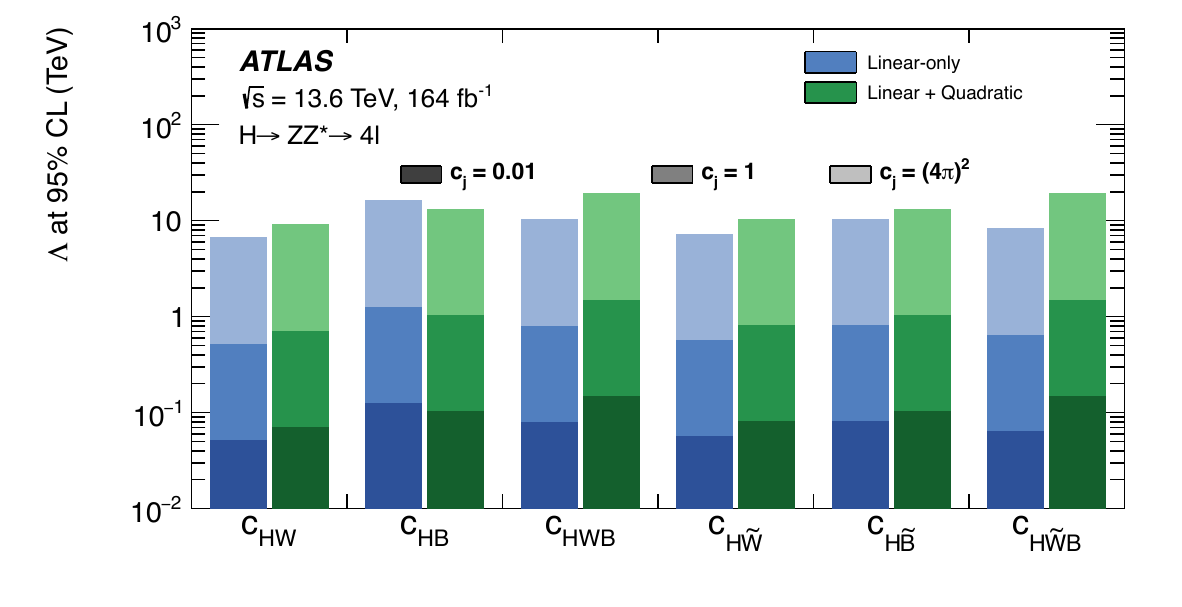}
\caption{Observed lower limits at 95\% confidence intervals (CL) on the new physics scale $\Lambda$ for the indicated benchmark values of a given Wilson coefficient, assuming all others are set to zero. Asymmetric uncertainties are symmetrised using the root mean square. With the sensitivity of the current results, the constraints for $c_j = 0.01$ are outside the validity of the EFT approach.}
\label{fig:lambdaUL}
\end{figure}

The corresponding limits on the new physics scale $\Lambda$ for benchmark coefficient values are shown in Figure~\ref{fig:lambdaUL}.

\begin{table}[htb!]
\centering
\caption{Expected and observed 95\% confidence intervals (CL) on the CP-even and CP-odd SMEFT Wilson coefficients for an integrated luminosity of \intLum at \sqrtS, assuming $\Lambda=1~\TeV$. Only one Wilson coefficient is fitted at a time while all others are set to zero. For each coefficient, results obtained in the linear-only and linear+quadratic SMEFT treatments are presented. $(\ptFourL,m_{34})$ observable is used for $c_{HW}$ results, while $\phi$ is used for the other operators.}
\renewcommand{\arraystretch}{1.25}
\setlength{\tabcolsep}{5pt}


\begin{tabular}{c|cc|cc|cc}
& \multicolumn{2}{c|}{Expected 95\% CL [$\TeV^{-2}$]}
& \multicolumn{2}{c|}{Observed 95\% CL [$\TeV^{-2}$]}
& \multicolumn{2}{c}{Best fit [$\TeV^{-2}$]} \\
\hline
& \multirow{2}*{Linear} & Linear +
& \multirow{2}*{Linear} & Linear +
& \multirow{2}*{Linear} & Linear +  \\
&        & Quadratic
&        & Quadratic
&        & Quadratic \\
\hline
$c_{HW}$
& $[-3.8,3.2]$ & $[-1.8,1.8]$
& $[-4.3,2.9]$ & $[-2.0,1.8]$
& -0.40 & -0.08 \\
\hline
$c_{HB}$
& $[-0.63,0.59]$ & $[-0.54,1.2]$
& $[-0.58,0.64]$ & $[-0.49,1.25]$
& 0.05 & 0.05 \\
\hline
$c_{HWB}$
& $[-1.5,1.4]$ & $[-0.42,0.48]$
& $[-1.6,1.3]$ & $[-0.40,0.45]$
& -0.08 & -0.02 \\
\midrule
\midrule
$c_{H\tilde{W}}$
& $[-3.0,3.0]$ & $[-1.6,1.6]$
& $[-2.1,3.9]$ & $[-1.3,1.6]$
& 0.88 & 0.38 \\
\hline
$c_{H\tilde{B}}$
& $[-1.5,1.5]$ & $[-1.0,0.96]$
& $[-1.1,1.6]$ & $[-0.80,0.96]$
& 0.39 & 0.20 \\
\hline
$c_{H\tilde{W}B}$
& $[-2.3,2.3]$ & $[-0.45,0.45]$
& $[-3.1,1.5]$ & $[-0.44,0.41]$
& -0.80 & -0.09 \\
\end{tabular}

 %
\label{tab:SMEFT_conf_intervals1D}
\end{table}

\FloatBarrier

\subsection{Constraints on Higgs boson self-coupling}
\label{sec:kappa_lambda}

The trilinear Higgs self-coupling, $\lambda_{3}$, enters single-Higgs production and decay at the one-loop level through electroweak corrections, and can therefore be constrained using precision measurements of single-Higgs observables. These effects modify both the overall rate and, in particular, the shape of kinematic distributions that probe hard radiation, making the Higgs boson transverse-momentum spectrum especially sensitive to $\lambda_{3}$~\cite{Maltoni:2017ims}. For this reason, both the differential measurement of $\ptFourL$ in the fiducial phase space and the STXS  measurements, which rely on different event categorisations and modelling assumptions, can be used to probe the effects of the self-coupling, allowing a comparison of the sensitivity achieved by the two complementary approaches.

The interpretation is performed in the $\kappa$-framework by introducing the modifier $\kappa_{\lambda}\equiv\lambda_{3}/\lambda_{3}^{\mathrm{SM}}$, with $\kappa_{\lambda}=1$ corresponding to the SM expectation and setting all other couplings to their SM expectations.

The dependence of an observable $\Sigma$ on the trilinear Higgs self-coupling is parametrised following Ref.~\cite{Maltoni:2017ims} as
\begin{equation}
\Sigma^{\mathrm{BSM}}_{\lambda_{3}}(\kappa_{\lambda})
= Z_{H}^{\mathrm{BSM}} \, \Sigma_{\mathrm{LO}}
\left(1 + \kappa_{\lambda} \, C_{1}(\{p_n\}) + \delta Z_{H}\right),
\label{eq:kLambda}
\end{equation}
where $\Sigma_{\mathrm{LO}}$ denotes the leading-order prediction, including factorisable higher-order QCD corrections. The factor $Z_{H}^{\mathrm{BSM}}$ represents the universal Higgs wave-function renormalisation, which accounts for corrections induced by the modified self-coupling. The coefficient $C_{1}(\{p_n\})$ encodes the finite one-loop contribution arising from the modified trilinear interaction and depends on the kinematics of the process, denoted collectively by $\{p_n\}$.
The factor $Z_{H}^{\mathrm{BSM}}$ is expressed in terms of $\delta Z_{H}$ and represents the NLO term ~\cite{Maltoni:2017ims}, as
\begin{equation}
Z_{H}^{\mathrm{BSM}}=\frac{1}{1-(\kappa_{\lambda}^{2}-1)\,\delta Z_{H}}\,.
\end{equation}
In this interpretation, all other coupling modifiers are fixed to their SM values, and the $\kappa_{\lambda}$ dependence is implemented as a reweighting of the expected signal yields in each analysis bin, using $C_{1}$ values evaluated in the relevant fiducial or STXS phase space. The $C_{1}$ coefficients are obtained with the automated reweighting procedure of Ref.~\cite{Maltoni:2017ims} for all production-modes supported by the tool, and supplemented for $\ggF$ using the dedicated calculation at $\sqrt{s}=13.6~\TeV$~\cite{Haisch:2024nzv}. The Higgs boson decay contribution is taken to be weakly dependent on kinematics~\cite{Maltoni:2017ims}, and the corresponding total-width contribution is taken from Ref.~\cite{Degrassi:2016wml}. The range of the validity for the given $C_{1}$ and $\delta Z_{H}$ is $|\kappa_{\lambda}| <$ 20.

\begin{figure}[htbp]
\centering
\includegraphics[width = 0.50 \textwidth]{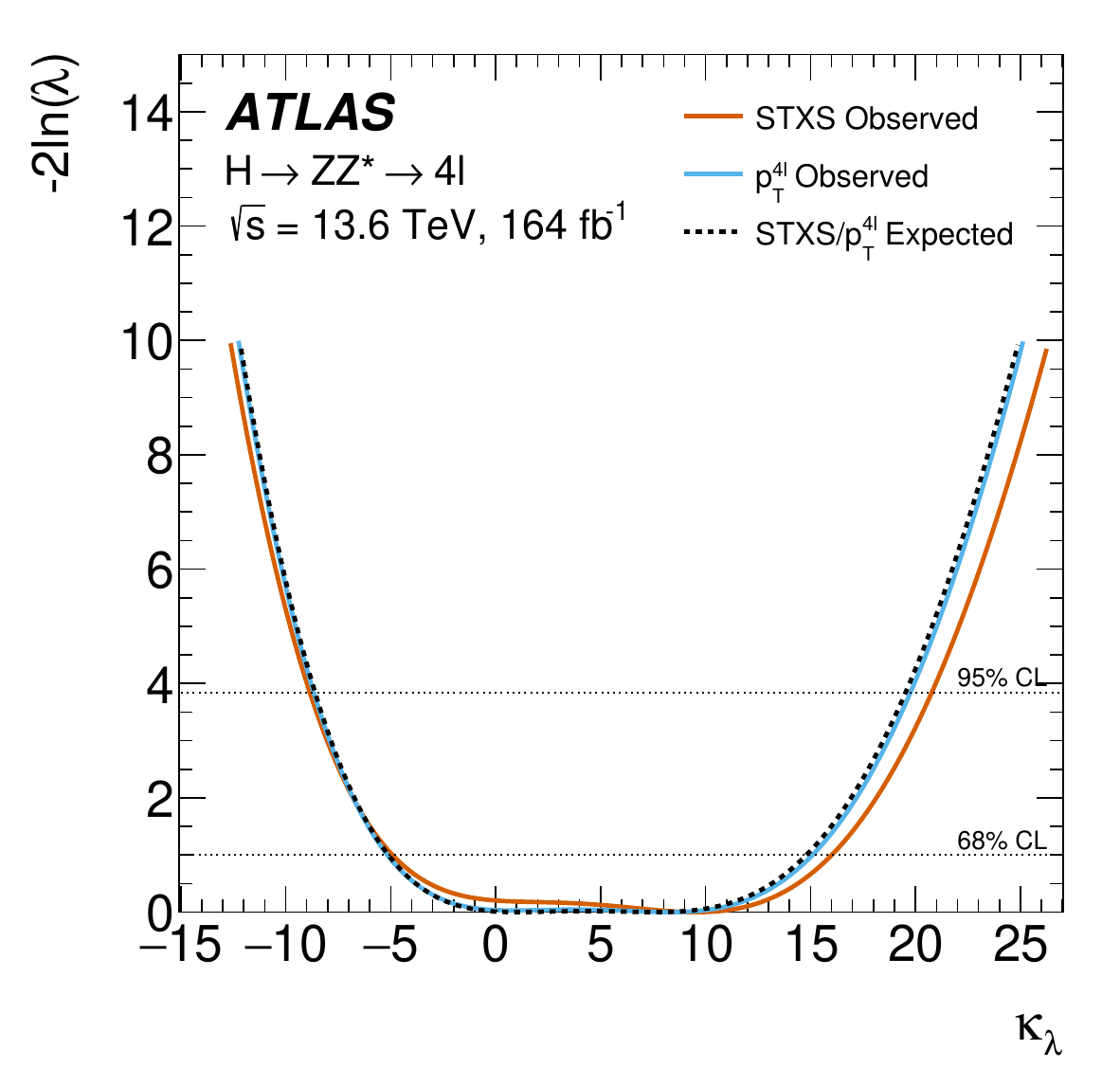}
\caption{Observed negative-log-likelihood scan as a function of the trilinear Higgs self-coupling modifier $\kappa_{\lambda}$, comparing the constraints obtained from the STXS production-mode measurement (orange line) and from the fiducial differential measurement based on $\ptFourL$ (light blue line). The expected results are plotted together for measurements due to their similar sensitivity (black dashed line). The curves are shown relative to their respective minima. The Standard Model prediction corresponds to $\kappa_{\lambda}=1$. \label{fig:kappa_lambda}}

\end{figure}

Complementary constraints on $\kappa_{\lambda}$ are derived from the differential $\ptFourL$ and the STXS production-mode measurements.
The corresponding likelihood scans are compared in Figure~\ref{fig:kappa_lambda}, yielding observed 95\% CL intervals of $[-8.8,\,19.9]$ and $[-9.0,\,20.9]$, respectively, compatible with expected 95\% CL intervals of $[-8.8,\,19.9]$ . The double-minimum structure of the likelihood arises from the quadratic dependence of the cross-section on $\kappa_{\lambda}$. The observed limit from the STXS measurement is slightly weaker than that obtained from the differential $\ptFourL$ measurement. This difference is driven by the strong sensitivity of the $\kappa_{\lambda}$ interpretation to the \ggF production mode and by the fact that the observed \ggF signal strength in the STXS measurement is lower than the inclusive signal strength.

\FloatBarrier


%
\section{Conclusion}
\label{sec:conclusion}

A measurement of inclusive and differential Higgs boson production cross-sections in the $\hfourl$ channel is presented using \intLum of \sqrtS \pp~collision data recorded by the ATLAS detector at the LHC between 2022 and 2024.
These results constitute the most comprehensive ATLAS measurements of Higgs boson production in the $\hfourl$ channel at $\sqrt{s}=13.6~\TeV$ to date, with a precision already comparable to that obtained using the full Run~2 dataset at $\sqrt{s}=13~\TeV$. The analysis exploits the data using improved analysis and ML techniques in two complementary frameworks, measuring fiducial cross-sections with minimal theoretical assumptions and production-mode and STXS cross-sections to separate the contributions from different Higgs boson production mechanisms.
The inclusive fiducial cross-section is measured to be \fidObsShort, in agreement with the Standard Model prediction of $\fidPred$, and is extrapolated to a total cross-section of \fidTotShort, also consistent with the Standard Model and previous results. Differential cross-sections are measured as a function of several Higgs boson production and decay observables and are found to be in agreement with state-of-the-art SM predictions.
Higgs boson candidate events are categorised into regions sensitive to different production mechanisms and are used to measure production-mode cross-sections and a Reduced STXS Stage~1.2 scheme. A product of inclusive production cross-section and branching ratio into $ZZ^\ast$ of \stxsIncObsShort is obtained for Higgs boson rapidities below $2.5$, consistent with the Standard Model prediction of \stxsIncPred.
The precision of both the differential and STXS measurements is comparable to that achieved by ATLAS at $\sqrt{s}=13~\TeV$ with Run~2 data~\cite{HIGG-2018-28,HIGG-2018-29}, and improves by approximately a factor of two relative to the first ATLAS results at $\sqrt{s}=13.6~\TeV$ in this channel~\cite{HIGG-2022-12}.
The differential measurements as a function of $\phi$ and $\pTmthreefour$ are used to constrain effective field theory operators and are found to be compatible with SM expectations.
The production-mode results are further interpreted in the $\kappa$ framework and combined with Run~2 measurements~\cite{HIGG-2018-28}, yielding best-fit values consistent with Run~2 and about 40\% (25\%) improvement on $\kappa_V$ ($\kappa_f$) limits.
Limits are also derived on the trilinear Higgs self-coupling, and are consistent with the Standard Model.


%

%
%
%

%
%

%
%
%
%

%

%
%
%

%
%
%
%

%

%
%
%
%
%
\section*{Acknowledgements}
%

%

%
%

%
%

We thank CERN for the very successful operation of the LHC and its injectors, as well as the support staff at
CERN and at our institutions worldwide without whom ATLAS could not be operated efficiently.

The crucial computing support from all WLCG partners is acknowledged gratefully, in particular from CERN, the ATLAS Tier-1 facilities at TRIUMF/SFU (Canada), NDGF (Denmark, Norway, Sweden), CC-IN2P3 (France), KIT/GridKA (Germany), INFN-CNAF (Italy), NL-T1 (Netherlands), PIC (Spain), RAL (UK) and BNL (USA), the Tier-2 facilities worldwide and large non-WLCG resource providers. Major contributors of computing resources are listed in Ref.~\cite{ATL-SOFT-PUB-2026-001}.

We gratefully acknowledge the support of ANPCyT, Argentina; YerPhI, Armenia; ARC, Australia; BMWFW and FWF, Austria; ANAS, Azerbaijan; CNPq and FAPESP, Brazil; NSERC, NRC and CFI, Canada; CERN; ANID, Chile; CAS, MOST, NSFC and FRFCU, China; Minciencias, Colombia; MEYS CR, Czech Republic; DNRF and DNSRC, Denmark; IN2P3-CNRS and CEA-DRF/IRFU, France; SRNSFG, Georgia; BMFTR, HGF and MPG, Germany; GSRI, Greece; RGC and Hong Kong SAR, China; ICHEP and Academy of Sciences and Humanities, Israel; INFN, Italy; MEXT and JSPS, Japan; CNRST, Morocco; NWO, Netherlands; RCN, Norway; MNiSW, Poland; FCT, Portugal; MNE/IFA, Romania; MSTDI, Serbia; MSSR, Slovakia; ARIS and MVZI, Slovenia; DSI/NRF, South Africa; MICIU/AEI, Spain; SRC and Wallenberg Foundation, Sweden; SERI, SNSF and Cantons of Bern and Geneva, Switzerland; NSTC, Taipei; TENMAK, T\"urkiye; STFC/UKRI, United Kingdom; DOE and NSF, United States of America.

Individual groups and members have received support from BCKDF, CANARIE, CRC and DRAC, Canada; CERN-CZ, FORTE and PRIMUS, Czech Republic; COST, ERC, ERDF, Horizon 2020 and Marie Sk{\l}odowska-Curie Actions, European Union; Investissements d'Avenir Labex, Investissements d'Avenir Idex and ANR, France; DFG and AvH Foundation, Germany; Herakleitos, Thales and Aristeia programmes co-financed by EU-ESF and the Greek NSRF, Greece; BSF-NSF and MINERVA, Israel; NCN and NAWA, Poland; La Caixa Banking Foundation, CERCA and AGAUR programs from Generalitat de Catalunya and PROMETEO and GenT Programmes Generalitat Valenciana, Spain; G\"{o}ran Gustafssons Stiftelse, Sweden; The Royal Society and Leverhulme Trust, United Kingdom; Eric and Wendy Schmidt Fund for Strategic Innovation, United States of America.

In addition, individual members wish to acknowledge support from Chile: Agencia Nacional de Investigaci\'on y Desarrollo (ANID FONDECYT reg. 1230987, FONDECYT 1230812, FONDECYT 1240864, Fondecyt 3240661, Fondecyt Regular 1240721); China: Fundamental Research Funds for the Central Universities (010-63263105), Chinese Ministry of Science and Technology (MOST-2023YFA1605700, MOST-2023YFA1609300), National Natural Science Foundation of China (NSFC 12275265, NSFC-W2543005); Czech Republic: Czech Science Foundation (GACR - 24-11373S), Ministry of Education Youth and Sports (ERC-CZ-LL2327, FORTE CZ.02.01.01/00/22\_008/0004632), PRIMUS Research Programme (PRIMUS/21/SCI/017); EU: H2020 European Research Council (ERC - 101002463); European Union: European Research Council (BARD No. 101116429, ERC - 101219398, ERC - 948254, ERC 101089007), European Regional Development Fund (HE COFUND GA No.101081355, ERDF), Marie Sklodowska-Curie Actions (GAP-101168829); France: Agence Nationale de la Recherche (ANR-21-CE31-0013, ANR-22-EDIR-0002, ANR-24-CE31-0504-01); Germany: Deutsche Forschungsgemeinschaft (DFG - 469666862); China: Research Grants Council (GRF); Italy: Istituto Nazionale di Fisica Nucleare (LHC-MIUR - 28003/2025), Ministero dell'Università e della Ricerca (NextGenEU 153D23001490006 M4C2.1.1, NextGenEU I53D23000820006 M4C2.1.1, NextGenEU I53D23001490006 M4C2.1.1, SOE2024\_0000023); Japan: Japan Society for the Promotion of Science (JSPS KAKENHI  JP25H0063, JSPS KAKENHI JP22H01227, JSPS KAKENHI JP22H04944, JSPS KAKENHI JP22KK0227, JSPS KAKENHI JP24K23939, JSPS KAKENHI JP24KK0251, JSPS KAKENHI JP25H00650, JSPS KAKENHI JP25H01291, JSPS KAKENHI JP25K01011, JSPS KAKENHI JP25K01023); Poland: Polish National Science Centre (NCN 2021/42/E/ST2/00350, NCN OPUS 2023/51/B/ST2/02507, NCN OPUS nr 2022/47/B/ST2/03059, NCN UMO-2019/34/E/ST2/00393, UMO-2022/47/O/ST2/00148, UMO-2023/49/B/ST2/04085, UMO-2023/51/B/ST2/00920, UMO-2024/53/N/ST2/00869); Spain: Agència de Gestió d'Ajuts Universitaris i de Recerca. (AGAUR - 2023 BP 00141), Ministry of Science and Innovation (RYC2019-028510-I, RYC2020-030254-I, RYC2021-031273-I, RYC2022-038164-I), Ministerio de Ciencia, Innovación y Universidades/Agencia Estatal de Investigaci\'on (EU NextGenerationEU (PRTR-C17.I1), PID2022-142604OB-C22); Sweden: Carl Trygger Foundation (Carl Trygger Foundation CTS 22:2312), Swedish Research Council (Swedish Research Council 2023-04654, VR 2021-03651, VR 2022-03845, VR 2022-04683, VR 2023-03403, VR 2024-05451, VR 2025-05940), Knut and Alice Wallenberg Foundation (KAW 2023.0366); Switzerland: Swiss National Science Foundation (SNSF - PCEFP2\_194658); United Kingdom: The Binks Trust, Royal Society (NIF-R1-231091); United States of America: U.S. Department of Energy (ECA DE-AC02-76SF00515), John Templeton Foundation (John Templeton Foundation 63206), Neubauer Family Foundation.

%
%


%
%
%
%
%
%

%
%

%
%
\printbibliography
\clearpage
 
\begin{flushleft}
\hypersetup{urlcolor=black}
{\Large The ATLAS Collaboration}

\bigskip

\AtlasOrcid[0000-0002-6665-4934]{G.~Aad}$^\textrm{\scriptsize 102}$,
\AtlasOrcid[0000-0001-7616-1554]{E.~Aakvaag}$^\textrm{\scriptsize 17}$,
\AtlasOrcid[0000-0002-5888-2734]{B.~Abbott}$^\textrm{\scriptsize 121}$,
\AtlasOrcid[0000-0002-0287-5869]{S.~Abdelhameed}$^\textrm{\scriptsize 83b}$,
\AtlasOrcid[0000-0002-1002-1652]{K.~Abeling}$^\textrm{\scriptsize 54}$,
\AtlasOrcid[0000-0001-5763-2760]{N.J.~Abicht}$^\textrm{\scriptsize 48}$,
\AtlasOrcid[0000-0002-8496-9294]{S.H.~Abidi}$^\textrm{\scriptsize 30}$,
\AtlasOrcid[0009-0003-6578-220X]{M.~Aboelela}$^\textrm{\scriptsize 44}$,
\AtlasOrcid[0000-0002-9987-2292]{A.~Aboulhorma}$^\textrm{\scriptsize 36e}$,
\AtlasOrcid[0000-0001-5329-6640]{H.~Abramowicz}$^\textrm{\scriptsize 154}$
% ...
\bigskip
\\

% ...
$^{17}$Department for Physics and Technology, University of Bergen, Bergen; Norway.\\
% ...
$^{30}$Physics Department, Brookhaven National Laboratory, Upton NY; United States of America.\\
% ...
$^{36}$$^{(a)}$Facult\'e des Sciences Ain Chock, Universit\'e Hassan II de Casablanca;$^{(b)}$Facult\'{e} des Sciences, Universit\'{e} Ibn-Tofail, K\'{e}nitra;$^{(c)}$Facult\'e des Sciences Semlalia, Universit\'e Cadi Ayyad, LPHEA-Marrakech;$^{(d)}$LPMR, Facult\'e des Sciences, Universit\'e Mohamed Premier, Oujda;$^{(e)}$Facult\'e des sciences, Universit\'e Mohammed V, Rabat;$^{(f)}$Institute of Applied Physics, Mohammed VI Polytechnic University, Ben Guerir; Morocco.\\
% ...
$^{44}$Physics Department, Southern Methodist University, Dallas TX; United States of America.\\
% ...
$^{48}$Fakult\"{a}t Physik, Technische Universit{\"a}t Dortmund, Dortmund; Germany.\\
% ...
$^{54}$II. Physikalisches Institut, Georg-August-Universit\"{a}t G\"ottingen, G\"ottingen; Germany.\\
% ...
$^{83}$$^{(a)}$Khalifa University of Science and Technology, Abu Dhabi;$^{(b)}$New York University Abu Dhabi, Abu Dhabi;$^{(c)}$United Arab Emirates University, Al Ain;$^{(d)}$University of Sharjah, Sharjah; United Arab Emirates.\\
% ...
$^{102}$CPPM, Aix-Marseille Universit\'e, CNRS/IN2P3, Marseille; France.\\
% ...
$^{121}$Homer L. Dodge Department of Physics and Astronomy, University of Oklahoma, Norman OK; United States of America.\\
% ...
$^{154}$Raymond and Beverly Sackler School of Physics and Astronomy, Tel Aviv University, Tel Aviv; Israel.\\
% ...
$^{a}$ Also at Affiliated with an institute formerly covered by a cooperation agreement with CERN.\\
$^{b}$ Also at An-Najah National University, Nablus; Palestine.\\
$^{c}$ Also at Borough of Manhattan Community College, City University of New York, New York NY; United States of America.\\
$^{d}$ Also at Center for Interdisciplinary Research and Innovation (CIRI-AUTH), Thessaloniki; Greece.\\
$^{e}$ Also at Centre of Physics of the Universities of Minho and Porto (CF-UM-UP); Portugal.\\
$^{f}$ Also at CERN, Geneva; Switzerland.\\
% ...

\end{flushleft}


%

%
%

%

%
%
%
%

\end{document}